# C O R E

## COMPENDIUM of RELATIONS

*Version 2.1*

V.I. Borodulin[1], R.N. Rogalyov[2], S.R. Slabospitsky[3]

*Institute for High Energy Physics*
*Protvino, Moscow Region, 142284*
*RUSSIA*

Protvino 1995

[1]E–mail: BORODULIN@mx.ihep.su
[2]E–mail: ROGALYOV@mx.ihep.su
[3]E–mail: SLABOSPITSKY@mx.ihep.su

# PREFACE

The present CORE 2.1 (COmpendium of RElations, *Version 2.1*) contains various formulas and relations used in the practical calculations in the Standard Model.

The properties of the Pauli, Dirac, and Gell–Mann matrices, widely used in calculations in the Standard Model, are considered in details. Properties of the wave functions of free fermions and gauge bosons are also discussed.

We present the full Lagrangian of the Standard Model and the corresponding Feynman rules. The method of the evaluation of the Feynman (loop) integrals and calculations in non-covariant gauges is considered.

We discuss in brief the relativistic kinematic and present a large number of the matrix elements of the various processes in the Standard Model.

Almost all of the presented relations and formulas can be found in literature. However, one can find different definitions, different normalizations, not widely used notations, etc. We try to collect various expressions in one place and make the notations and normalizations consistent.

We hope that the present CORE 2.1 will be useful for practical calculations in the Standard Model, especially for post–graduates and young physicists.

We wish to thank A.V. Razumov for multiple and useful discussions.

*Protvino, 1995*

# Contents

# 1 NOTATION AND DEFINITIONS

• Throughout this article units are used in which $\hbar = c = 1$.
• Everywhere the *repeated* indexes imply the summation.
• $\delta^{ik} = \delta_{ik} = \delta^i_k = \delta^k_i$ $(i, k = 1, \dots, n)$ is Kronecker symbol,

$$\delta^{ik} = 0 \quad i \neq k, \quad \delta^{11} = \delta^{22} = \dots \delta^{nn} = 1.$$

Metric tensor in Minkowski space $g^{\mu\nu} = g_{\mu\nu}$ $(\mu, \nu = 0, 1, \dots, n)$ equals:

$$g^{\mu\nu} = 0 \quad \mu \neq \nu, \quad g^{00} = 1, \quad g^{11} = g^{22} = \dots = g^{nn} = -1.$$

The tensor $g^{\mu\nu}$ is used for raising and lowering of the Lorentz subscripts and superscripts.
• The scalar products of any two $p$ and $q$ vectors (both in Euclidean and in Minkowski spaces) is denoted as follows:

$$p \cdot q \quad \text{or} \quad (pq), \quad \text{i.e.} \quad p \cdot q = (pq).$$

The scalar products of any two $p$ and $q$ Euclidean vectors would be also denoted as:

$$\vec{p}\vec{q}, \quad \text{i.e.} \quad \vec{p}\vec{q} = p \cdot q = (pq).$$

• 4–vector $p^\mu$ in Minkowski space is given by

$$p^\mu \equiv (E, \vec{p}) \;=\; (p_0, p_1, p_2, p_3), \;\; p_\mu = g_{\mu\nu} p^\nu \;=\; (E, -\vec{p}).$$

The scalar product of any two vectors $p$ and $q$ in Minkowski space is given by

$$p^\mu g_{\mu\nu} q^\nu \;=\; p^\mu q_\mu \;=\; p_\nu q^\nu \;=\; p_0 q_0 - p_1 q_1 - p_2 q_2 - p_3 q_3.$$

The products of the 4–vector $p^\mu$ with Dirac $\gamma^\mu$ matrix denotes as usual

$$\hat{p} \;\equiv\; p^\mu g_{\mu\nu} \gamma^\nu \;=\; p^\mu \gamma_\mu \;=\; p_\nu \gamma^\nu.$$

• **Totally antisymmetric tensor** $\varepsilon^{AB\dots N}$.
• *$\varepsilon$-symbol in two dimensions*: $\varepsilon^{AB}$ $(A, B = 1, 2)$:

$$\varepsilon_{12} = \varepsilon^{12} = 1; \; \varepsilon_{21} = \varepsilon^{21} = -1; \; \varepsilon_{AB} \;=\; \begin{pmatrix} 0 & 1 \\ -1 & 0 \end{pmatrix};$$

$$\varepsilon^{AB} = \varepsilon_{AB}, \; \varepsilon_{BA} = -\varepsilon_{AB}, \; \varepsilon_{AB}\varepsilon^{AB} \;=\; 2; \; \varepsilon_{AB}\varepsilon^{BC} \;=\; -\delta^C_A;$$

$$\varepsilon_{AB}\varepsilon^{CD} \;=\; \delta^C_A \delta^D_B \;-\; \delta^D_A \delta^C_B;$$

$$\varepsilon_{AB}\varepsilon_{CD} + \varepsilon_{AC}\varepsilon_{DB} + \varepsilon_{AD}\varepsilon_{BC} = \; 0.$$

$\varepsilon^{AB}$–symbol is used for rising and lowering of the spinor indexes (see Subsection 2.5).

• *$\varepsilon$-symbol in three dimensions*: $\varepsilon^{ijk}$ $(i,j,k=1,2,3)$:

$$\varepsilon^{123}=\varepsilon_{123}=1, \quad \varepsilon^{ijk}=\varepsilon_{ijk}, \quad \varepsilon_{ijk}\varepsilon^{lmn}=\begin{vmatrix} \delta^l_i & \delta^l_j & \delta^l_k \\ \delta^m_i & \delta^m_j & \delta^m_k \\ \delta^n_i & \delta^n_j & \delta^n_k \end{vmatrix}$$

$$\varepsilon_{ijk}\varepsilon^{lmk}=\delta^l_i\delta^m_j-\delta^m_i\delta^l_j, \quad \varepsilon_{ijk}\varepsilon^{ljk}=2\,\delta^l_i, \quad \varepsilon_{ijk}\varepsilon^{ijk}=6.$$

*Schouten identity.* For any 3–vector $p^i$ one has:

$$p_{i_1}\varepsilon_{i_2i_3i_4}-p_{i_2}\varepsilon_{i_1i_3i_4}+p_{i_3}\varepsilon_{i_1i_2i_4}-p_{i_4}\varepsilon_{i_1i_2i_3} \;=\; 0.$$

• *$\varepsilon$-symbol in four–dimensional Minkowski space*: $\varepsilon^{\alpha\beta\mu\nu}$ $(\alpha,\dots\nu=0,1,2,3)$:

$$\varepsilon^{0123} \;=\; -\,\varepsilon_{0123} \;=\; 1.$$

$$\varepsilon_{\mu\nu\alpha\beta}\varepsilon^{\lambda\rho\sigma\tau}=-\begin{vmatrix} \delta^\lambda_\mu & \delta^\lambda_\nu & \delta^\lambda_\alpha & \delta^\lambda_\beta \\ \delta^\rho_\mu & \delta^\rho_\nu & \delta^\rho_\alpha & \delta^\rho_\beta \\ \delta^\sigma_\mu & \delta^\sigma_\nu & \delta^\sigma_\alpha & \delta^\sigma_\beta \\ \delta^\tau_\mu & \delta^\tau_\nu & \delta^\tau_\alpha & \delta^\tau_\beta \end{vmatrix}, \quad \varepsilon_{\mu\nu\alpha\beta}\varepsilon^{\lambda\rho\sigma\beta}=-\begin{vmatrix} \delta^\lambda_\mu & \delta^\lambda_\nu & \delta^\lambda_\alpha \\ \delta^\rho_\mu & \delta^\rho_\nu & \delta^\rho_\alpha \\ \delta^\sigma_\mu & \delta^\sigma_\nu & \delta^\sigma_\alpha \end{vmatrix},$$

$$\varepsilon_{\mu\nu\alpha\beta}\varepsilon^{\lambda\rho\alpha\beta} \;=\; -2(\delta^\lambda_\mu\delta^\rho_\nu-\delta^\rho_\mu\delta^\lambda_\nu), \quad \varepsilon_{\mu\nu\alpha\beta}\varepsilon^{\lambda\nu\alpha\beta} \;=\; -6\delta^\lambda_\mu, \quad \varepsilon_{\mu\nu\alpha\beta}\varepsilon^{\mu\nu\alpha\beta} \;=\; -24.$$

*Schouten identity.* For any 4–vector $p_\mu$ one has:

$$p_{\mu_1}\varepsilon_{\mu_2\mu_3\mu_4\mu_5}+p_{\mu_2}\varepsilon_{\mu_3\mu_4\mu_5\mu_1}+p_{\mu_3}\varepsilon_{\mu_4\mu_5\mu_1\mu_2}+p_{\mu_4}\varepsilon_{\mu_5\mu_1\mu_2\mu_3}+p_{\mu_5}\varepsilon_{\mu_1\mu_2\mu_3\mu_4}=0.$$

• *Generalized Kronecker deltas*

Sometimes one can make no difference between a vector and index. For example, one can write:

$$\varepsilon^{p_1p_2p_3p_4} \text{ or } \varepsilon(p_1,p_2,p_3,p_4) \quad \text{instead of } \varepsilon_{\mu\nu\rho\sigma}p_1^\mu p_2^\nu p_3^\rho p_4^\sigma.$$

These notation can be used in operations with generalized Kronecker deltas:

$$\delta^{j_1\dots j_n}_{i_1\dots i_n}\equiv\begin{vmatrix} \delta^{j_1}_{i_1} & \dots & \delta^{j_1}_{i_n} \\ \dots & \dots & \dots \\ \delta^{j_1}_{i_n} & \dots & \delta^{j_n}_{i_n} \end{vmatrix}, \text{ or } \delta^{q_1\dots q_n}_{p_1\dots p_n}\equiv\begin{vmatrix} p_1\cdot q_1 & \dots & p_n\cdot q_1 \\ \dots & \dots & \dots \\ p_1\cdot q_n & \dots & p_n\cdot q_n \end{vmatrix}.$$

In $n$-dimensional Euclidean space one has:

$$\delta^{q_1\ldots q_m}_{p_1\ldots p_m} = \frac{1}{(n-m)!}\varepsilon^{q_1\ldots q_m\alpha_{m+1}\ldots\alpha_n}\varepsilon_{p_1\ldots p_m\alpha_{m+1}\ldots\alpha_n}.$$

In Minkowski space the minus sign appears:

$$\delta^{q_1q_2q_3}_{p_1p_2p_3} = -\varepsilon_{p_1p_2p_3\mu}\varepsilon^{q_1q_2q_3\mu}, \quad \delta^{q_1q_2}_{p_1p_2} = -\frac{1}{2}\varepsilon_{p_1p_2\mu\nu}\varepsilon^{q_1q_2\mu\nu}.$$

• *Matrices*

For any matrix $A = (a_{ik}) \quad (i,k = 1,\ldots n)$ we use the following notation:
$I$ is the *unit* matrix, i.e. $I = \delta_{ik}$ (sometimes, the unit matrix will be denote just 1);
$A^{-1}$ is the *inverse* matrix, i.e. $A^{-1}A = AA^{-1} = I$;
$A^\top$ is the *transposed* matrix, i.e. $a^\top_{ik} = a_{ki}$;
$A^*$ is the *complex conjugated* matrix, i.e. $(a^*)_{ik} = (a_{ik})^*$;
$A^\dagger$ is the *Hermitian conjugated* matrix, i.e. $a^\dagger_{ik} = a^*_{ki}$;
$H$ – *Hermitian* and $U$ – *unitary* matrices should satisfy the following conditions:

$$H^\dagger = H,$$
$$U = (U^\dagger)^{-1}, \quad \text{hence} \quad U^\dagger = U^{-1}, \quad UU^\dagger = U^\dagger U = I.$$

$\det A$ is the *determinant* of matrix $A$

$$\det A = \varepsilon^{i_1i_2\ldots i_n}a_{i_11}a_{i_22}\cdots a_{i_nn} = \frac{1}{n!}\varepsilon^{i_1i_2\ldots i_n}\varepsilon^{k_1k_2\ldots k_n}a_{i_1k_1}a_{i_2k_2}\cdots a_{i_nk_n}.$$

$\mathrm{Tr}A$ is the *trace* of matrix $A$ : $\mathrm{Tr}A = a_{ii}\,(= \sum_{i=1}^n a_{ii})$. The chief properties of the trace are as follows (below $\lambda$ and $\mu$ are parameters):

$$\mathrm{Tr}(\lambda A + \mu B) = \lambda\mathrm{Tr}A + \mu\mathrm{Tr}B,$$
$$\mathrm{Tr}A^\top = \mathrm{Tr}A, \quad \mathrm{Tr}A^* = \mathrm{Tr}A^\dagger = (\mathrm{Tr}A)^*, \quad \mathrm{Tr}I = n,$$
$$\mathrm{Tr}(AB) = \mathrm{Tr}(BA), \qquad \det(e^A) = e^{\mathrm{Tr}A}.$$

For any two matrices $A$ and $B$ the *commutator* $[A,B]$ and *anticommutator* $\{A,B\}$ are denoted as usual:

$$[A,B] \equiv AB - BA, \quad \{A,B\} \equiv AB + BA.$$

# 2 PAULI MATRICES

## 2.1 *Main Properties*

The Pauli matrices $\sigma_i$ ($i = 1, 2, 3$) are generators of the group $SU(2)$. The $\sigma_i$ are equal [1, 2, 3, 4]:

$$\sigma_1 = \begin{pmatrix} 0 & 1 \\ 1 & 0 \end{pmatrix}, \quad \sigma_2 = \begin{pmatrix} 0 & -i \\ i & 0 \end{pmatrix}, \quad \sigma_3 = \begin{pmatrix} 1 & 0 \\ 0 & -1 \end{pmatrix}.$$

The main properties of $\sigma_i$ are as follows:

$$\begin{aligned} \sigma_i^\dagger &= \sigma_i, \quad \mathrm{Tr}\sigma_i = 0, \quad \det\sigma_i = -1, \\ \sigma_i\sigma_k &= i\varepsilon_{ikj}\sigma_j + \delta_{ik}. \end{aligned} \tag{2.1}$$

Using relation (2.1), one gets:

$$\begin{aligned} &\sigma_i^2 = I, \quad [\sigma_i, \sigma_k] = 2i\varepsilon_{ikj}\sigma_j, \quad \{\sigma_i, \sigma_k\} = 2\delta_{ik}, \\ &\sigma_i\sigma_k\sigma_l = i\varepsilon_{ikl}I + \delta_{ik}\sigma_l - \delta_{il}\sigma_k + \delta_{kl}\sigma_i, \\ &\mathrm{Tr}(\sigma_i\sigma_k) = 2\delta_{ik}, \quad \mathrm{Tr}(\sigma_i\sigma_k\sigma_l) = 2i\varepsilon_{ikl}, \\ &\mathrm{Tr}(\sigma_i\sigma_k\sigma_l\sigma_m) = 2(\delta_{ik}\delta_{lm} + \delta_{im}\delta_{kl} - \delta_{il}\delta_{km}). \end{aligned}$$

## 2.2 *Fiertz Identities*

The Fiertz identities for the Pauli matrices have the form:

$$\sigma^i_{AB}\sigma^i_{CD} = 2\delta_{AD}\delta_{CB} - \delta_{AB}\delta_{CD}, \tag{2.2}$$

$$\sigma^i_{AB}\sigma^i_{CD} = \frac{3}{2}\delta_{AD}\delta_{CB} - \frac{1}{2}\sigma^i_{AD}\sigma^i_{CB}. \tag{2.3}$$

Using (2.2), one can obtain the following relations:

$$\begin{aligned} \delta_{AB}\sigma^i_{CD} &= \frac{1}{2}[\delta_{AD}\sigma^i_{CB} + \sigma^i_{AD}\delta_{CB} + i\varepsilon^{ikl}\sigma^k_{AD}\sigma^l_{CB}], \\ \sigma^i_{AB}\delta_{CD} &= \frac{1}{2}[\delta_{AD}\sigma^i_{CB} + \sigma^i_{AD}\delta_{CB} - i\varepsilon^{ikl}\sigma^k_{AD}\sigma^l_{CB}], \\ \delta_{AB}\sigma^i_{CD} &+ \sigma^i_{AB}\delta_{CD} = \sigma^i_{AD}\delta_{CB} + \delta_{AD}\sigma^i_{CB}, \\ \sigma^i_{AB}\sigma^k_{CD} &= \frac{1}{2}[\sigma^i_{AD}\sigma^k_{CB} + \delta^{ik}\delta_{AD}\delta_{CB} - \delta^{ik}\sigma^l_{AD}\sigma^l_{CB} + \\ &+ i\varepsilon^{ikl}\sigma^l_{AD}\delta_{CB} - i\varepsilon^{ikl}\delta_{AD}\sigma^l_{CB}]. \end{aligned}$$

## 2.3 $\sigma_+$ and $\sigma_-$ Matrices

The $\sigma_+$ and $\sigma_-$ matrices are defined as follows:

$$\sigma_+ \equiv \frac{1}{2}(\sigma_1 + i\sigma_2) = \begin{pmatrix} 0 & 1 \\ 0 & 0 \end{pmatrix}, \quad \sigma_- \equiv \frac{1}{2}(\sigma_1 - i\sigma_2) = \begin{pmatrix} 0 & 0 \\ 1 & 0 \end{pmatrix}.$$

The relations for these matrices are given by

$$\begin{aligned}
&(\sigma_\pm)^\dagger = \sigma_\mp, \quad \mathrm{Tr}\sigma_\pm = 0, \quad \det\sigma_\pm = 0, \\
&[\sigma_\pm, \sigma_1] = \pm\sigma_3, \quad [\sigma_\pm, \sigma_2] = i\sigma_3, \quad [\sigma_\pm, \sigma_3] = \mp 2\sigma_\pm, \quad [\sigma_+, \sigma_-] = \sigma_3, \\
&\{\sigma_\pm, \sigma_1\} = I, \quad \{\sigma_\pm, \sigma_2\} = \pm iI, \quad \{\sigma_\pm, \sigma_3\} = 0, \quad \{\sigma_+, \sigma_-\} = I, \\
&\sigma_+^2 = \sigma_-^2 = 0, \quad \sigma_+\sigma_3 = -\sigma_+, \quad \sigma_3\sigma_+ = \sigma_+, \\
&\sigma_+\sigma_- = \frac{1}{2}(I + \sigma_3), \quad \sigma_-\sigma_+ = \frac{1}{2}(I - \sigma_3), \\
&(\sigma_+\sigma_-)^n = \sigma_+\sigma_-, \quad (\sigma_-\sigma_+)^n = \sigma_-\sigma_+.
\end{aligned}$$

For any parameter $\xi$ one gets:

$$exp(\xi\frac{\sigma_3}{2})\sigma_\pm exp(-\xi\frac{\sigma_3}{2}) = \sigma_\pm exp(\pm\xi).$$

If $f(\sigma_+\sigma_-)$ (or $f(\sigma_-\sigma_+)$) is an arbitrary function of $\sigma_+\sigma_-$ (or of $\sigma_-\sigma_+$), and this function can be expanded into power series with respect to $\sigma_+\sigma_-$ (or with respect to $\sigma_-\sigma_+$), then

$$\begin{aligned}
f(\sigma_+\sigma_-) &= f(0) + [f(1) - f(0)]\sigma_+\sigma_-, \\
f(\sigma_-\sigma_+) &= f(0) + [f(1) - f(0)]\sigma_-\sigma_+.
\end{aligned}$$

## 2.4 Various Relations

Any $2 \times 2$ matrix $A$ can be expanded over the set $\{I, \sigma_i\}$:

$$A = a_0 I + a_i\sigma_i,$$

where $a_0 = \frac{1}{2}\mathrm{Tr}A$, and $a_i = \frac{1}{2}\mathrm{Tr}(\sigma_i A)$.
Let $\alpha_i$ be the 3–vector. Then

$$e^{\alpha_i\sigma_i} = \cosh\sqrt{\vec{\alpha}^2} + \frac{\sinh\sqrt{\vec{\alpha}^2}}{\sqrt{\vec{\alpha}^2}}(\alpha_i\sigma_i) = p_0 + p_i\sigma_i. \tag{2.4}$$

The components of the 4–vector $p^\mu$ equal:

$$p_0 = \cosh\sqrt{\vec{\alpha}^2}, \quad p_i = \frac{\sinh\sqrt{\vec{\alpha}^2}}{\sqrt{\vec{\alpha}^2}}\alpha_i, \quad p_0^2 - \vec{p}^{\,2} = p^2 = 1, \tag{2.5}$$

and we have

$$\alpha_i = \frac{p_i}{\sqrt{\vec{p}^{\,2}}}\ln(p_0 + \sqrt{\vec{p}^{\,2}}). \tag{2.6}$$

Let $p$ and $q$ be two 4–vectors, and $p^2 = q^2 = 1$, then

$$(p_0 + p_i\sigma_i)(q_0 + q_k\sigma_k) = a_0 + a_l\sigma_l = e^{\beta_i\sigma_i}, \tag{2.7}$$

where $a_0 = p_0q_0 + (\vec{p}\vec{q})$, $\quad a_j = p_0q_j + p_jq_0 + i\varepsilon^{jkl}p_kq_l$, and the 3–vector $\beta_i$ in the relation (2.7) is expressed through $a_0$ and $\vec{a}$ as in (2.6).

## 2.5 *4–dimensional $\sigma^\mu$ Matrices*

Here we present the various properties of $2\times 2$ matrices $\sigma^\mu$ and $\bar{\sigma}^\mu$ ($\mu = 0,1,2,3$):

$$\sigma^\mu_{A\dot{B}} \equiv (I, -\sigma_i); \quad \bar{\sigma}^{\mu\dot{A}B} \equiv (I, \sigma_i), \quad \mu = 0,\ 1,\ 2,\ 3, \tag{2.8}$$

where $\sigma_i$ are Pauli matrices.
With the help of $\sigma^\mu$–matrices any tensor in Minkowski space can be unambiguously rewritten in spinorial form. In order to deal only with Lorentz–covariant expressions one should clearly distinguish between dot and undot, lower and upper Weyl indices. The $\varepsilon$–symbol (see Section 1) used here for rising and lowering indices.
The chief properties of the $\sigma^\mu$ matrices are as follows:

$$\bar{\sigma}^{\mu\dot{A}A} = \varepsilon^{\dot{A}\dot{B}}\varepsilon^{AB}\sigma^\mu_{B\dot{B}}, \quad \sigma^\mu_{\dot{A}A} = \varepsilon_{AB}\varepsilon_{\dot{A}\dot{B}}\bar{\sigma}^{\mu\dot{B}B},$$
$$(\sigma^\mu)^\dagger = \sigma^\mu, \quad (\bar{\sigma}^\mu)^\dagger = \bar{\sigma}^\mu, \quad \det\sigma^\mu = \det\bar{\sigma}^\mu = 1(-1), \text{ for } \mu = 0(1,2,3).$$

For any 4–vector $p^\mu$ one has:

$$\det p_\mu\sigma^\mu = \det p_\mu\bar{\sigma}^\mu = p^2.$$

Various products of $\sigma^\mu$ matrices have the form:

$$
\begin{aligned}
&\sigma^\mu_{A\dot{C}}\bar{\sigma}^{\nu\dot{C}B} + \sigma^\nu_{A\dot{C}}\bar{\sigma}^{\mu\dot{C}B} = 2g^{\mu\nu}\delta_A{}^B, \\
&\bar{\sigma}^{\mu\dot{A}C}\sigma^\nu_{C\dot{B}} + \bar{\sigma}^{\nu\dot{A}C}\sigma^\mu_{C\dot{B}} = 2g^{\mu\nu}\delta^{\dot{A}}{}_{\dot{B}}, \\
&\sigma^\mu_{A\dot{C}}\bar{\sigma}^{\dot{C}B}_\mu = 4\delta_A{}^B, \quad \bar{\sigma}^{\mu\dot{A}C}\sigma_{\mu C\dot{B}} = 4\delta^{\dot{A}}{}_{\dot{B}}, \\
&\sigma^\mu_{A\dot{A}}\sigma_{\mu B\dot{B}}\varepsilon^{\dot{A}\dot{B}} = 4\varepsilon_{AB}, \quad \sigma^\mu_{A\dot{A}}\sigma_{\mu B\dot{B}}\varepsilon^{AB} = 4\varepsilon_{\dot{A}\dot{B}}, \\
&\bar{\sigma}^{\mu\dot{A}A}\bar{\sigma}^{\dot{B}B}_\mu\varepsilon_{\dot{A}\dot{B}} = 4\varepsilon^{AB}, \quad \bar{\sigma}^{\mu\dot{A}A}\bar{\sigma}^{\dot{B}B}_\mu\varepsilon_{AB} = 4\varepsilon^{\dot{A}\dot{B}}, \\
&\sigma^\mu_{A\dot{A}}\sigma^\nu_{B\dot{B}}\varepsilon^{AB}\varepsilon^{\dot{A}\dot{B}} \;=\; \bar{\sigma}^{\mu\dot{A}A}\bar{\sigma}^{\nu\dot{B}B}\varepsilon_{AB}\varepsilon_{\dot{A}\dot{B}} \;=\; 2g^{\mu\nu}, \\
&\sigma^\mu\bar{\sigma}^\lambda\sigma^\nu = g^{\mu\lambda}\sigma^\nu + g^{\nu\lambda}\sigma^\mu - g^{\mu\nu}\sigma^\lambda - i\varepsilon^{\mu\lambda\nu\rho}\sigma^\rho, \\
&\bar{\sigma}^\mu\sigma^\lambda\bar{\sigma}^\nu = g^{\mu\lambda}\sigma^\nu + g^{\nu\lambda}\sigma^\mu - g^{\mu\nu}\sigma^\lambda + i\varepsilon^{\mu\lambda\nu\rho}\sigma^\rho, \\
&\varepsilon^{\mu\nu\rho\lambda} = i\sigma^{\mu\dot{A}A}\sigma^{\nu\dot{B}B}\sigma^{\rho\dot{C}C}\sigma^{\lambda\dot{D}D}(\varepsilon_{AC}\varepsilon_{BD}\varepsilon_{\dot{A}\dot{D}}\varepsilon_{\dot{B}\dot{C}} - \varepsilon_{AD}\varepsilon_{BC}\varepsilon_{\dot{A}\dot{C}}\varepsilon_{\dot{B}\dot{D}}).
\end{aligned}
$$

The commutators of $\sigma^\mu$ and $\bar{\sigma}^\mu$ matrices have the special notation:

$$
\sigma^{\mu\nu B}_A \equiv \frac{1}{4}(\sigma^\mu_{A\dot{C}}\bar{\sigma}^{\nu\dot{C}B} - \sigma^\nu_{A\dot{C}}\bar{\sigma}^{\mu\dot{C}B}), \quad \bar{\sigma}^{\mu\nu\dot{A}}{}_{\dot{B}} \equiv \frac{1}{4}(\bar{\sigma}^{\mu\dot{A}C}\sigma^\nu_{C\dot{B}} - \bar{\sigma}^{\nu\dot{A}C}\sigma^\mu_{C\dot{B}}).
$$

The main properties of $\sigma^{\mu\nu}$ are as follows:

$$
\begin{aligned}
&\sigma^{0i} = \frac{1}{2}\sigma^i, \; \sigma^{ik} = -\frac{i}{2}\varepsilon^{ikl}\sigma^l, \quad \bar{\sigma}^{0i} = -\frac{1}{2}\sigma^i, \; \bar{\sigma}^{ik} = \sigma^{ik}, \\
&\sigma^{\mu\nu} = -\sigma^{\nu\mu}, \quad \bar{\sigma}^{\mu\nu} = -\bar{\sigma}^{\nu\mu}, \\
&(\sigma^\mu\bar{\sigma}^\nu)_A{}^B = g^{\mu\nu}\delta_A{}^B + 2\sigma^{\mu\nu B}_A, \quad (\bar{\sigma}^\mu\sigma^\nu)^{\dot{A}}{}_{\dot{B}} = g^{\mu\nu}\delta^{\dot{A}}{}_{\dot{B}} + 2\bar{\sigma}^{\mu\nu\dot{A}}{}_{\dot{B}}, \\
&\sigma^{\mu\nu K}_A\varepsilon_{KB} = \sigma^{\mu\nu K}_B\varepsilon_{KA}, \quad \bar{\sigma}^{\mu\nu\dot{A}}{}_{\dot{K}}\varepsilon^{\dot{K}\dot{B}} = \bar{\sigma}^{\mu\nu\dot{B}}{}_{\dot{K}}\varepsilon^{\dot{K}\dot{A}}, \\
&\varepsilon^{\mu\nu\lambda\rho}\sigma_{\lambda\rho} = -2i\sigma^{\mu\nu}, \quad \varepsilon^{\mu\nu\lambda\rho}\bar{\sigma}_{\lambda\rho} = 2i\bar{\sigma}^{\mu\nu}.
\end{aligned}
$$

## 2.6 *Traces of $\sigma^\mu$ Matrices*

$$
\begin{aligned}
&\mathrm{Tr}\sigma^\mu = \mathrm{Tr}\bar{\sigma}^\mu \;=\; 2(0) \quad \text{for} \quad \mu = 0\,(1,2,3), \\
&\mathrm{Tr}\sigma^{\mu\nu} = \mathrm{Tr}\bar{\sigma}^{\mu\nu} = 0, \quad \mathrm{Tr}(\sigma^\mu\bar{\sigma}^\nu) = \mathrm{Tr}(\bar{\sigma}^\mu\sigma^\nu) = 2g^{\mu\nu}, \\
&\mathrm{Tr}(\sigma^\mu\bar{\sigma}^\nu\sigma^\lambda\bar{\sigma}^\rho) = 2(g^{\mu\nu}g^{\lambda\rho} + g^{\mu\rho}g^{\nu\lambda} - g^{\mu\lambda}g^{\nu\rho} - i\varepsilon^{\mu\nu\lambda\rho}), \\
&\mathrm{Tr}(\sigma^{\mu\nu}\sigma^{\lambda\rho}) = \mathrm{Tr}(\bar{\sigma}^{\mu\nu}\bar{\sigma}^{\lambda\rho}) = \frac{1}{2}(g^{\mu\rho}g^{\nu\lambda} - g^{\mu\lambda}g^{\nu\rho} - i\varepsilon^{\mu\nu\lambda\rho}).
\end{aligned}
$$

## 2.7 *Fiertz Identities for $\sigma^\mu$ Matrices*

The Fiertz identities for $\sigma^\mu$ equal:

$$\sigma^\mu_{A\dot{A}}\bar{\sigma}^{\dot{B}B}_\mu = 2\delta_A{}^B\delta_{\dot{A}}{}^{\dot{B}}, \quad \sigma^\mu_{A\dot{A}}\sigma_{\mu B\dot{B}} = 2\varepsilon_{AB}\varepsilon_{\dot{A}\dot{B}}, \quad \bar{\sigma}^{\mu\dot{A}A}\bar{\sigma}^{\dot{B}B}_\mu = 2\varepsilon^{\dot{A}\dot{B}}\varepsilon_{AB}. \quad (2.9)$$

From the relations (2.9) one gets:

$$\begin{aligned}
\sigma^\mu_{A\dot{A}}\bar{\sigma}^{\nu\dot{B}B} &= \frac{1}{2}g^{\mu\nu}\delta_A{}^B\delta_{\dot{A}}{}^{\dot{B}} - \delta_A{}^B\bar{\sigma}^{\mu\nu\dot{B}}{}_{\dot{A}} + \sigma^{\mu\nu B}_A\delta_{\dot{A}}{}^{\dot{B}} + 2\sigma^{\nu\lambda B}_A\bar{\sigma}^{\mu\lambda\dot{B}}{}_{\dot{A}}, \\
\sigma^\mu_{A\dot{A}}\sigma^\nu_{B\dot{B}} &= \frac{1}{2}g^{\mu\nu}\varepsilon_{AB}\varepsilon_{\dot{A}\dot{B}} + \varepsilon_{AB}(\varepsilon_{\dot{A}\dot{C}}\bar{\sigma}^{\mu\nu\dot{C}}{}_{\dot{B}}) + (\sigma^{\mu\nu C}_A\varepsilon_{CB})\varepsilon_{\dot{A}\dot{B}} \\
&\quad -2(\sigma^{\mu\lambda C}_A\varepsilon_{CB})(\varepsilon_{\dot{A}\dot{C}}\bar{\sigma}^{\nu\lambda\dot{C}}{}_{\dot{B}}), \\
\bar{\sigma}^{\mu\dot{A}A}\bar{\sigma}^{\nu\dot{B}B} &= \frac{1}{2}g^{\mu\nu}\varepsilon^{AB}\varepsilon^{\dot{A}\dot{B}} + (\varepsilon^{AC}\sigma^{\mu\nu B}_C)\varepsilon^{\dot{A}\dot{B}} + \varepsilon^{AB}(\bar{\sigma}^{\mu\nu\dot{A}}{}_{\dot{C}}\varepsilon^{\dot{C}\dot{B}}) \\
&\quad -2(\varepsilon^{AC}\sigma^{\mu\lambda C}_B)(\bar{\sigma}^{\nu\lambda\dot{A}}{}_{\dot{C}}\varepsilon^{\dot{C}\dot{B}}),
\end{aligned}$$

$$(\varepsilon^{AC}\sigma^{\mu\lambda C}_B)(\bar{\sigma}^{\nu\lambda\dot{A}}{}_{\dot{C}}\varepsilon^{\dot{C}\dot{B}}) = (\varepsilon^{AC}\sigma^{\nu\lambda C}_B)(\bar{\sigma}^{\mu\lambda\dot{A}}{}_{\dot{C}}\varepsilon^{\dot{C}\dot{B}}).$$

# 3 DIRAC MATRICES

## 3.1 *Main Properties*

The main properties of the Dirac $\gamma$-matrices are as follows [1, 2, 3, 4, 5]:

$$\gamma^\mu\gamma^\nu + \gamma^\nu\gamma^\mu = 2g^{\mu\nu}, \tag{3.1}$$
$$(\gamma^0)^2 = I, \quad (\gamma^i)^2 = -I, \quad ,(\gamma^0)^\dagger = \gamma^0, \quad (\gamma^i)^\dagger = -\gamma^i. \tag{3.2}$$

The commutator of the $\gamma^\mu$–matrices has a special notation:

$$\sigma^{\mu\nu} \ \equiv \ \frac{1}{2}(\gamma^\mu\gamma^\nu - \gamma^\nu\gamma^\mu), \quad \sigma^{\mu\nu} \ = \ -\sigma^{\nu\mu} \tag{3.3}$$

The $\gamma^5$ matrix is defined as follows:

$$\gamma^5 \equiv i\gamma^0\gamma^1\gamma^2\gamma^3 \ = \ -\frac{i}{4!}\varepsilon_{\alpha\beta\mu\nu}\gamma^\alpha\gamma^\beta\gamma^\mu\gamma^\nu. \tag{3.4}$$

The $\gamma^5$ matrix has the following properties:

$$\gamma^5\gamma^\mu + \gamma^\mu\gamma^5 = 0, \quad (\gamma^5)^2 \ = \ I, \quad (\gamma^5)^\dagger \ = \ \gamma^5, \quad \{\gamma^\mu, \gamma^5\} = 0.$$

The *Dirac conjugation* of any $4 \times 4$–matrix $A$ is defined as follows:

$$\bar{A} \equiv \gamma^0 A^\dagger \gamma^0. \tag{3.5}$$

From (3.5) one gets:

$$\overline{\gamma^\mu} = \gamma^\mu, \quad \overline{\gamma^5} = -\gamma^5, \quad \overline{\gamma^\alpha\gamma^\beta\cdots\gamma^\lambda} = \gamma^\lambda\cdots\gamma^\beta\gamma^\alpha,$$
$$\overline{\gamma^\alpha\gamma^\beta\cdots\gamma^\mu\gamma^5\cdots\gamma^\lambda} = \gamma^\lambda\cdots(-\gamma^5)\gamma^\mu\cdots\gamma^\beta\gamma^\alpha = \gamma^\lambda\cdots\gamma^\mu\gamma^5\cdots\gamma^\beta\gamma^\alpha.$$

In this Section for the string of the $\gamma$–matrices we shall use the special notation:

$$S = S^n \equiv \gamma^{\alpha_1}\gamma^{\alpha_2}\cdots\gamma^{\alpha_n}, \quad S_R = S_R^n \equiv \gamma^{\alpha_n}\cdots\gamma^{\alpha_2}\gamma^{\alpha_1}. \tag{3.6}$$

Odd– and even–numbered string of $\gamma$–matrices will be denoted as follows:

$$S^{odd} \equiv \gamma^{\alpha_1}\gamma^{\alpha_2}\cdots\gamma^{\alpha_{2k+1}}, \quad S^{even} \equiv \gamma^{\alpha_1}\gamma^{\alpha_2}\cdots\gamma^{\alpha_{2k}}. \tag{3.7}$$

## 3.2 *Representations of the Dirac Matrices*

The non–singular transformation $\gamma \to U\gamma U^\dagger$ connects the different representations of the $\gamma$–matrices (Pauli lemma). Here we present three representations of the Dirac matrices.

• **Dirac (standard)** representation

$$\gamma_D^0 = \begin{pmatrix} 1 & 0 \\ 0 & -1 \end{pmatrix}, \quad \gamma_D^i = \begin{pmatrix} 0 & \sigma^i \\ -\sigma^i & 0 \end{pmatrix}, \quad \gamma_D^5 = \begin{pmatrix} 0 & 1 \\ 1 & 0 \end{pmatrix}.$$

• **Chiral (spinorial)** representation

$$\gamma_C^\mu = U_C \gamma_D^\mu U_C^\dagger, \qquad U_C = \frac{1}{\sqrt{2}}(\gamma_D^0 + \gamma_D^5) = \frac{1}{\sqrt{2}}\begin{pmatrix} 1 & 1 \\ 1 & -1 \end{pmatrix},$$

$$\gamma^\mu = \begin{pmatrix} 0 & \sigma^\mu \\ \bar{\sigma}^\mu & 0 \end{pmatrix},$$

$$\gamma^0 = \begin{pmatrix} 0 & 1 \\ 1 & 0 \end{pmatrix}, \quad \gamma^i = \begin{pmatrix} 0 & -\sigma^i \\ \sigma^i & 0 \end{pmatrix}, \quad \gamma^5 = \begin{pmatrix} 1 & 0 \\ 0 & -1 \end{pmatrix},$$

$$1 + \gamma^5 = 2\begin{pmatrix} 1 & 0 \\ 0 & 0 \end{pmatrix}, \quad 1 - \gamma^5 = 2\begin{pmatrix} 0 & 0 \\ 0 & 1 \end{pmatrix}.$$

where matrices $\sigma^\mu$ and $\bar{\sigma}^\mu$ were defined in (2.8).

• **Majorana** representation

$$\gamma_M^\mu = U_M \gamma_D^\mu U_M^\dagger, \qquad U_M = \frac{1}{\sqrt{2}}\begin{pmatrix} 1 & \sigma_2 \\ \sigma_2 & -1 \end{pmatrix},$$

$$\gamma^0 = \begin{pmatrix} 0 & \sigma_2 \\ \sigma_2 & 0 \end{pmatrix}, \quad \gamma^1 = \begin{pmatrix} i\sigma_3 & 0 \\ 0 & i\sigma_3 \end{pmatrix}, \quad \gamma^2 = \begin{pmatrix} 0 & -\sigma_2 \\ \sigma_2 & 0 \end{pmatrix},$$

$$\gamma^3 = \begin{pmatrix} -i\sigma_1 & 0 \\ 0 & -i\sigma_1 \end{pmatrix}, \quad \gamma^5 = \begin{pmatrix} \sigma_2 & 0 \\ 0 & -\sigma_2 \end{pmatrix}.$$

## 3.3 *Expansion of* $4 \times 4$ *Matrices*

The following 16 matrices $\Gamma_A \quad (A = 1, \ldots, 16)$

$$I, \quad \gamma^5, \quad \gamma^\mu, \quad \gamma^5\gamma^\mu, \quad \sigma^{\mu\nu} \tag{3.8}$$

are the full set of $4 \times 4$–matrices.
The main properties of $\Gamma_A$ are as follows:

$$\mathrm{Tr}I = 4, \quad \mathrm{Tr}\gamma^5 = \mathrm{Tr}\gamma^\mu = \mathrm{Tr}\gamma^5\gamma^\mu = \mathrm{Tr}\sigma^{\mu\nu} = 0 \tag{3.9}$$

Any $4 \times 4$–matrix $A$ can be expanded over set of the $\Gamma_A$-matrices:

$$A = a_0 I + a_5\gamma^5 + v_\mu\gamma^\mu + a_\mu\gamma^5\gamma^\mu + T_{\mu\nu}\sigma^{\mu\nu}, \tag{3.10}$$

where the coefficients could be found from the following relations:

$$a_0 = \frac{1}{4}\mathrm{Tr}A, \quad a_5 = \frac{1}{4}\mathrm{Tr}(\gamma^5 A), \quad v^\mu = \frac{1}{4}\mathrm{Tr}(\gamma^\mu A),$$
$$a^\mu = -\frac{1}{4}\mathrm{Tr}(\gamma^5\gamma^\mu A), \quad T^{\mu\nu} = -T^{\nu\mu} = -\frac{1}{8}\mathrm{Tr}(\sigma^{\mu\nu}A).$$

For the expansion of a matrix $A$ one can use another set of $\Gamma'_A$ ($\Gamma'_A = X,\ Y,\ U^\mu,\ V^\mu,\ \sigma^{\mu\nu}$):

$$X = I + \gamma^5, \quad Y = I - \gamma^5, \quad U^\mu = (I + \gamma^5)\gamma^\mu, \quad V^\mu = (I - \gamma^5)\gamma^\mu,$$
$$X^2 = 2X, \quad Y^2 = 2Y.$$

These matrices have the following properties:

$$U^2 = V^2 = XY = YX = XU^\mu = YV^\mu = 0,$$
$$\mathrm{Tr}X = \mathrm{Tr}Y = 4,\ \mathrm{Tr}U^\mu = \mathrm{Tr}U^\mu = \mathrm{Tr}\sigma^{\mu\nu} = 0.$$

The expansion of any $4 \times 4$–matrix $A$ over set of $\Gamma'$-matrices has the form:

$$A = a_x X + a_y Y + b_\mu U^\mu + c_\mu V^\mu + T_{\mu\nu}\sigma^{\mu\nu},$$

where

$$a_x = \frac{1}{8}\mathrm{Tr}(XA), \quad a_y = \frac{1}{8}\mathrm{Tr}(YA), \quad b^\mu = \frac{1}{8}\mathrm{Tr}(V^\mu A), \quad c^\mu = \frac{1}{8}\mathrm{Tr}(U^\mu A),$$
$$T^{\mu\nu} = -\frac{1}{8}\mathrm{Tr}(\sigma^{\mu\nu}A).$$

## 3.4 *Products of the Dirac Matrices*

$$\begin{aligned}
\gamma^\mu\gamma^\nu &= g^{\mu\nu}+\sigma^{\mu\nu}, \quad \sigma^{\mu\nu}=\gamma^\mu\gamma^\nu-g^{\mu\nu}=-\gamma^\nu\gamma^\mu+g^{\mu\nu},\\
\gamma^5\gamma^\mu\gamma^\nu &= g^{\mu\nu}\gamma^5+\frac{i}{2}\varepsilon^{\mu\nu\alpha\beta}\sigma_{\alpha\beta}, \quad \gamma^5\sigma^{\mu\nu}=+\frac{i}{2}\varepsilon^{\mu\nu\alpha\beta}\sigma_{\alpha\beta},\\
\gamma^\lambda\sigma^{\mu\nu} &= (g^{\mu\lambda}\gamma^\nu-g^{\nu\lambda}\gamma^\mu)-i\varepsilon^{\lambda\mu\nu\alpha}\gamma^5\gamma_\alpha,\\
\sigma^{\mu\nu}\gamma^\lambda &= -(g^{\mu\lambda}\gamma^\nu-g^{\nu\lambda}\gamma^\mu)-i\varepsilon^{\lambda\mu\nu\alpha}\gamma^5\gamma_\alpha,\\
\gamma^5\gamma^\lambda\sigma^{\mu\nu} &= (g^{\mu\lambda}\gamma^5\gamma^\nu-g^{\nu\lambda}\gamma^5\gamma^\mu)-i\varepsilon^{\lambda\mu\nu\alpha}\gamma_\alpha,\\
\sigma^{\mu\nu}\gamma^5\gamma^\lambda &= -(g^{\mu\lambda}\gamma^5\gamma^\nu-g^{\nu\lambda}\gamma^5\gamma^\mu)-i\varepsilon^{\lambda\mu\nu\alpha}\gamma_\alpha,\\
\sigma^{\alpha\beta}\sigma^{\mu\nu} &= g^{\alpha\nu}g^{\beta\mu}-g^{\alpha\mu}g^{\beta\nu}-i\varepsilon^{\alpha\beta\mu\nu}\gamma^5\\
&\quad +(g^{\alpha\nu}g^{\beta\lambda}g^{\mu\sigma}-g^{\alpha\mu}g^{\beta\lambda}g^{\nu\sigma}-g^{\beta\nu}g^{\alpha\lambda}g^{\mu\sigma}+g^{\beta\mu}g^{\alpha\lambda}g^{\nu\sigma})\sigma_{\lambda\sigma},\\
\sigma^{\alpha\beta}\sigma^{\mu\nu} &+ \sigma^{\mu\nu}\sigma^{\alpha\beta}=2(g^{\alpha\nu}g^{\beta\mu}-g^{\alpha\mu}g^{\beta\nu}-i\varepsilon^{\alpha\beta\mu\nu}\gamma^5).
\end{aligned}$$

The totally antisymmetric tensor $\gamma^{\mu\nu\lambda}$ is defined as follows:

$$\begin{aligned}
&\gamma^{\mu\nu\lambda}\equiv\frac{1}{6}(\gamma^\mu\gamma^\nu\gamma^\lambda+\gamma^\nu\gamma^\lambda\gamma^\mu+\gamma^\lambda\gamma^\mu\gamma^\nu-\gamma^\nu\gamma^\mu\gamma^\lambda-\gamma^\lambda\gamma^\nu\gamma^\mu-\gamma^\mu\gamma^\lambda\gamma^\nu),\\
&\gamma^\mu\gamma^\nu\gamma^\lambda=\gamma^{\mu\nu\lambda}+g^{\mu\nu}\gamma^\lambda-g^{\mu\lambda}\gamma^\nu+g^{\nu\lambda}\gamma^\mu,\\
&\gamma^{\mu\nu\lambda}=-i\varepsilon^{\mu\nu\lambda\alpha}\gamma^5\gamma_\alpha, \quad \gamma^5\gamma^\alpha=\frac{i}{6}\varepsilon^{\alpha\mu\nu\lambda}\gamma_{\mu\nu\lambda}.
\end{aligned}$$

The products of the type $\sum_{A=1}^{16}\Gamma^A\Gamma^B\Gamma^A$ are presented in the Table 3.1.
**Table 3.1.**

| $\Gamma^B$ | $\gamma^5\Gamma^B\gamma^5$ | $\gamma^\nu\Gamma^B\gamma^\nu$ | $\gamma^5\gamma^\nu\Gamma^B\gamma^5\gamma^\nu$ | $\sigma^{\mu\nu}\Gamma^B\sigma^{\mu\nu}$ |
|---|---|---|---|---|
| $I$ | $I$ | $4$ | $-4$ | $-1$ |
| $\gamma^5$ | $\gamma^5$ | $-4\gamma^5$ | $-4\gamma^5$ | $-12\gamma^5$ |
| $\gamma^\alpha$ | $-\gamma^\alpha$ | $-2\gamma^\alpha$ | $-2\gamma^\alpha$ | $0$ |
| $\gamma^5\gamma^\alpha$ | $-\gamma^5\gamma^\alpha$ | $2\gamma^5\gamma^\alpha$ | $2\gamma^5\gamma^\alpha$ | $0$ |
| $\sigma^{\alpha\beta}$ | $\sigma^{\alpha\beta}$ | $0$ | $0$ | $4\sigma^{\alpha\beta}$ |

So–called Chisholm identities are given by [5]:

$$\gamma^\mu S^{odd}\gamma_\mu=-2S_R^{odd}, \tag{3.11}$$
$$\gamma^\mu S^{even}\gamma_\mu=\gamma^\mu\gamma^\lambda S'^{odd}\gamma_\mu=2\gamma^\lambda S_R'^{odd}+2S'^{odd}\gamma^\lambda, \tag{3.12}$$

where in the last relation $S^{even} = \gamma^\lambda S'^{odd}$.
Using the relations (3.11) and (3.12), one gets:

$$\begin{aligned}
\gamma^\mu S^{even}\gamma_\mu &= \mathrm{Tr}(S^{even})I - \mathrm{Tr}(\gamma^5 S^{even})\gamma^5, \\
\hat{p}S^{even}\hat{p} &= -p^2 S_R^{even} + \frac{1}{2}\mathrm{Tr}(\hat{p}\gamma^\alpha S_R^{even})\gamma_\alpha \hat{p}, \\
\hat{p}S^{even}\hat{p} &= -\hat{p}S_R^{even}\hat{p}, \quad \text{for} \quad p^2 = 0, \\
\gamma^\mu S^{odd}\gamma_\mu &= -\frac{1}{2}\mathrm{Tr}(\gamma^\alpha S^{odd})\gamma_\alpha + \frac{1}{2}\mathrm{Tr}(\gamma^5\gamma^\alpha S^{odd})\gamma_\alpha\gamma^5, \\
\hat{p}S^{odd}\hat{p} &= -p^2 S_R^{odd} + \frac{1}{2}\mathrm{Tr}(\hat{p}S_R^{odd})\hat{p} + \frac{1}{2}\mathrm{Tr}(\gamma^5\hat{p}S_R^{odd})\hat{p}\gamma^5, \\
S^{odd} &= \frac{1}{4}\mathrm{Tr}(\gamma^\alpha S^{odd})\gamma_\alpha + \frac{1}{4}\mathrm{Tr}(\gamma^5\gamma^\alpha S^{odd})\gamma_\alpha\gamma^5, \\
S^{odd} &+ S_R^{odd} = \frac{1}{2}\mathrm{Tr}(\gamma^\alpha S^{odd})\gamma_\alpha.
\end{aligned}$$

Using (3.11) and (3.12), one can write also the well known relations for $S^1$, $S^2$, $S^3$, and $S^4$:

$$\begin{aligned}
\gamma^\mu\gamma^\alpha\gamma_\mu &= -2\gamma^\alpha, \quad \gamma^\mu\gamma^\alpha\gamma^\beta\gamma^\delta\gamma_\mu = -2\gamma^\delta\gamma^\beta\gamma^\alpha, \\
\gamma^\mu\gamma^\alpha\gamma^\beta\gamma_\mu &= 4g^{\alpha\beta}, \\
\gamma^\mu\gamma^{\alpha_1}\gamma^{\alpha_2}\gamma^{\alpha_3}\gamma^{\alpha_4}\gamma_\mu &= 2(\gamma^{\alpha_4}\gamma^{\alpha_1}\gamma^{\alpha_2}\gamma^{\alpha_3} + \gamma^{\alpha_3}\gamma^{\alpha_2}\gamma^{\alpha_1}\gamma^{\alpha_4}) \\
&= 2(\gamma^{\alpha_1}\gamma^{\alpha_4}\gamma^{\alpha_3}\gamma^{\alpha_2} + \gamma^{\alpha_2}\gamma^{\alpha_3}\gamma^{\alpha_4}\gamma^{\alpha_1}), \\
\gamma^\mu\sigma^{\alpha\beta}\gamma_\mu &= 0, \quad \gamma^\mu\sigma^{\alpha\beta}\gamma^\delta\gamma_\mu = 2\gamma^\delta\sigma^{\alpha\beta}, \quad \gamma^\mu\gamma^\delta\sigma^{\alpha\beta}\gamma_\mu = 2\sigma^{\alpha\beta}\gamma^\delta.
\end{aligned}$$

## 3.5 *Fiertz Identities*

Fiertz identities for $\gamma$–matrices could be obtained from the basic formula:

$$\delta_{ij}\delta_{kl} = \frac{1}{4}[\delta_{il}\delta_{kj} + \gamma^5_{il}\gamma^5_{kj} + \gamma^\mu_{il}\gamma_{\mu\, kj} - (\gamma^5\gamma^\mu)_{il}(\gamma^5\gamma_\mu)_{kj} - \frac{1}{2}\sigma^{\mu\nu}_{il}(\sigma_{\mu\nu})_{kj}]. \quad (3.13)$$

Using (3.13) one can obtain the well known relations:

$$\Gamma^M_{ij}\Gamma^M_{kl} = \sum_{N=1}^{16} C_{MN}\Gamma^N_{il}\Gamma^N_{kj}, \quad (3.14)$$

The coefficients $C_{MN}$ are presented in Table 3.2, where we use the traditional notations:

$$S = I, \quad P = \gamma^5, \quad V = \gamma^\mu, \quad A = \gamma^5\gamma^\mu, \quad T = \sigma^{\mu\nu}.$$

**Table 3.2.**

*

| | $N=S$ | $V$ | $T$ | $A$ | $P$ |
|---|---|---|---|---|---|
| $M=I$ | $\frac{1}{4}$ | $\frac{1}{4}$ | $-\frac{1}{8}$ | $-\frac{1}{4}$ | $\frac{1}{4}$ |
| $V$ | 1 | $-\frac{1}{2}$ | 0 | $-\frac{1}{2}$ | $-1$ |
| $T$ | $-3$ | 0 | $-\frac{1}{2}$ | 0 | $-3$ |
| $A$ | $-1$ | $-\frac{1}{2}$ | 0 | $-\frac{1}{2}$ | 1 |
| $P$ | $\frac{1}{4}$ | $-\frac{1}{4}$ | $-\frac{1}{8}$ | $\frac{1}{4}$ | $\frac{1}{4}$ |

Using relation (3.13) one gets:

$$(1\pm\gamma^5)_{ij}\delta_{kl} =$$
$$\frac{1}{8}[2(1\pm\gamma^5)_{il}(1\pm\gamma^5)_{kj} + 2((1\pm\gamma^5)\gamma^\mu)_{il}((1\mp\gamma^5)\gamma_\mu)_{kj} - ((1\pm\gamma^5)\sigma_{\mu\nu})_{il}\sigma^{\mu\nu}_{kj}],$$
$$\delta_{ij}(1\pm\gamma^5)_{kl} =$$
$$\frac{1}{8}[2(1\pm\gamma^5)_{il}(1\pm\gamma^5)_{kj} + 2((1\mp\gamma^5)\gamma^\mu)_{il}((1\pm\gamma^5)\gamma_\mu)_{kj} - ((1\pm\gamma^5)\sigma_{\mu\nu})_{il}\sigma^{\mu\nu}_{kj}],$$
$$(1\pm\gamma^5)_{ij}(1\pm\gamma^5)_{kl} = \frac{1}{4}[2(1\pm\gamma^5)_{il}(1\pm\gamma^5)_{kj} - ((1\pm\gamma^5)\sigma_{\mu\nu})_{il}\sigma^{\mu\nu}_{kj}],$$
$$(1\pm\gamma^5)_{ij}(1\mp\gamma^5)_{kl} = \frac{1}{2}[(1\pm\gamma^5)\gamma^\mu]_{il}[(1\mp\gamma^5)\gamma_\mu]_{kj},$$
$$[(1\pm\gamma^5)\gamma^\mu]_{ij}[(1\pm\gamma^5)\gamma_\mu]_{kl} = -[(1\pm\gamma^5)\gamma^\mu]_{il}[(1\pm\gamma^5)\gamma_\mu]_{kj},$$
$$[(1\pm\gamma^5)\gamma^\mu]_{ij}[(1\mp\gamma^5)\gamma_\mu]_{kl} = 2(1\pm\gamma^5)_{il}(1\mp\gamma^5)_{kj},$$
$$(\gamma^\mu)_{ij}(\gamma_\mu)_{kl} + (\gamma^5\gamma^\mu)_{ij}(\gamma^5\gamma_\mu)_{kl} = -[(\gamma^\mu)_{il}(\gamma_\mu)_{kj} + (\gamma^5\gamma^\mu)_{il}(\gamma^5\gamma_\mu)_{kj}].$$

## 3.6 *Traces of the $\gamma$-matrices*

The trace of any odd–numbered string of $\gamma$–matrices (including any number of $\gamma^5$ matrices) and trace of the $\gamma^5\gamma^\mu\gamma^\nu$ product are equal to zero:

$$\mathrm{Tr}S^{odd} = \mathrm{Tr}(S^{odd}(\cdots\gamma^5\cdots)) = \mathrm{Tr}(\gamma^5\gamma^\mu\gamma^\nu) = 0.$$

In this Subsection we use the following notation:

$$T^{\mu_1\mu_2\ldots\mu_n} \equiv \frac{1}{4}\mathrm{Tr}(\gamma^{\mu_1}\gamma^{\mu_2}\ldots\gamma^{\mu_n}).$$

Then

$$
\begin{aligned}
& T^{\mu\nu} = g^{\mu\nu}, \qquad T^{\alpha\beta\delta\sigma} = g^{\alpha\beta}g^{\delta\sigma} + g^{\alpha\sigma}g^{\beta\delta} - g^{\alpha\delta}g^{\beta\sigma}, \\
& T^{\alpha\beta\delta\lambda\rho\sigma} = g^{\alpha\beta}T^{\delta\lambda\rho\sigma} - g^{\alpha\delta}T^{\beta\lambda\rho\sigma} + g^{\alpha\lambda}T^{\beta\delta\rho\sigma} - g^{\alpha\rho}T^{\beta\delta\lambda\sigma} + g^{\alpha\sigma}T^{\beta\delta\lambda\rho}, \\
& \mathrm{Tr}(\gamma^5) = 0, \quad \mathrm{Tr}(\gamma^5\gamma^\mu\gamma^\nu) = 0, \quad \mathrm{Tr}(\gamma^5\gamma^\alpha\gamma^\beta\gamma^\delta\gamma^\lambda) = -4i\varepsilon^{\alpha\beta\delta\lambda}, \\
& \mathrm{Tr}(\gamma^5\gamma^{\alpha_1}\gamma^{\alpha_2}\gamma^{\alpha_3}\gamma^{\alpha_4}\gamma^{\alpha_5}\gamma^{\alpha_6}) = 4i(g^{\alpha_1\alpha_2}\varepsilon^{\alpha_3\alpha_4\alpha_5\alpha_6} - g^{\alpha_1\alpha_3}\varepsilon^{\alpha_2\alpha_4\alpha_5\alpha_6} \\
& + g^{\alpha_2\alpha_3}\varepsilon^{\alpha_1\alpha_4\alpha_5\alpha_6} + g^{\alpha_4\alpha_5}\varepsilon^{\alpha_1\alpha_2\alpha_3\alpha_6} - g^{\alpha_4\alpha_6}\varepsilon^{\alpha_1\alpha_2\alpha_3\alpha_5} + g^{\alpha_5\alpha_6}\varepsilon^{\alpha_1\alpha_2\alpha_3\alpha_4}), \\
& \mathrm{Tr}\sigma^{\alpha\beta}\sigma^{\mu\nu} = 4(g^{\alpha\nu}g^{\beta\mu} - g^{\alpha\mu}g^{\beta\nu}).
\end{aligned}
$$

Using the relation (3.13), one can rewrite the trace of the product of two $4 \times 4$ matrices $A$ and $B$ as follows:

$$
\begin{aligned}
4\mathrm{Tr}(AB) &= \mathrm{Tr}(A)\mathrm{Tr}(B) + \mathrm{Tr}(\gamma^5 A)\mathrm{Tr}(\gamma^5 B) + \mathrm{Tr}(\gamma^\mu A)\mathrm{Tr}(\gamma_\mu B) \\
&- \mathrm{Tr}(\gamma^5\gamma^\mu A)\mathrm{Tr}(\gamma^5\gamma_\mu B) - \frac{1}{2}\mathrm{Tr}(\sigma^{\mu\nu}A)\mathrm{Tr}(\sigma_{\mu\nu}B).
\end{aligned}
$$

The additional equation can be obtained using the Chisholm identities (3.11) and (3.12):

$$
\mathrm{Tr}(A\gamma^\mu B)\mathrm{Tr}(\gamma_\mu S^{odd}) = 2\Big[\mathrm{Tr}(AS^{odd}B) + \mathrm{Tr}(AS^{odd}_R B)\Big].
$$

## 3.7 *Dirac Matrices Algebra in n–dimensions*

In the framework of dimensional regularization one gets:

$$
\begin{aligned}
\mathrm{Tr}\ I &= f(n), /; /; f(4) = 4, \quad g^{\mu\nu}g_{\mu\nu} = n, \\
\gamma^\mu\gamma^\nu + \gamma^\nu\gamma^\mu &= 2g^{\mu\nu}, \\
\gamma_\mu\gamma^\alpha\gamma^\mu &= (2-n)\gamma^\alpha, \\
\gamma_\mu\gamma^\alpha\gamma^\beta\gamma^\mu &= 4g^{\alpha\beta} + (n-4)\gamma^\alpha\gamma^\beta, \\
\gamma_\mu\gamma^\alpha\gamma^\beta\gamma^\delta\gamma^\mu &= -2\gamma^\delta\gamma^\beta\gamma^\alpha + (4-n)\gamma^\alpha\gamma^\beta\gamma^\delta.
\end{aligned}
$$

# 4 THEORY OF SPINORS IN $N$ DIMENSIONS

Let $E_n$ be $n$-dimensional complex linear normalized space. The cases of odd and even $n$ should be considered separately.

## 4.1 *The Odd–dimensional Case :* $n = 2\nu + 1$

• *Isotropic planes*

Let vector $x$ has coordinates $x = (x_0,\ x_1,\ ..., x_{2\nu})$. The basis in $E_{2\nu+1}$ can always be chosen so that the the norm squared of the vector $x$ looks as follows:

$$x \cdot x \ = \ x_0^2 + x_1 x_{\nu+1} + ... + x_\nu x_{2\nu}. \tag{4.1}$$

A (hyper)plane is called isotropic if each vector in it has zero norm.

**Isotropic planes in $E_{2\nu+1}$ have not more than$\nu$ dimensions.** (4.2)

So, one has the following natural decomposition of $E_{2\nu+1}$ into the sum of three spaces:

$$E_{2\nu+1} \ = \ R^I \oplus E_\nu^I \oplus E_\nu^{II}, \tag{4.3}$$

i.e., if $x \in E_{2\nu+1}$ then

$$x \ = \ \mathbf{x} + \mathbf{x}' + x_0 e_0, \quad \text{where} \ \ \mathbf{x} \in E_\nu^I,\ \mathbf{x}' \in E_\nu^{II}. \tag{4.4}$$

$$x \cdot x \ = \ \sum_{i=1}^{\nu} \mathbf{x}_i {\mathbf{x}'}_i + x_0^2.$$

$E_\nu^I$ and $E_\nu^{II}$ are isotropic $\nu$–planes: $\mathbf{x} \cdot \mathbf{x} = 0$ and $\mathbf{x}' \cdot \mathbf{x}' = 0$.

• *Grassman algebras*

Let one consider Grassman algebra $G_\nu$ (algebra of external forms) on some $\nu$–dimensional vector space $F_\nu$. An element $\xi \in G_\nu$ has the form:

$$\xi \ = \ \sum_{p=0}^{\nu} \xi_{i_1 ... i_p} e^{i_1} \wedge ... \wedge e^{i_p}. \tag{4.5}$$

$\xi_{i_1 ... i_p}$ here are components of antisymmetric rank $p$ tensors. There are three important operations on $G_\nu$:

**A.** External product.

Let $x \in F_\nu$, $\xi \in G_\nu$, then $x \wedge \xi \in G_\nu$, the components of $x \wedge \xi$ look as follows:

$$\begin{aligned}(x\wedge\xi)_{i_1\ldots i_{p+1}} &= (-1)^p x_{i_1}\xi_{i_2\ldots i_{p+1}} + x_{i_2}\xi_{i_3\ldots i_{p+1}i_1} + \\ &\quad +(-1)^p x_{i_3}\xi_{i_4\ldots.i_{p+1}i_1i_2} + x_{i_{p+1}}\xi_{i_1\ldots.i_p}. \end{aligned} \tag{4.6}$$

An element $\Omega^{(p)} \in G_\nu$ is called $p$-form if its decomposition (4) has only $p$-th term.

$$\Omega^{(p)} \wedge \Omega^{(q)} = (-1)^{pq}\, \Omega^{(q)} \wedge \Omega^{(p)}.$$

**B.** $\forall \xi \in G_\nu$ there exists the dual element $*\xi$ which has the components:

$$(*\xi)_{i_1\ldots.i_{\nu-p}} = \frac{1}{p!}\varepsilon_{i_1\ldots.i_{\nu-p}\lambda_1\ldots.\lambda_p}\xi_{\lambda_1\ldots.\lambda_p}. \tag{4.7}$$

**C.** If $x \in F_\nu, \xi \in G_\nu$ then $id(x)\circ\xi \ \in G_\nu$ and has the following components:

$$(id(x)\circ\xi)_{i_1\ldots.i_{p-1}} = (-1)^{p-1}\xi_{j\ i_1\ldots.i_{p-1}}x_j \tag{4.8}$$

These operations satisfy the following relations [6]:

$$x\wedge(x\wedge\xi) = 0, \tag{4.9}$$

$$id(x)\circ(id(x)\circ\xi) = 0, \tag{4.10}$$

$$id(x)\circ(y\wedge\xi) + y\wedge(id(x)\circ\xi) = (x_iy_i)\xi, \tag{4.11}$$

here $x_i$ and $y_i$ are components of $x, y \ \in F_\nu$.

$$\begin{aligned} *\ \ (*\xi) &= \sum_{p=0}^{\nu}(-1)^{p(\nu-p)}\xi_{i_1\ldots.i_p}e^{i_1}\wedge\ldots\wedge e^{i_p}, \\ *\ \ (id(x)\circ(\Omega^{(p)})) &= (-1)^{\nu(p+1)-1}x\wedge\Omega^{(p)}, \\ *\ \ (x\wedge\Omega^{(p)}) &= (-1)^{\nu(p+1)}id(x)\circ\Omega^{(p)}. \end{aligned}$$

It is also useful to introduce the operation $H_0$:

$$H_0\xi = \sum_{p=0}^{\nu}(-1)^p\xi_{i_1\ldots.i_p}e^{i_1}\wedge\ldots\wedge e^{i_p}. \tag{4.12}$$

It satisfies the properties:

$$H_0(x\wedge\xi) + x\wedge(H_0\xi) = 0, \tag{4.13}$$

$$H_0(id(x)\circ\xi) + id(x)\circ(H_0\xi) = 0. \tag{4.14}$$

• *Definition of a spinor*

For any $x \in E_{2\nu+1}$ one can define the operation $Cliff(x)$ on Grassman algebra $G_\nu$:

$$Cliff(x) \circ \xi \ = \ id(\mathbf{x}) \circ \xi + \mathbf{x}' \wedge \xi + x_0 H_0(\xi), \qquad (4.15)$$

where $\mathbf{x}$, $\mathbf{x}'$ and $x_0$ are defined in (4.4). From the properties (4.9 – 4.11), (4.13), and (4.14) and the the expression for the norm squared (4.1) one can easily derive the following property of operation $Cliff$:

$$Cliff(x) \circ (Cliff(x) \circ \xi) \ = \ (x \cdot x)\ \xi. \qquad (4.16)$$

Let now an element $\xi \in G_\nu$ be written as the column vector:

$$\xi \ = \ \begin{pmatrix} \xi_0 \\ \xi_{i_1} \\ \xi_{i_1 i_2} \\ \cdots \\ \xi_{i_1 \dots . i_\nu} \end{pmatrix} \qquad \begin{matrix} \text{scalar} \\ \text{one-form (covector)} \\ \text{two-form (bivector)} \\ \dots\dots \\ \nu\text{-form } (\nu\text{-vector}) \end{matrix}$$

- Such $2^\nu$-component column vectors are called spinors in the space $E_{2\nu+1}$

One can now construct the matrix representation of the operation $Cliff$: the column vector corresponding to $Cliff(x) \circ \xi$ should be written in the form $\hat{x}\xi$, where $\hat{x}$ is $2^\nu \times 2^\nu$ matrix. So one has the representation

$$x \longrightarrow \hat{x} \qquad (4.17)$$

of vectors $x \in E_{2\nu+1}$ by $2^\nu \times 2^\nu$ matrices which satisfies the fundamental property:

$$\hat{x}\hat{x} \ = \ x \cdot x, \qquad (4.18)$$

or, if one substitutes vector $x + y$ instead of $x$ here one can rewrite it in the form:

$$\hat{x}\hat{y} + \hat{y}\hat{x} \ = \ 2x \cdot y. \qquad (4.19)$$

• $O(2\nu + 1)$*-transformations*

Let us start with the observation that

$$q'_\mu \ = \ q_\mu + 2q \cdot l\, l_\mu \qquad (4.20)$$

where $q'$ is obtained from $q$ by the reflection with respect to the $2\nu$-plane which is orthogonal to a unit space–like vector $l: \; l^2 = -1$. The representation (4.17) of this formula looks as follows:

$$\hat{q}' \equiv S_l(q) \; = \; \hat{l}\hat{q}\hat{l}. \tag{4.21}$$

Let $R_{a_1 a_2} \equiv \rho^2$, where $\rho$ is the boost (or rotation) in $(a_1, \, a_2)$ plane $(a_1^2 = a_2^2 = -1)$, which transforms $a_1$ into $a_2$. Since any rotation is a composition of an even number of reflections, one can easily derive the relations:

$$\begin{aligned} R_{a_1 a_2} &\equiv \; S_{a_2}(S_{a_1}(q)) \; = \; \hat{a}_2\hat{a}_1\hat{q}\hat{a}_1\hat{a}_2 \; = \; A\hat{q}A^{-1}, \\ A \; &= \; \hat{a}_2\hat{a}_1. \end{aligned} \tag{4.22}$$

Hence the action of reflections and rotations on spinors looks as follows:

$$S_l\xi \; = \; \pm\hat{l}\xi, \;\; R_{l_2 l_1}\xi \; = \; \pm\hat{l}_2\hat{l}_1\xi. \tag{4.23}$$

• *Geometrical interpretation: spinors are $\nu$-forms in $E_{2\nu+1}$.*

**The equation**

$$\hat{x}\psi \; = \; 0 \tag{4.24}$$

**determines $\nu$–dimensional isotropic plane for each spinor $\xi$.**

Indeed, if $\hat{x}\psi = 0$ and $\hat{y}\psi = 0$ then, due to (4.19), $x \cdot y = 0$. Hence this hyperplane is isotropic. That it has not more than $\nu$ dimensions is seen from the (4.2). $\nu$–dimensional plane can be manifestly constructed for the case when only $\xi_0$ is not equal to zero. (Strictly speaking, this plane has exactly $\nu$ dimensions only for so called simple spinors (spinor $\xi$ is called simple, if rang of the system (4.24) is $\nu + 1$). However, each spinor is a sum of simple spinors, for $\nu = 1, 2$ each spinor is simple. In general simple spinors in $E_{2\nu+1}$ lie in a manifold in the space of all spinors, which is determined by $N = 2^{\nu-1}(2^\nu + 1) - C^\nu_{2\nu+1}$ equations).

**The components of $\psi$ can be interpreted as elements of the Grassman algebra on this $\nu$-plane.**

One can also consider isotropic $\nu$-forms instead isotropic $\nu$-planes, that is equivalent. The $\nu$-form corresponding to some isotropic $\nu$-plane is the external product of the $\nu$ linearly independent vectors from this $\nu$-plane.

Proofs of all the statements listed here can be found in [7].

• *An analog with a fermion system* [8].

One can also consider the Grassman algebra $G_\nu$ as the Fock space $H(F)$ of a fermion system with $\nu$ degrees of freedom. Let now $a_i^\dagger$ and $a_i$ be the creation and annihilation operators which satisfy the relations:

$$\{a_i, a_j\} = 0, \quad \{a_i^\dagger, a_j^\dagger\} = 0, \quad \{a_i, a_j^\dagger\} = \delta_{ij}.$$

An arbitrary vector $|\xi> \in H(F)$ can be written in the form:

$$|\xi> = \sum_{p=0}^{\nu} \xi_{i_1 \ldots i_p} a_{i_1}^\dagger \ldots a_{i_p}^\dagger |0> . \tag{4.25}$$

The operators $\wedge$ and $id$ can be represented by the action of creation and annihilation operators:

$$\begin{aligned} x \wedge \xi &\to x_i a_i^\dagger |\xi> \\ id(x) \circ \xi &\to x_i a_i |\xi>, \end{aligned}$$

where $x_i$ are components of vector $x \in F_\nu$.
The $\gamma$-matricies in $2\nu + 1$ dimensional space have the following form:

$$\gamma_j = a_j^\dagger + a_j \tag{4.26}$$

$$\gamma_{j+\nu} = \frac{a_j^\dagger - a_j}{i} \tag{4.27}$$

$$\gamma_{2\nu+1} = (-1)^{N_F}, \quad \text{where } N_F = \sum_{i=1}^{\nu} a_i^\dagger a_i \tag{4.28}$$

The $\sigma_{kl} = (\gamma_k \gamma_l - \gamma_l \gamma_k)/2$ are generators of $SO(2\nu + 1)$, so one has natural realization of $SO(2\nu + 1)$-symmetry in $H(F)$ in addition to usual $U_\nu$ whose generators are $a_k^\dagger a_l$.

• *Invariant Forms*

A generalization of the matrix of charge conjugation can be defined as the matrix representation of the operation $*$ in the space of spinors:

$$C\xi = (-1)^{p\nu - \frac{p(p-1)}{2}} (*\xi) \tag{4.29}$$

Its main properties are as follows:

$$C\hat{x} = (-1)^\nu \hat{x}^T C, \quad C^2 = (-1)^{\frac{\nu(\nu+1)}{2}}.$$

Then

$$\eta^T C \xi$$

is scalar in $E_{2\nu+1}$ (i.e. with respect to $O(2\nu+1)$) if $\nu$ is even, and pseudoscalar ($2\nu + 1$-form) if $\nu$ is odd.

## 4.2 *The Even–Dimensional Case:* $n = 2\nu$

The space $E_{2\nu}$ can be considered as $2\nu$-dimensional hyperplane in $E_{2\nu+1}$, which is orthogonal to some vector $a$. Each spinor $\psi \in E_{2\nu+1}$ can be represented as the sum $\psi = \varphi + \chi$, where $\varphi$ and $\chi$ are eigenspinors of $\hat{a}$:

$$\hat{a}\varphi = \varphi, \quad \hat{a}\chi = \ -\chi.$$

Only the $\nu$-planes corresponding to eigenspinors of $\hat{a}$ lie in $E_{2\nu}$; an arbitrary isotropic $\nu$-plane intersects $E_{2\nu}$ over some $(\nu - 1)$-plane. In the language of external forms this idea can be formulated as follows. An arbitrary $\nu$-form $T$ on $E_{2\nu+1}$ can be uniquely decomposed into two components:

$$T \ = \ R + a \wedge S,$$

where $R$ is $\nu$-form, $S$ is $(\nu - 1)$-form, both are defined on $E_{2\nu}$. The two forms $R$ and $S$ give the geometrical interpretation of a spinor in $E_{2\nu}$.

The fact that there exist just the two types of isotropic $\nu$-planes in $E_{2\nu}$ (corresponding to spinors of the type $\varphi$ and $\chi$) is the consequence of the theorem that the representation of $SO(2\nu)$ on $\nu$-forms can be decomposed into two irreducible parts (self–dual and antiself–dual forms).

The most widely used case in high energy physics is $\nu = 2$, $\hat{a} = \gamma^5$, $\xi = u(p, n)$. Concerning the vector $\omega_+$ (see (6.11 – 6.13)) one can say that it determines the line which is the intersection of the isotropic 2-plane corresponding to a spinor $u(p, n)$ with (complexified) momentum space.

# 5 VECTOR ALGEBRA

Let $\{p_1, \ldots, p_n\}$ be some basis and scalar products $p_i \cdot p_j$ define a matrix $M$: $M_{ij} = p_i \cdot p_j$. The dual basis is the set of vectors $\{\xi_1, ..., \xi_n\}$, which satisfy the conditions:

$$\xi_i \cdot p_j = \delta_{ij}, \quad \xi_i \cdot \xi_j = (M^{-1})_{ij}.$$

Then

$$\xi_i^\alpha = \delta^{p_1,\ldots,p_{i-1},\alpha,p_{i+1},\ldots,p_n}_{p_1,\ldots\ldots\ldots\ldots,p_n} / \Delta_n,$$

where $\Delta_n = \delta^{p_1,\ldots,p_n}_{p_1,\ldots,p_n}$. Sometimes one needs to represent some vector $Q$ in the form [9]:

$$Q_\alpha = \mathcal{P}_\alpha + V_\alpha,$$

where $\mathcal{P}_\alpha$ is a linear combination of $p_1, ..., p_m$ $(m < n)$, and $V \cdot p_i = 0$ for $i = 1, ..., m$.

$m = 1$
$$\mathcal{P}_\alpha = \frac{p_1 \cdot Q}{p_1 \cdot p_1} p_{1\alpha} \qquad V_\alpha = \frac{1}{p_1 \cdot p_1} \delta^{p_1 \alpha}_{p_1 Q}$$

$m = 2$
$$\mathcal{P}_\alpha = \frac{1}{\Delta_2} \delta^{Q\mu}_{p_1 p_2} \delta^{\mu\alpha}_{p_1 p_2} = \frac{1}{\Delta_2} \left( \delta^{Q p_2}_{p_1 p_2} p_{1\alpha} + \delta^{p_1 Q}_{p_1 p_2} p_{2\alpha} \right)$$
$$V_\alpha = \frac{1}{\Delta_2} \delta^{p_1 p_2 \alpha}_{p_1 p_2 Q}$$

$m = 3$
$$\mathcal{P}_\alpha = \frac{1}{2\Delta_3} \delta^{Q\mu\nu}_{p_1 p_2 p_3} \delta^{\alpha\mu\nu}_{p_1 p_2 p_3} = \frac{1}{\Delta_3} \left( \delta^{Q p_2 p_3}_{p_1 p_2 p_3} p_{1\alpha} + \delta^{p_1 Q p_3}_{p_1 p_2 p_3} p_{2\alpha} + \delta^{p_1 p_2 Q}_{p_1 p_2 p_3} p_{3\alpha} \right)$$
$$V_\alpha = \frac{1}{\Delta_3} \delta^{p_1 p_2 p_3 \alpha}_{p_1 p_2 p_3 Q}.$$

## 5.1 *Representation of 3–dimensional Vectors, Reflections and Rotations Using the Pauli Matrices*

• **Vectors**

Any vector $\vec{x}$ in 3–dimensional Euclidean space can be represented in the matrix form:

$$\begin{aligned} \hat{x} \ &\equiv \ \vec{x} \cdot \vec{\sigma} = x^i \sigma^i = \begin{pmatrix} x^3 & x^1 - i x^2 \\ x^1 + i x^2 & -x^3 \end{pmatrix}, \\ \det \hat{x} \ &= \ -\vec{x} \cdot \vec{x} \ = \ -(x_1^2 + x_2^2 + x_3^2), \end{aligned}$$

where $\sigma^i$ is the Pauli matrices (see Section 2).
The fundamental property of this representation is

$$(\hat{x})^2 = \vec{x} \cdot \vec{x}\, I, \quad \text{hence} \quad \hat{x}\hat{y} + \hat{y}\hat{x} = 2\, (\vec{x} \cdot \vec{y})\, I. \tag{5.1}$$

One should also note that

$$\hat{x}\hat{y} - \hat{y}\hat{x} \;=\; 2i\,\widehat{\vec{x}\times\vec{y}}.$$

If components of $\vec{x}$ are real, then $\hat{x}^\dagger = \hat{x}$. However, in some practically important cases $(\hat{x})^2 = 0$ and, hence, $\vec{x}\cdot\vec{x} = 0$, components of $\vec{x}$ are complex, say, $x^2 = ix^0$. Then the matrix

$$\hat{x} = \begin{pmatrix} x^3 & x^1 + x^0 \\ x^1 - x^0 & -x^3 \end{pmatrix}$$

represents a vector from 3–dimensional space–time, in that case $\hat{x}^\dagger \neq \hat{x}$.

- **Reflections**

Let $\vec{x}$ be an arbitrary vector and $S$ be the plane orthogonal to some unit vector $\vec{s}$. Then, vector $\vec{x}'$ which results from $\vec{x}$ after the reflection in the plane $S$ is equal to:

$$\vec{x}' = \vec{x} - 2(\vec{x}\vec{s})\vec{s},$$

or, in the considered matrix representation

$$\hat{x}' = -\hat{s}\hat{x}\hat{s}.$$

- **Rotations**

Let $\vec{p}$ and $\vec{q}$ be the two unit vectors with the angle $\theta/2$ between them:

$$\vec{p}^{\,2} = \vec{q}^{\,2} = 1, \quad (\vec{p}\vec{q}) = \cos(\theta/2).$$

Since any spatial rotation is a composition of two reflections, the rotation by the angle $\theta$ in the direction from $\vec{p}$ to $\vec{q}$ is given by the matrix

$$M \;=\; \hat{q}\hat{p},$$

i.e. an arbitrary vector $\vec{x}$ transforms as follows:

$$\hat{x}' = M\hat{x}M^{-1} = \hat{q}\hat{p}\hat{x}\hat{p}\hat{q}.$$

The matrix $M$ can be rewritten in widely used form:

$$M = \hat{q}\hat{p} = \vec{q}\cdot\vec{p}\,I + i\varepsilon_{qpr}\hat{r} = \cos(\theta/2)I \;-\; i\,\vec{n}\vec{\sigma}\;\sin(\theta/2), \tag{5.2}$$

where $\vec{n}\;\sin(\theta/2) \;=\; \vec{p}\times\vec{q}$, $\;-\pi < \theta < \pi$, **positive values of $\theta$ correspond to counterclockwise rotations if one sees from the head of vector $\vec{n}$.** So, we get the two forms of representation of a spatial rotation:

- The rotation by angle $\theta$ about a unit vector $\vec{n}$ is given by

$$M = \cos(\theta/2)I \; - \; i\,\vec{n}\vec{\sigma}\, \sin(\theta/2).$$

- The rotation in the plane of unit vectors $\vec{p}$ and $\vec{q}$ which transforms $\vec{p}$ into $\vec{q}$ is represented by

$$M \;=\; \frac{I \;+\; \hat{q}\hat{p}}{\sqrt{2(1+\vec{q}\cdot\vec{p})}}$$

## 5.2 *Representation of 4–dimensional Vectors, Reflections and Rotations Using the Dirac Matrices*

**• Vectors**

$4 \times 4$ matrix $\hat{x}$, which represents the 4–vector $x^\mu$ in Minkowski space looks as follows:

$$\hat{x} \equiv x^\mu \gamma_\mu = \begin{pmatrix} 0 & 0 & -x^0 - x^3 & -x^1 + ix^2 \\ 0 & 0 & -x^1 - ix^2 & -x^0 + x^3 \\ -x^0 + x^3 & x^1 - ix^2 & 0 & 0 \\ x^1 + ix^2 & -x^0 - x^3 & 0 & 0 \end{pmatrix}.$$

This matrix satisfies the fundamental property

$$(\hat{x})^2 = x \cdot x I \quad \Rightarrow \quad \hat{x}\hat{y} + \hat{y}\hat{x} = 2\, x \cdot y. \qquad (5.3)$$

**• Reflections**

Using the relation (5.3) one can easily derive formulas for the reflections in 3–hyperplanes. Let $x^\mu$ be an arbitrary vector and $S$ be the 3-hyperplane orthogonal to some unit vector $s$. Then, vector $x'$ which results from $x$ after the reflection in the hyperplane $S$ is equal to

$$x' = x - 2\; x \cdot s\; s,$$

or, in the considered matrix representation

$$\hat{x}' = -\hat{s}\hat{x}\hat{s}.$$

**• Lorentz transformations**

Let $p$ and $q$ be the two unit space-like vectors:

$$p \cdot p = q \cdot q = -1.$$

Lorentz transformation, which is a composition of the reflections in 3–hyper-planes determined by the vectors $p$ and $q$ is given by the matrix:

$$M = \hat{q}\hat{p},$$

i.e. an arbitrary vector $x$ transforms as follows:

$$\hat{x}' = M\hat{x}M^{-1} = \hat{q}\hat{p}\hat{x}\hat{p}\hat{q}.$$

The Lorentz transformation in the 2-plane (defined by the vectors $p$ and $q$), which transforms $p$ into $q$, is represented by

$$M = \frac{I + \hat{q}\hat{p}}{\sqrt{2(1 + q \cdot p)}}.$$

# 6 2–COMPONENT SPINORS

## 6.1 *General Properties*

The representation (5.2) of spatial rotations acts on two-component column–vectors:

$$u = \begin{pmatrix} u_1 \\ u_2 \end{pmatrix}$$

which are called **spinors** (see [10] for details).
So, the matrix $M$ in (5.2) can be written in the form

$$M = \begin{pmatrix} \cos(\frac{\theta}{2}) - in_3 \sin(\frac{\theta}{2}) & (n_2 - in_1)\sin(\frac{\theta}{2}) \\ -(n_2 + in_1)\sin(\frac{\theta}{2}) & \cos(\frac{\theta}{2}) + in_3 \sin(\frac{\theta}{2}) \end{pmatrix} = \begin{pmatrix} \alpha & \beta \\ -\beta^* & \alpha^* \end{pmatrix},$$

$$M^\dagger = M^{-1} \text{ hence } M^* = (M^T)^{-1}.$$

The $\varepsilon$-tensor (see Section 1), which defines $SU(2)$-invariant scalar product on spinors has the form:

$$\varepsilon = \begin{pmatrix} 0 & 1 \\ -1 & 0 \end{pmatrix}.$$

The main properties of the $\varepsilon$-tensor are as follows:

$$\begin{aligned} \varepsilon\varepsilon^T &= I, \quad \varepsilon^2 = -I, \\ \varepsilon\hat{x} &= \hat{x}^T\varepsilon. \end{aligned}$$

For each matrix $\hat{x}$ corresponding to a vector $x$ and for each matrix $M$ corresponding to a rotation, one has

$$\varepsilon M = (M^T)^{-1}\varepsilon. \tag{6.1}$$

$M \in SU(2) \Rightarrow (M^T)^{-1} = M^*$ and this relation implies that this representation is equivalent to its complex conjugate, i.e. the conjugate spinor $u' = i\varepsilon u^*$ is transformed with the same matrix $M$ as $u$.
So, the bilinear form on spinors

$$(u, v) = u^\dagger v = u_1^* v_1 + u_2^* v_2 \tag{6.2}$$

is Hermitian $SU(2)$-invariant form, while the form

$$\langle u, v\rangle = u^T\varepsilon v = u_0 v_1 - u_1 v_0 \tag{6.3}$$

is also $SU(2)$-invariant, but not Hermitian. Both these forms are *not invariant* with respect to reflections.

## 6.2 *Spinors and Vectors*

The object $uv^T\varepsilon$ transforms as a vector under the reflections and rotations provided that $u$ and $v$ are spinors:

$$\begin{aligned} &u \rightarrow Mu, \quad v \rightarrow Mv, \\ &uv^T\varepsilon \rightarrow Muv^T\varepsilon M^{-1}. \end{aligned} \tag{6.4}$$

Hence, for any spinor $u$ the relation $\hat{\xi} = uu^T\varepsilon$ defines the vector $\vec{\xi}$. ( $\mathrm{Tr}\,\hat{\xi} = uu^T\varepsilon = u^T\varepsilon u = 0$). The basic property which connects spinor $u$ and vector $\vec{\xi}$ is

$$\hat{\xi}u = 0 \tag{6.5}$$

Since $\hat{\xi}(\hat{\xi}u) = (\xi \cdot \xi)u = 0 \;\; \Rightarrow \;\; \xi \cdot \xi = 0$, vector $\vec{\xi}$ should have complex components. So, this relation expresses

> **equivalence between 2-component spinors and (complex) zero-norm vectors.**

$3 \times 3$ matrix of spatial rotations $O(n, \theta)$ looks as follows:

$$O = \cos\theta\, I + (1 - \cos\theta) \begin{pmatrix} n_1^2 & n_1 n_2 & n_1 n_3 \\ n_2 n_1 & n_2^2 & n_2 n_3 \\ n_3 n_1 & n_3 n_2 & n_3^2 \end{pmatrix} + \sin\theta \begin{pmatrix} 0 & -n_3 & n_2 \\ n_3 & 0 & -n_1 \\ -n_2 & n_1 & 0 \end{pmatrix}.$$

Eigenvalues of this matrix are 1 and $e^{\pm i\theta}$. Eigenvector $(n_1, n_2, n_3)$, corresponding to the eigenvalue 1 directs the axes of the rotation. The other two eigenvectors have zero norm and correspond to eigenspinors of matrix (5.2).

## 6.3 *Representation of Lorentz Transformations by $2 \times 2$ Matrices*

4–vectors of Minkowski space can also be represented by $2 \times 2$ matrices:

$$\begin{aligned} \mathbf{x} \;&\equiv\; x^\mu \sigma^\mu = \begin{pmatrix} x^0 + x^3 & x^1 - ix^2 \\ x^1 + ix^2 & x^0 - x^3 \end{pmatrix}, \\ \bar{\mathbf{x}} \;&\equiv\; x^\mu \bar{\sigma}^\mu = \begin{pmatrix} x^0 - x^3 & -x^1 + ix^2 \\ -x^1 - ix^2 & x^0 + x^3 \end{pmatrix}, \\ \det \mathbf{x} = \det \bar{\mathbf{x}} \;&=\; (x^0)^2 - (x^1)^2 - (x^2)^2 - (x^3)^2 \;=\; x \cdot x, \;\; \mathbf{x}^\dagger = \mathbf{x}. \end{aligned}$$

For each $L:$ det $L=1$ matrix $\mathbf{x}' = L\mathbf{x}L^\dagger$ is Hermitian if $\mathbf{x}$ is Hermitian, and det $\mathbf{x}' =$ det $\mathbf{x}$. Hence $x' = \Lambda x$, where $\Lambda$ is some Lorentz transformation. So, $2\times 2$ matrices

$$L = \begin{pmatrix} a & b \\ c & d \end{pmatrix} : \ ad - bc = 1$$

give the representation of Lorentz group on two–component spinors ($SO(3,1) \sim SL(2,C)/Z_2$, both $L$ and $-L$ correspond to the same Lorentz transformation).
The relation

$$\varepsilon\mathbf{x} = \mathbf{x}^T\varepsilon$$

is also valid in four dimensions, as well as

$$\varepsilon L = (L^T)^{-1}\varepsilon \quad \forall L \in SL(2,C).$$

However, in contrast to $SU(2)$, $\quad (L^T)^{-1} \neq L^*$, and

> **the $SL(2,C)$ representation of Lorentz group is not equivalent to its complex conjugate one.**

Subsequently, one should differentiate 2-spinors, which are transformed with the matrix $L$ from that which are transformed with the matrix $L^*$. Here and below they will be denoted with the help of undot and dot indices, correspondingly:

$$u_A \to L_A^{\ B} u_B, \quad u_{\dot{A}} \to L^{*\ \dot{B}}_{\dot{A}} u_{\dot{B}}$$

The form

$$\langle u, v\rangle = u^T\varepsilon v = u_0 v_1 - u_1 v_0$$

is Lorentz ($SL(2,C)$) - invariant, while a Hermitian $SL(2,C)$-invariant form on two-component spinors does not exist.

## 6.4 *Self–dual and Anti–self–dual Tensors*

The representation of Lorentz group on antisymmetric second rank tensors is **reducible** and can be decomposed into two components. Any antisymmetric tensor $F_{\mu\nu}$ can be written as

$$F_{\mu\nu} = F^+_{\mu\nu} + F^-_{\mu\nu}, \qquad (6.6)$$

where

$$F^{\pm}_{\mu\nu} = \frac{1}{2}\left(F_{\mu\nu} \pm \frac{i}{2}\varepsilon_{\mu\nu\rho\sigma}F^{\rho\sigma}\right).$$

The $F^{+}_{\mu\nu}$ is self–dual, while $F^{-}_{\mu\nu}$ is anti–self–dual:

$$\frac{1}{2}\varepsilon_{\mu\nu\rho\sigma}F^{(\pm)\rho\sigma} = \mp iF^{(\pm)}_{\mu\nu}. \tag{6.7}$$

Each of these irreducible tensors has three independent components which transform through each other under Lorentz transformations. The corresponding second rank spinors look as follows:

$$\phi_{AB} = F^{+}_{\mu\nu}\sigma^{\mu\nu}_{AB}, \quad \bar{\phi}_{\dot{A}\dot{B}} = F^{-}_{\mu\nu}\bar{\sigma}^{\mu\nu}_{\dot{A}\dot{B}}. \tag{6.8}$$

where $\sigma^{\mu\nu}$ and $\bar{\sigma}^{\mu\nu}$ matrices are defined in Section 2. Note, that

$$F^{+}_{\mu\nu}\bar{\sigma}^{\mu\nu}_{AB} = 0, \quad F^{-}_{\mu\nu}\sigma^{\mu\nu}_{\dot{A}\dot{B}} = 0.$$

For example, selfdual electromagnetic field tensor describes photons with negative helicity ($\vec{E} = -i\vec{B}$), while anti–selfdual electromagnetic field tensor describes photons with positive helicity ($\vec{E} = i\vec{B}$). A selfdual tensor in matrix form looks as follows:

$$F_{\mu\nu} = \begin{pmatrix} 0 & -iF_{23} & iF_{13} & -iF_{12} \\ iF_{23} & 0 & F_{12} & F_{13} \\ -iF_{13} & -F_{12} & 0 & F_{23} \\ iF_{12} & -F_{13} & -F_{12} & 0 \end{pmatrix}.$$

## 6.5 *Correspondence Between 2–spinors and 4–bivectors*

Using a 2–spinor $u_A$ one can construct symmetric traceless second rank spinor $\phi_{AB} = u_A u_B$. Then one can transform it into the selfdual tensor $\phi^{\mu\nu}$ with the help of $\sigma^{\mu\nu}$-matricies:

$$\phi^{\mu\nu} \;=\; \sigma^{\mu\nu}_{AB}\phi^{AB}. \tag{6.9}$$

This tensor can be represented in the form

$$\phi^{\mu\nu} = k^{\mu}\Omega^{\nu} - \Omega^{\mu}k^{\nu}$$

here $\hat{k} = uu^\dagger$, $\hat{\Omega} = u\tau^\dagger \Rightarrow k \cdot k = \Omega \cdot \Omega = k \cdot \Omega = 0$, where the gauge spinor $\tau$ is determined by the requirement

$$u_A \tau_B - \tau_A u_B = \epsilon_{AB}. \tag{6.10}$$

The gauge spinor $\tau_A$ is defined up to the transformations $\tau_A \to \tau_A + \lambda u_A$, hence the vector $\Omega^\mu$ is defined up to the transformations $\Omega^\mu \to \Omega^\mu + \lambda k^\mu$. Tensor $\phi^{\mu\nu}$ does not depend on the arbitrariness in spinor $\tau_A$ and vector $\Omega^\mu$.

For example, $\Omega^\mu$ can serve as a (gauge dependent) polarization vector of a photon with momentum $k$, $\phi^{\mu\nu}$ – as electromagnetic field tensor describing this photon.

## 6.6 *Isotropic Tetrads in Minkowski Space*

Any ordered pair of light-like vectors $k_1^\mu$ and $k_2^\mu$:

$$k_1^2 = 0,\ k_2^2 = 0,\ k_1 \cdot k_2 = 1/2,$$

determines another pair of light-like (complex) vectors $\omega_+^\mu$ and $\omega_-^\mu$, which are orthogonal to both $k_1^\mu$ and $k_2^\mu$:

$$\left.\begin{array}{lll} \omega_+ \cdot \omega_+ = 0, & \omega_- \cdot \omega_- = 0, & \omega_+ \cdot \omega_- = -1, \\ \omega_+ \cdot k_i = 0, & \omega_- \cdot k_i = 0. & \end{array}\right\} \tag{6.11}$$

It is sufficient to require that the antisymmetric tensors

$$(k_1^\mu \omega_+^\nu - k_1^\nu \omega_+^\mu) \text{ and } (k_2^\mu \omega_-^\nu - k_2^\nu \omega_-^\mu)$$

are self–dual:

$$\hat{k}_1 \hat{\omega}_+ (1 - \gamma^5) = \hat{k}_2 \hat{\omega}_- (1 - \gamma^5) = 0, \tag{6.12}$$

while tensors

$$(k_1^\mu \omega_-^\nu - k_1^\nu \omega_-^\mu) \text{ and } (k_2^\mu \omega_+^\nu - k_2^\nu \omega_+^\mu)$$

are anti–self–dual:

$$\hat{k}_1 \hat{\omega}_- (1 + \gamma^5) = \hat{k}_2 \hat{\omega}_+ (1 + \gamma^5) = 0. \tag{6.13}$$

One should note that

$$\omega_+(k_2, k_1) \sim \omega_-(k_1, k_2) \text{ and } \omega_-(k_2, k_1) \sim \omega_+(k_1, k_2).$$

• **Vectors $\omega_+$ and $\omega_-$ are determined by $k_1$, $k_2$ and requirements (6.11 – 6.13) up to a factor.**
The vector $\omega_+(\omega_-)$ can be interpreted as the polarization vector of the photon with momentum $k_1$ and *positive* (*negative*) helicity (in the gauge $(k_2 A) = 0$). The explicit expression for $\omega_\pm$ can be obatained by using another arbitrary vector $q$:

$$\omega_\pm^\mu = \frac{1}{\sqrt{2\Delta_3}} (k_1 \cdot k_2 q^\mu - q \cdot k_2 k_1^\mu - q \cdot k_1 k_2^\mu \pm i\varepsilon^{\mu q k_1 k_2}), \tag{6.14}$$

where

$$\Delta_3 = \delta^{k_1 k_2 q}_{k_1 k_2 q} = k_1 \cdot k_2 (2\, k_1 \cdot q\; k_2 \cdot q - q \cdot q\; k_1 \cdot k_2),$$

and $\Delta_3 \neq 0$.
In spite of the presence of an arbitrary vector $q$ in (6.14) both $\omega_+$ and $\omega_-$ are almost independent on $q$: if $q$ varies they acquire only a phase factor. Formula (6.14) can be rewritten in the equivalent form (compare with [11]):

$$\hat{\omega}_\pm = -\frac{1}{\sqrt{2\Delta_3}} \left( \hat{k}_1 \hat{q} \hat{k}_2 \frac{1 \mp \gamma^5}{2} + \hat{k}_2 \hat{q} \hat{k}_1 \frac{1 \pm \gamma^5}{2} \right). \tag{6.15}$$

The following relations are also valid:

$$\omega_+^\mu \omega_-^\nu = k_1^\mu k_2^\nu + k_2^\mu k_1^\nu - g^{\mu\nu} k_1 \cdot k_2 - i\varepsilon^{k_1 k_2 \mu\nu}, \tag{6.16}$$

$$\varepsilon^{\alpha\beta\mu\nu} (k_1)_\alpha (k_2)_\beta (\omega_+)_\mu (\omega_-)_\nu = -\frac{i}{2}, \tag{6.17}$$

$$\hat{\omega}_+ \hat{\omega}_- \gamma^5 = I - \gamma^5 - 2\hat{k}_1 \hat{k}_2. \tag{6.18}$$

For an arbitrary vector $p^\mu$ one has:

$$\begin{aligned} \varepsilon^{k_1 k_2 \omega_- p} &= -\frac{i}{2} \omega_- \cdot p, \quad \varepsilon^{k_1 k_2 \omega_+ p} = \frac{i}{2} \omega_+ \cdot p, \\ \varepsilon^{k_1 \omega_+ \omega_- p} &= -i k_1 \cdot p, \quad \varepsilon^{k_2 \omega_+ \omega_- p} = i k_2 \cdot p, \end{aligned}$$

or, in equivalent form

$$\begin{aligned} \hat{k}_1 \hat{k}_2 \hat{\omega}_- &= \hat{\omega}_- \frac{1+\gamma^5}{2}, \quad \hat{k}_1 \hat{k}_2 \hat{\omega}_+ = \hat{\omega}_+ \frac{1-\gamma^5}{2}, \\ \hat{k}_1 \hat{\omega}_+ \hat{\omega}_- &= -\hat{k}_1 \frac{1-\gamma^5}{2}, \quad \hat{k}_2 \hat{\omega}_+ \hat{\omega}_- = -\hat{k}_2 \frac{1+\gamma^5}{2}. \end{aligned}$$

Let $\mathbf{o}$ and $\iota$ are two arbitrary 2-component spinors which satisfy the relation

$$\mathbf{o}_A \iota_B - \iota_A \mathbf{o}_B = \epsilon_{AB}.$$

Then the vectors

$$k_1 = \mathbf{o}_A \mathbf{o}^\dagger_{\dot{A}}, \quad k_2 = \iota_A \iota^\dagger_{\dot{A}}, \quad \omega_+ = \sqrt{2}\mathbf{o}_A \iota^\dagger_{\dot{A}}, \quad \omega_- = \sqrt{2}\iota_A \mathbf{o}^\dagger_{\dot{A}}$$

satisfy all the above mentioned properties.
There always exists a basis, where all these vectors and spinors look as follows:

$$\mathbf{o} = \begin{pmatrix} 1 \\ 0 \end{pmatrix}, \quad \iota = \begin{pmatrix} 0 \\ 1 \end{pmatrix},$$

$$k_1 = \frac{1}{2}(1,\ 0,\ 0,\ 1), \quad k_2 = \frac{1}{2}(1,\ 0,\ 0,\ -1),$$

$$\omega_+ = \frac{1}{\sqrt{2}}(0,\ 1,\ i,\ 0), \quad \omega_- = \frac{1}{\sqrt{2}}(0,\ 1,\ -i,\ 0).$$

# 7 DIRAC SPINORS

## 7.1 *General Properties*

Dirac spinors $u(p,n)$ and $v(p,n)$ describe the solutions of the Dirac equation with positive and negative energy:

$$(\hat{p}-m)\,u(p,n)=0, \quad (\hat{p}+m)\,v(p,n)=0. \tag{7.1}$$

They are functions of 4-momentum $p^\mu$ on the mass shell $p^0 = \sqrt{\vec{p}^{\,2}+m^2}$. The *conjugated* spinors are defined as follows:

$$\bar{u} = u^\dagger\gamma^0, \quad \bar{v} = v^\dagger\gamma^0,$$

$$\bar{u}(p,n)(\hat{p}-m)=0, \quad \bar{v}(p,n)(\hat{p}+m)=0,$$

and the normalization condition is chosen, so that:

$$\begin{aligned}\bar{u}(p,n)u(p,n) &= +2m,\\ \bar{v}(p,n)v(p,n) &= -2m.\end{aligned}$$

Symbol $n$ stands for the polarization of the fermion. The axial–vector $n^\mu$ of the fermion spin is defined by the relations:

$$\begin{aligned}&\bar{u}(p,n)\gamma^\mu\gamma^5 u(p,n) = m\,n^\mu,\\ &n\cdot n=-1, \quad n\cdot p=0.\end{aligned}$$

The spinor $u(p,n)$ describes a fermion with momentum $p$ and the vector of spin $n$. The spinor $v(p,n)$ describes an antifermion with momentum $p$ and the vector of spin $-n$. (One should note, that axial vector $n$ describing spin of a fermion has only spatial non-zero components in the rest frame of this fermion. However, it transforms together with the vector $p$ under Lorentz transformations.)
Spinors $u(p,n)$ and $v(p,n)$ satisfy the following relations:

$$u(p,n)\,\bar{u}(p,n) = \frac{(\hat{p}+m)\,(1+\gamma^5\hat{n})}{2}, \tag{7.2}$$

$$v(p,n)\,\bar{v}(p,n) = \frac{(\hat{p}-m)\,(1+\gamma^5\hat{n})}{2}, \tag{7.3}$$

$$\hat{n}\gamma^5 u(p,n) = u(p,n), \qquad \hat{n}\gamma^5 v(p,n) = v(p,n), \tag{7.4}$$

as well as the Gordon identities:

$$
\begin{aligned}
\bar{u}(p_1,n_1) \quad \gamma^\mu \quad & u(p_2,n_2) \\
= \quad & \frac{1}{2m}\bar{u}(p_1,n_1)\left[(p_1+p_2)^\mu + \sigma^{\mu\nu}(p_1-p_2)_\nu\right] u(p_2,n_2), \\
\bar{u}(p_1,n_1) \quad \gamma^\mu \quad & \gamma^5 u(p_2,n_2) \\
= \quad & \frac{1}{2m}\bar{u}(p_1,n_1)\left[(p_1-p_2)^\mu\gamma^5 + \sigma^{\mu\nu}(p_1+p_2)_\nu\gamma^5\right] u(p_2,n_2),
\end{aligned}
$$

Both $p^\mu + mn^\mu$ and $p^\mu - mn^\mu$ are light-like vectors. Another couple of light-like (complex) vectors $\omega_+$ and $\omega_-$ determined by $p^\mu + mn^\mu$ and $p^\mu - mn^\mu$, are described in the Subsection 6.6.
Using the vectors $\omega^\pm$ one can obtain the following identities:

$$
\begin{aligned}
\hat{\omega}_+ u(p,-n) &= -\sqrt{2}\, v(p,-n), \quad \hat{\omega}_+ v(p,n) = -\sqrt{2}\, u(p,n), && (7.5) \\
\hat{\omega}_- u(p,n) &= \sqrt{2}\, v(p,n), \quad \hat{\omega}_- v(p,-n) = \sqrt{2}\, u(p,-n), && (7.6) \\
\hat{\omega}_+ u(p,n) &= \hat{\omega}_+ v(p,-n) = \hat{\omega}_- u(p,-n) = \hat{\omega}_- v(p,n) = 0. && (7.7)
\end{aligned}
$$

Let us now introduce the spinors:

$$
\begin{aligned}
\varphi_\pm(p,n) &\equiv u(p,n) \pm v(p,-n), \\
\psi_\pm(p,n) &\equiv u(p,-n) \pm v(p,n).
\end{aligned}
$$

They satisfy the following relations:

$$
\begin{aligned}
(\hat{p}+m\hat{n})\,\varphi_+ &= 2\varphi_-, \quad (\hat{p}-m\hat{n})\,\varphi_- = 2\varphi_+, \\
(\hat{p}+m\hat{n})\,\varphi_- &= (\hat{p}-m\hat{n})\,\varphi_+ = 0, \\
(\hat{p}+m\hat{n})\,\psi_+ &= 2\psi_-, \quad (\hat{p}-m\hat{n})\,\psi_- = 2\psi_+, \\
(\hat{p}+m\hat{n})\,\psi_- &= (\hat{p}-m\hat{n})\,\psi_+ = 0, \\
\hat{\omega}_+\varphi_+ &= \hat{\omega}_+\varphi_- = 0, \\
\hat{\omega}_-\varphi_+ &= \sqrt{2}\psi_+, \quad \hat{\omega}_-\varphi_- = -\sqrt{2}\psi_-, \\
\hat{\omega}_-\psi_+ &= \hat{\omega}_-\psi_- = 0, \\
\hat{\omega}_+\psi_+ &= -\sqrt{2}\varphi_+, \quad \hat{\omega}_+\psi_- = \sqrt{2}\varphi_-, \\
(1+\gamma^5)\,\varphi_+ &= (1-\gamma^5)\,\varphi_- = 0, \\
(1+\gamma^5)\,\psi_- &= (1-\gamma^5)\,\psi_+ = 0.
\end{aligned}
$$

Since $p^\mu + mn^\mu$ and $p^\mu - mn^\mu$ are light-like vectors, spinors $\varphi_\pm$ and $\psi_\pm$ satisfy the massless Dirac equation. $\varphi_+$ and $\psi_-$ describe its right-handed solutions,

$\varphi_-$ and $\psi_+$ - left-handed solutions. In chiral representation they look as follows:

$$\varphi_+ = \begin{pmatrix} 0 \\ \varphi_+^{\dot B} \end{pmatrix}, \ \psi_= \begin{pmatrix} 0 \\ \psi_-^{\dot B} \end{pmatrix}, \ \varphi_- = \begin{pmatrix} \varphi_{-\ A} \\ 0 \end{pmatrix}, \ \psi_+ = \begin{pmatrix} \psi_{+\ A} \\ 0 \end{pmatrix}.$$

where $\varphi_+^{\dot B}$, $\psi_-^{\dot B}$, $\varphi_{-\ A}$ and $\psi_{+\ A}$ are two-component spinors.

## 7.2 *Bilinear Combination Decomposition*

Using the vectors $\omega_\pm$ one can write down the expressions for the bilinear spinor combinations of the type:

$$u(p_1, n_1)\bar u(p_2, n_2) = S + V_\mu \gamma^\mu + T_{\mu\nu}\sigma^{\mu\nu} + A_\mu \gamma^\mu \gamma^5 + P\gamma^5. \qquad (7.8)$$

There is very important case, when all the four vectors $p_1$, $p_2$, $n_1$ and $n_2$ lie in a single plane (this case corresponds to spin–flip and spin–non–flip amplitudes).

Let us denote the polarization vectors which lie in the 2–plane of the vectors $p_1$ and $p_2$ through $N_1$ and $N_2$. They satisfy the conditions

$$N_1 \cdot p_1 = N_2 \cdot p_2 = 0, \ N_1^2 = N_2^2 = -1,$$

and look as follows:

$$N_1^\mu = \frac{1}{\Delta}\left(\frac{p_1 \cdot p_2}{m_1} p_1^\mu - m_1 p_2^\mu\right), \quad N_2^\mu = \frac{1}{\Delta}\left(\frac{p_1 \cdot p_2}{m_2} p_2^\mu - m_2 p_1^\mu\right), \qquad (7.9)$$

where

$$\Delta = \sqrt{(p_1 \cdot p_2)^2 - m_1^2 m_2^2}.$$

A fermion with the momentum $p_i$ and an arbitrary polarization vector $n$, in general, can be represented as a superposition

$$u(p_i, n) = \alpha\, u(p_i, N_i) + \beta\, u(p_i, -N_i),$$

where the coefficients $\alpha$ and $\beta$ depend on the polarization $n$. So, to describe the full set of helicity amplitudes one can restrict the consideration to the following four cases:

$$n_1 = \pm N_1, \ \ n_2 = \pm N_2.$$

Then one has:

$$u\bar{u}(\pm,\pm) = \left( j_+ \frac{1 \pm \gamma^5}{2} - j_- \ \frac{1 \mp \gamma^5}{2} + m_1 \hat{k}_2 - m_2 \hat{k}_1 \right) \frac{\hat{\omega}^\pm}{\sqrt{2}}, \qquad (7.10)$$

$$u\bar{u}(\pm,\mp) = (j_+ \hat{k}_1 + m_1)\hat{k}_2 \frac{1 \pm \gamma^5}{2} + (j_- \hat{k}_2 + m_2)\hat{k}_1 \frac{1 \mp \gamma^5}{2}, \qquad (7.11)$$

where

$$j_\pm = \frac{\sqrt{p_1 \cdot p_2 + m_1 m_2} \pm \sqrt{p_1 \cdot p_2 - m_1 m_2}}{\sqrt{2}},$$

$$k_1^\mu = \frac{1}{2\Delta}\left( j_+ p_1^\mu - \frac{m_1}{m_2} j_- p_2^\mu \right), \ k_2^\mu = \frac{1}{2\Delta}\left( j_+ p_2^\mu - \frac{m_2}{m_1} j_- p_1^\mu \right), \ k_1 \cdot k_2 = \frac{1}{2}.$$

Left hand sides of the equations (7.10) and (7.11) are given in a brief form: for instance, the symbol $u\bar{u}(+,-)$ denotes $u(p_1, n_1 = +N_1)\bar{u}(p_2, n_2 = -N_2)$. The vectors $\omega_+$ and $\omega_-$ are determined by $p_1$ and $N_1$ with the help of relations (7.5 - 7.7) [12, 13].

# 8 GELL–MANN MATRICES

## 8.1 *Main Properties*

The Gell-Mann $3 \times 3$ matrices $\lambda_i (i = 1, \ldots, 8)$ are generators of the group $SU(3)$. Their properties were presented elsewhere [3, 4, 14, 15, 16].
Usually in QCD instead of $\lambda_i$ one deals with matrices $t_i$:

$$t_i \equiv \frac{1}{2} \lambda_i.$$

Eight $\lambda_i$ matrices equal:

$$\lambda_1 = \begin{pmatrix} 0 & 1 & 0 \\ 1 & 0 & 0 \\ 0 & 0 & 0 \end{pmatrix}, \quad \lambda_2 = \begin{pmatrix} 0 & -i & 0 \\ i & 0 & 0 \\ 0 & 0 & 0 \end{pmatrix}, \quad \lambda_3 = \begin{pmatrix} 1 & 0 & 0 \\ 0 & -1 & 0 \\ 0 & 0 & 0 \end{pmatrix},$$

$$\lambda_4 = \begin{pmatrix} 0 & 0 & 1 \\ 0 & 0 & 0 \\ 1 & 0 & 0 \end{pmatrix}, \quad \lambda_5 = \begin{pmatrix} 0 & 0 & -i \\ 0 & 0 & 0 \\ i & 0 & 0 \end{pmatrix}, \quad \lambda_6 = \begin{pmatrix} 0 & 0 & 0 \\ 0 & 0 & 1 \\ 0 & 1 & 0 \end{pmatrix},$$

$$\lambda_7 = \begin{pmatrix} 0 & 0 & 0 \\ 0 & 0 & -i \\ 0 & i & 0 \end{pmatrix}, \quad \lambda_8 = \frac{1}{\sqrt{3}} \begin{pmatrix} 1 & 0 & 0 \\ 0 & 1 & 0 \\ 0 & 0 & -2 \end{pmatrix}.$$

The main properties of $t_i$ (or $\lambda_i$) are as follows:

$$t_i^\dagger = t_i, \quad \det t_i = 0, \quad (i = 1, \ldots, 7), \quad \det t_8 = -\frac{1}{12\sqrt{3}},$$

$$[t^a, t^b] = i f^{abc} t^c, \qquad \{t^a, t^b\} = \frac{1}{3} \delta^{ab} + d^{abc} t^c, \tag{8.1}$$

where $d^{abc} (f^{abc})$ is totally symmetric (anti-symmetric) tensor. The non-zero elements of $f^{abc}$ and $d^{abc}$ are equal to:

$$f_{123} = 1, \; f_{147} = -f_{156} = f_{246} = f_{257} = f_{345} = -f_{367} = \frac{1}{2}, \; f_{458} = f_{678} = \frac{\sqrt{3}}{2},$$

$$d_{146} = d_{157} = -d_{247} = d_{256} = d_{344} = d_{355} = -d_{366} = -d_{377} = \frac{1}{2},$$

$$d_{118} = d_{228} = d_{338} = -d_{888} = \frac{1}{\sqrt{3}}, \; d_{448} = d_{558} = d_{668} = d_{778} = -\frac{1}{2\sqrt{3}}.$$

Throughout this Section we use two additional notations:

$$h^{abc} = d^{abc} + if^{abc}, \quad h^{abc} = h^{bca} = h^{cab}, \quad h^{aab} = 0,$$
$$S(a_1 a_2 \ldots a_n) \equiv t^{a_1} t^{a_2} \ldots t^{a_n}, \; S_R(a_1 a_2 \ldots a_n) \equiv t^{a_n} \ldots t^{a_2} t^{a_1}.$$

Thus from (8.1) one has:

$$t^a t^b = \frac{1}{6}\delta^{ab} + \frac{1}{2}(d^{abk} + if^{abk})t^k \; = \frac{1}{6}\delta^{ab} + \frac{1}{2}h^{abk}t^k. \tag{8.2}$$

## 8.2 *Traces of the $t^a$–matrices*

Trace of any string of $t^a$ matrices can be evaluated recursively using the relation (8.2):

$$\begin{aligned} \mathrm{Tr}S(a_1 a_2 \ldots a_n) \;&=\; \frac{1}{6}\delta^{a_{n-1}a_n}\mathrm{Tr}S(a_1 a_2 \ldots a_{n-2}) \\ &+\; \frac{1}{2}h^{a_{n-1}a_n k}\mathrm{Tr}S(a_1 a_2 \ldots a_{n-2}k). \end{aligned} \tag{8.3}$$

Using (8.1) and (8.3) one gets:

$$\mathrm{Tr}(t^a) = 0, \quad \mathrm{Tr}(t^a t^b) = \frac{1}{2}\delta^{ab}, \quad \mathrm{Tr}(t^a t^b t^c) = \frac{1}{4}(d^{abc} + if^{abc}) = \frac{1}{4}h^{abc},$$
$$\mathrm{Tr}(t^a t^b t^c t^d) = \frac{1}{12}\delta^{ab}\delta^{cd} + \frac{1}{8}h^{abn}h^{ncd},$$
$$\mathrm{Tr}(t^a t^b t^c t^d t^e) = \frac{1}{24}h^{abc}\delta^{de} + \frac{1}{24}\delta^{ab}h^{cde} + \frac{1}{16}h^{abn}h^{nck}h^{kde}.$$

## 8.3 *Fiertz Identity*

The Fiertz identity for $t^a$ has the form:

$$t^a_{ik}t^a_{jl} = \frac{1}{2}(\delta_{il}\delta_{kj} - \frac{1}{3}\delta_{ik}\delta_{jl}). \tag{8.4}$$

Any $3 \times 3$ matrix $A$ can be expanded over set $\{I, t^a\}$:

$$A = a_0 I + a^i t^i, \quad \text{where} \quad a_0 = \frac{1}{3}\mathrm{Tr}A, \quad a^i = 2\,\mathrm{Tr}(t^i A).$$

Decomposition of the two $u_i$ and $\bar{u}_i$ color spinors products into color–singlet and color–octet parts has the form:

$$u_i \bar{u}_j = \frac{\delta_{ij}}{\sqrt{3}} + \sqrt{2}\varepsilon^k t^k_{ij}, \quad \varepsilon^k \varepsilon^l = \delta^{kl}.$$

## 8.4 *Products of the $t^a$–matrices*

The product of $n$ matrices $t^a$ could be written in the form $a_0 + a_i t^i$ using the following relations (see (8.2)):

$$S(a_1 a_2 \dots a_n) = \frac{1}{6}\delta^{a_{n-1}a_n} S(a_1 a_2 \dots a_{n-2}) + \frac{1}{2} h^{a_{n-1}a_n k} S(a_1 a_2 \dots a_{n-2} k). \quad (8.5)$$

Thus, the products of two, three, and four matrices equal:

$$\begin{aligned} t^a t^b &= \frac{1}{6}\delta^{ab} + \frac{1}{2}(d^{abk} + i f^{abk}) t^k = \frac{1}{6}\delta^{ab} + \frac{1}{2} h^{abk} t^k, \\ t^a t^b t^c &= \frac{1}{6}\delta^{ab} t^c + \frac{1}{12} h^{abc} + \frac{1}{4} h^{abk} h^{kcn} t^n, \\ t^a t^b t^c t_d &= \frac{1}{36}\delta^{ab}\delta^{cd} + \frac{1}{24} h^{abk} h^{kcd} + \frac{1}{12}[h^{abk}\delta^{cd} + \delta^{ab} h^{cdk}] t^k \\ &+ \frac{1}{8} h^{abn} h^{cdk} h^{nkp} t^p. \end{aligned}$$

The products of the type $t^k S t^k$ have the form:

$$\begin{aligned} & t^k S t^k = \frac{1}{2}\mathrm{Tr}(S) - \frac{1}{6} S, \\ & t^k t^k = \frac{4}{3} I, \quad t^k t^a t^k = -\frac{1}{6} t^a, \quad t^k t^a t^b t^k = \frac{1}{4}\delta^{ab} - \frac{1}{6} t^a t^b, \\ & t^k t^a t^b t^c t^k = \frac{1}{8} h^{abc} - \frac{1}{6} t^a t^b t^c, \\ & t^k t^a t^b t^c t^d t^k = -\frac{1}{6} t^a t^b t^c t^d + \frac{1}{24}\delta^{ab}\delta^{cd} + \frac{1}{16} h^{abn} h^{ncd}. \end{aligned}$$

The products of the type $SS^{\Pi}$ (here $S^{\Pi}$ is denoted any permutation of the $t^{a_i}$–matrices) are given by

$$\begin{aligned} & t^{a_1} t^{a_2} \dots t^{a_n} \; t^{a_n} \dots t^{a_2} t^{a_1} = \left(\frac{4}{3}\right)^n, \\ & t^a t^b t^a t^b = -\frac{2}{9} I, \;\; t^a t^b t^b t^a = \frac{16}{9} I. \end{aligned}$$

The products of the $S(abc)S^{\Pi}(abc)$ and $S(abcd)S^{\Pi}(abcd)$ are presented on the following tables (in these tables symbol $(abc)$ stands for $t^a t^b t^c$, etc). **Table 8.1.**

The products of the $(abc)$ on the $(abc)^{\Pi}$. All products are contain the common factor $\frac{1}{27}I$.

| | | | | | |
|---|---|---|---|---|---|
| $(abc)$ | 10 | $(bac)$ | 1 | $(cab)$ | $-8$ |
| $(acb)$ | 1 | $(bca)$ | $-8$ | $(cba)$ | 64 |

**Table 8.2.**
The products of the $(abcd)$ on the $(abcd)^{\Pi}$. All products are contain the common factor $\frac{1}{81}I$.

| | | | | | | | |
|---|---|---|---|---|---|---|---|
| $(abcd)$ | $-14$ | $(bacd)$ | $+31$ | $(cabd)$ | $-5$ | $(dabc)$ | $+40$ |
| $(abdc)$ | $+31$ | $(badc)$ | $+\frac{71}{2}$ | $(cadb)$ | $-\frac{1}{2}$ | $(dacb)$ | $+4$ |
| $(acbd)$ | $+31$ | $(bcad)$ | $-5$ | $(cbad)$ | $-\frac{1}{2}$ | $(dbac)$ | $+4$ |
| $(acdb)$ | $-5$ | $(bcda)$ | $+40$ | $(cbda)$ | $+4$ | $(dbca)$ | $-32$ |
| $(adbc)$ | $-5$ | $(bdac)$ | $-\frac{1}{2}$ | $(cdab)$ | $+4$ | $(dcab)$ | $-32$ |
| $(adcb)$ | $-\frac{1}{2}$ | $(bdca)$ | $+4$ | $(cdba)$ | $-32$ | $(dcba)$ | $+256$ |

## 8.5 *Convolutions of $d^{abc}$ and $f^{abc}$ with $t^a$*

The convolutions of the coefficients $d^{abc}$ and $f^{abc}$ with the $t^a$–matrices equal:

$$d^{abc}t^c = t^a t^b + t^b t^a - \frac{1}{3}\delta^{ab}, \quad f^{abc}t^c = i(t^b t^a - t^a t^b),$$
$$h^{abc}t^c = 2t^a t^b - \frac{1}{2}\delta^{ab},$$
$$d^{abk}d^{kcl}t^l = (t^a t^b t^c + t^b t^a t^c + t^c t^a t^b + t^c t^b t^a) - \frac{1}{3}d^{abc}I - \frac{2}{3}\delta^{ab}t^c,$$
$$d^{abk}f^{kcl}t^l = i(-t^a t^b t^c - t^b t^a t^c + t^c t^a t^b + t^c t^b t^a),$$
$$f^{abk}d^{kcl}t^l = i(-t^a t^b t^c + t^b t^a t^c - t^c t^a t^b + t^c t^b t^a) - \frac{1}{3}f^{abc}I,$$
$$f^{abk}f^{kcl}t^l = (-t^a t^b t^c + t^b t^a t^c + t^c t^a t^b - t^c t^b t^a),$$
$$d^{abc}t^a t^b t^c = \frac{10}{9}I, \; f^{abc}t^a t^b t^c = 2iI, \; h^{abc}t^a t^b t^c = -\frac{8}{9}I,$$
$$d^{abc}t^a t^b = \frac{5}{6}t^c, \; f^{abc}t^a t^b = \frac{3}{2}it^c, \; h^{abc}t^a t^b = -\frac{2}{3}t^c.$$

The Jacobi identities for the coefficients $f^{abc}$ and $d^{abc}$ equal:

$$f_{abk}f_{kcl} + f_{bck}f_{kal} + f_{cak}f_{kbl} = 0,$$
$$d_{abk}f_{kcl} + d_{bck}f_{kal} + d_{cak}f_{kbl} = 0.$$

The various relations of a such type were presented in [15]:

$$d_{abk}d_{kcl} + d_{bck}d_{kal} + d_{cak}d_{kbl} = \frac{1}{3}(\delta_{ab}\delta_{cl} + \delta_{ac}\delta_{bl} + \delta_{al}\delta_{bc}),$$
$$f_{abk}f_{kcl} = \frac{2}{3}(\delta_{ac}\delta_{bl} - \delta_{al}\delta_{bc}) + d_{ack}d_{blk} - d_{alk}d_{bck},$$
$$3d_{abk}d_{kcl} = \delta_{ac}\delta_{bl} + \delta_{al}\delta_{bc} - \delta_{ab}\delta_{cl} + f_{ack}f_{blk} + f_{alk}f_{bck},$$
$$d_{aac} = f_{aac} = d_{abc}f_{abm} = 0.$$

$$\begin{aligned}
f_{akl}f_{bkl} &= 3\delta_{ab}, \qquad d_{akl}d_{bkl} = \frac{5}{3}\delta_{ab},\\
f_{pak}f_{kbl}f_{lcp} &= -\frac{3}{2}f_{abc}, \quad d_{pak}f_{kbl}f_{lcp} = -\frac{3}{2}d_{abc},\\
d_{pak}d_{kbl}f_{lcp} &= \frac{5}{6}f_{abc}, \quad d_{pak}d_{kbl}d_{lcp} = -\frac{1}{2}d_{abc},\\
d_{piq}d_{qjm}d_{mkt}d_{tlp} &= \frac{1}{36}(13\delta_{ij}\delta_{kl} - 7\delta_{ik}\delta_{jl} + 13\delta_{il}\delta_{jk} - d_{ikm}d_{mjl}),\\
d_{piq}d_{qjm}d_{mkt}f_{tlp} &= \frac{1}{12}(-7d_{ijm}f_{mkl} + d_{ikm}f_{mjl} + 9d_{ilm}f_{mjk}),\\
d_{piq}d_{qjm}d_{mkt}d_{tlp} &= \frac{1}{36}(-21\delta_{ij}\delta_{kl} + 19\delta_{ik}\delta_{jl} - \delta_{il}\delta_{jk})\\
&+ \frac{1}{6}(d_{ikm}d_{mjl} - 4d_{ilm}d_{mjk}),\\
d_{piq}f_{qjm}f_{mkt}d_{tlp} &= \frac{3}{4}(d_{ikm}f_{mil} + d_{ilm}f_{mkj}),\\
f_{piq}f_{qjm}f_{mkt}f_{tlp} &= \frac{1}{4}(5\delta_{ij}\delta_{kl} + \delta_{ik}\delta_{jl} + 5\delta_{il}\delta_{jk} - 6d_{ikm}d_{mjl}).
\end{aligned}$$

$$\begin{aligned}
&d^{abc}d^{abc} = \tfrac{40}{3}, && f^{abc}f^{abc} = 24,\\
&h^{abc}h^{abc} = -\tfrac{32}{3}, && h^{abc}h^{bac} = \tfrac{112}{3},\\
&d^{abk}d^{klc}d^{cbn}d^{nla} = -\tfrac{20}{3}, && d^{abk}d^{klc}d^{cbn}f^{nla} = 0,\\
&d^{abk}d^{klc}f^{cbn}f^{nla} = 20, && d^{abk}f^{klc}d^{cbn}f^{nla} = -20,\\
&d^{abk}f^{klc}f^{cbn}f^{nla} = 0, && f^{abk}f^{klc}f^{cbn}f^{nla} = 36,\\
&h^{abk}h^{klc}h^{cbn}h^{nla} = -\tfrac{32}{3}.
\end{aligned}$$

## 8.6 *Invariant $SU(3)$ Tensors*

Following the results of [15] we present here the $SU(3)$–invariant tensors of the third, forth, and fifth ranks.

There are two independent tensors of the third rank $(H_{ikl})$:

$$H^{(1)}_{ikl} = d_{ikl} \quad \text{and} \quad H^{(2)}_{ikl} = f_{ikl}.$$

There are eight independent tensors of the forth rank:

$$\begin{array}{llll} H^{(1)}_{ijkl} = \delta_{ij}\delta_{kl}, & H^{(2)}_{ijkl} = \delta_{ik}\delta_{jl}, & H^{(3)}_{ijkl} = \delta_{il}\delta_{jk}, & H^{(4)}_{ijkl} = d_{ijm}d_{klm}, \\ H^{(5)}_{ijkl} = d_{ikm}d_{jlm}, & H^{(6)}_{ijkl} = d_{ijm}f_{klm}, & H^{(7)}_{ijkl} = d_{ikm}f_{jlm}, & H^{(8)}_{ijkl} = d_{ilm}f_{jkm}. \end{array}$$

The convolutions of these tensors of the type $Q^{\alpha\beta} = H^{(\alpha)}_{ijkl}H^{(\beta)}_{ijkl}$ are presented by following matrix:

$$Q^{\alpha\beta} = \begin{pmatrix} 64 & 8 & 8 & 0 & \frac{40}{3} & 0 & 0 & 0 \\ 8 & 64 & 8 & \frac{40}{3} & 0 & 0 & 0 & 0 \\ 8 & 8 & 64 & \frac{40}{3} & \frac{40}{3}0 & 0 & 0 & \\ 0 & \frac{40}{3} & \frac{40}{3} & \frac{200}{9} & -\frac{20}{3} & 0 & 0 & 0 \\ \frac{40}{3} & 0 & \frac{40}{3} & -\frac{20}{3} & \frac{200}{9} & 0 & 0 & 0 \\ 0 & 0 & 0 & 0 & 0 & 40 & -20 & 20 \\ 0 & 0 & 0 & 0 & 0 & -20 & 40 & -20 \\ 0 & 0 & 0 & 0 & 0 & 20 & -20 & 40 \end{pmatrix}.$$

There are 32 independent tensors of the fifth rank $(H_{ijklm})$:

$$\begin{array}{lll} H^{(1)}_{ijklm} = \delta_{ij}d_{klm}, & H^{(2)}_{ijklm} = \delta_{ik}d_{jlm}, & H^{(3)}_{ijklm} = \delta_{ik}d_{ilm}, \\ H^{(4)}_{ijklm} = \delta_{il}d_{jkm}, & H^{(5)}_{ijklm} = \delta_{jl}d_{ikm}, & H^{(6)}_{ijklm} = \delta_{im}d_{jkl}, \\ H^{(7)}_{ijklm} = \delta_{jm}d_{ikl}, & H^{(8)}_{ijklm} = \delta_{kl}d_{ijm}, & H^{(9)}_{ijklm} = \delta_{km}d_{ijl}, \\ H^{(10)}_{ijklm} = \delta_{lm}d_{ijk}, & H^{(11)}_{ijklm} = \delta_{ij}f_{klm}, & H^{(12)}_{ijklm} = \delta_{ik}f_{jlm}, \\ H^{(13)}_{ijklm} = \delta_{jk}f_{ilm}, & H^{(14)}_{ijklm} = \delta_{il}f_{ikm}, & H^{(15)}_{ijklm} = \delta_{jl}f_{ikm}, \\ H^{(16)}_{ijklm} = \delta_{im}f_{jkl}, & H^{(17)}_{ijklm} = \delta_{jm}f_{ikl}, & H^{(18)}_{ijklm} = \delta_{kl}f_{ijm}, \\ H^{(19)}_{ijklm} = \delta_{km}f_{ijl}, & H^{(20)}_{ijklm} = \delta_{lm}f_{iik}, & H^{(21)}_{ijklm} = f_{ijp}f_{pkt}f_{tlm}, \\ H^{(22)}_{ijklm} = f_{ikp}f_{pjt}f_{tlm}, & H^{(23)}_{ijklm} = f_{ijp}f_{plt}f_{tkm}, & H^{(24)}_{ijklm} = f_{ikp}f_{plt}f_{tjm}, \\ H^{(25)}_{ijklm} = f_{kjp}f_{plt}f_{tim}, & H^{(26)}_{ijklm} = f_{ilp}f_{pjt}f_{tkm}, & H^{(27)}_{ijklm} = f_{ijp}d_{pkt}f_{tlm}, \\ H^{(28)}_{ijklm} = d_{ijp}d_{pkt}d_{tlm}, & H^{(29)}_{ijklm} = d_{ikp}d_{pjt}d_{tlm}, & H^{(30)}_{ijklm} = d_{ijp}d_{plt}d_{tkm}, \\ H^{(31)}_{ijklm} = d_{ikp}d_{plt}d_{tjm}, & H^{(32)}_{ijklm} = d_{ilp}d_{pjt}d_{tkm}. & \end{array}$$

# 9 STANDARD MODEL LAGRANGIAN

In this Section we present the basic Lagrangian of the Standard Model(SM), corresponding to the $SU(3) \times SU(2) \times U(1)$ local gauge symmetry (see, for example, [3, 4, 17]). The algebra of the semisimple group $SU(3) \times SU(2) \times U(1)$ is generated by Gell-Mann matrices $t^a = \frac{1}{2}\lambda^a$ (a =1,...8) (Section 8), Pauli matrices $\tau^i = \sigma^i/2$ (Section 2) and hypercharge $Y$ with the following commutation relations

$$
\begin{aligned}
\left[t^a, t^b\right] &= i\, f^{abc} t^c, \\
\left[\tau^i, \tau^j\right] &= i\, \epsilon^{ijk} \tau^k, \\
\left[\tau^i, Y\right] &= \left[t^a, \tau^j\right] = [t^a, Y] = 0.
\end{aligned}
$$

The full SM Lagrangian has the form [3, 4]:

$$
\mathcal{L} = \mathcal{L}_{\mathcal{G}} + \mathcal{L}_{\mathcal{F}} + \mathcal{L}_{\mathcal{H}} + \mathcal{L}_{\mathcal{M}} + \mathcal{L}_{\mathcal{GF}} + \mathcal{L}_{\mathcal{FP}}. \tag{9.1}
$$

Here $\mathcal{L}_{\mathcal{G}}$ is the Yang-Mills Lagrangian without matter fields

$$
\mathcal{L}_{\mathcal{G}} = -\frac{1}{4} F^i_{\mu\nu}(W) F^{\mu\nu}_i(W) - \frac{1}{4} F^{\mu\nu}(W^0) F^{\mu\nu}(W^0) - \frac{1}{4} F^a_{\mu\nu}(G) F^{\mu\nu}_a(G), \tag{9.2}
$$

where $F^i_{\mu\nu}(W)$, $F^a_{\mu\nu}(G)$, $F_{\mu\nu}(W^0)$ are given by

$$
\begin{aligned}
F^i_{\mu\nu}(W) &= \partial_\mu W^i_\nu - \partial_\nu W^i_\mu + g\, \epsilon^{ijk} W^j_\mu W^k_\nu, \\
F_{\mu\nu}(W^0) &= \partial_\mu W^0_\nu - \partial_\nu W^0_\mu, \\
F^a_{\mu\nu}(G) &= \partial_\mu G^a_\nu - \partial_\nu G^a_\mu + g_s\, f^{abc} G^b_\mu G^c_\nu,
\end{aligned}
$$

with $W^i_\mu$, $W^0_\mu$ the $SU(2) \times U(1)$ original gauge fields and $G^a_\mu$ the gluon fields. The infinitesimal gauge transformations of these fields are given by

$$
\begin{aligned}
\delta W^0_\mu &= \partial_\mu \theta(x), \\
\delta W^i_\mu &= \partial_\mu \theta^i - g \epsilon^{ijk} \theta^j W^k_\mu = \mathcal{D}^{ij}_\mu(W) \theta^j \\
\delta G^a_\mu &= \partial_\mu \epsilon^a - g_s f^{abc} \epsilon^b G^c_\mu = \mathcal{D}^{ab}_\mu(G) \epsilon^b
\end{aligned}
$$

Here $\mathcal{D}^{ij}_\mu(W)$ and $\mathcal{D}^{ab}_\mu(G)$ stand for the covariant derivatives, $g_s$ and $g$ are the $SU(3)$ and $SU(2)$ gauge coupling constants, respectively, $\epsilon$ and $\theta^{a(i)}$ are an arbitrary functions depending on the space-time coordinates. It can be easily checked that Lagrangian (9.2) is invariant under these gauge transformations.

Lagrangian $\mathcal{L}_{\mathcal{F}}$ describes coupling of fermions with gauge fields. For simplicity we shall consider one lepton generation, say $e^-$ and $\nu_e$, and three quark generations. Fermions constitute only doublets and singlets in $SU(2)\times U(1)$

$$
\begin{aligned}
R &= e^-_R, \quad L = \begin{pmatrix} \nu_L \\ e^-_L \end{pmatrix} \\
R_I &= (q_I)_R, \quad R_i = (q_i)_R, \quad L_I = \begin{pmatrix} q_I \\ \\ V^{-1}_{Ii} q_i \end{pmatrix}
\end{aligned}
$$

where $L$ and $R$ denote left- and right-handed components of the spinors, respectively:

$$
e_{R,L} = \frac{1 \pm \gamma_5}{2} e.
$$

The neutrino is assumed to be left-handed, while right-handed components of both up- and down-quarks enter in the $\mathcal{L}_{\mathcal{F}}$. Indices $I$ and $i$ numerate three quark generations: $I, i = 1, 2, 3$, and $I(i)$ refers to the up (down) quarks. A possible mixing of quark generations was taken into account by introduction of Kobayashi-Maskava matrix $V_{iI}$ (see, for example, [4, 18] for details). The infinitesimal gauge transformations of fermion fields looks as follows:

$$
\begin{aligned}
\delta\psi_{lep} &= \left( ig'\frac{Y}{2}\theta(x) + ig\frac{\sigma^i}{2}\theta^i(x) \right) \psi_{lep}, \\
\delta\psi_{quark} &= \left( ig'\frac{Y}{2}\theta(x) + ig\frac{\sigma^i}{2}\theta^i(x) + ig_s t^a \theta^a(x) \right) \psi_{quark},
\end{aligned}
$$

where $g'$ is $U(1)$ gauge coupling constant. Obviously, lepton and quark fields belong to the fundamental representation of the $SU(3)\times SU(2)\times U(1)$. Under the requirements of the $SU(3) \times SU(2) \times U(1)$ local gauge symmetry and renormalizability of the theory, the Lagrangian $\mathcal{L}_{\mathcal{F}}$ acquires the following expression

$$
\mathcal{L}_{\mathcal{F}} = i\bar{L}\hat{D}_L L + i\bar{R}\hat{D}_R R + i\sum_I \left( \bar{L}_I \hat{D}^q_L L_I + \bar{R}_I \hat{D}^q_R R_I \right) + i\sum_i \bar{R}_i \hat{D}^q_R R_i, \quad (9.3)
$$

where covariant derivatives are given by

$$
\begin{aligned}
D_{L\,\mu} &= \partial_\mu - ig'\frac{Y}{2}W^0_\mu - ig\frac{\sigma^i}{2}W^i_\mu, \\
D_{R\,\mu} &= \partial_\mu - ig'\frac{Y}{2}W^0_\mu, \\
D^q_{L\,\mu} &= \partial_\mu - ig'\frac{Y}{2}W^0_\mu - ig\frac{\sigma^i}{2}W^i_\mu - ig_s t^a G^a_\mu, \\
D^q_{R\,\mu} &= \partial_\mu - ig'\frac{Y}{2}W^0_\mu - ig_s t^a G^a_\mu.
\end{aligned}
$$

We remind that the value of hypercharge $Y$ is determined by the following relation $Q = \tau_3 + Y/2$ with $Q$ being the charge operator.

Both the gauge fields and fermion ones described above have zero mass, while in the reality all charged fermions are massive and intermediate bosons are known to be very heavy. To make the weak bosons massive one can use Higgs mechanism of spontaneous breakdown of the $SU(2) \times U(1)$ symmetry to the $U(1)$ symmetry. The widely accepted way to do that consists in the introduction of the Higgs $SU(2)$ doublet $\Phi$ (with $Y = 1$). This doublet acquires the nonzero vacuum expectation value:

$$
< \Phi >= \begin{pmatrix} 0 \\ \dfrac{v}{\sqrt{2}} \end{pmatrix}.
$$

The potential term $V(\Phi)$, which can give rise to the symmetry violation, reads

$$
V(\Phi) = -\mu^2 \Phi^+\Phi + \lambda \left(\Phi^+\Phi\right)^2 .
$$

One can easily verify that the vacuum expectation value satisfies to the conditions:

$$
\begin{aligned}
\tau^i < \Phi > &= \frac{1}{2}\sigma^i < \Phi > \neq 0, \\
Q < \Phi > &= \frac{1}{2}\left(\sigma_3 + Y\right) = 0.
\end{aligned}
$$

It means, that only the symmetry generated by $Q$ is not broken on this vacuum. Let us choose the Lagrangian for the Higgs field interaction with gauge fields in the form:

$$\mathcal{L}_{\mathcal{H}} = (D_{L\,\mu}\Phi)^{+}\,(D_L^{\mu}\Phi) - V(\Phi). \tag{9.4}$$

Then one finds that only gauge boson coupling to $Q$ ( i.e. photon) remains massless, while other bosons acquire masses. Diagonalization of the mass matrix gives

$$\begin{aligned}
W_\mu^{\pm} &= \frac{1}{\sqrt{2}}(W_\mu^1 \mp iW_\mu^2),\;\; M_W = \frac{1}{2}gv, && (9.5)\\
Z_\mu &= \frac{1}{\sqrt{g^2+g'^2}}(gW_\mu^3 - g'W_\mu^0),\;\; M_Z = \frac{1}{2}\sqrt{g^2+g'^2}v, && (9.6)\\
A_\mu &= \frac{1}{\sqrt{g^2+g'^2}}(g'W_\mu^3 + gW_\mu^0),\;\; M_A = 0, && (9.7)
\end{aligned}$$

where $W_\mu^{\pm}, Z_\mu$ are charged and neutral weak bosons, $A_\mu$ is the photon. It is suitable to introduce rotation angle $\vartheta_W$ between $(W^3, W^0)$ and $(Z, A)$, which is called the *Weinberg angle*

$$\sin\vartheta_W \equiv g'/\sqrt{g^2+g'^2}. \tag{9.8}$$

The relation of constants $g, g'$ with electromagnetic coupling constants $e$ follows from (9.3). Since the photon coupling with charged particles is proportional to $gg'/\sqrt{g^2+g'^2}$, we should identify this quantity with the electric charge $e$:

$$e = \frac{gg'}{\sqrt{g^2+g'^2}}. \tag{9.9}$$

In order to find mass spectrum in the Higgs sector, let us express doublet $\Phi$ in the form

$$\Phi = \begin{pmatrix} i\omega^{+} \\ \frac{1}{\sqrt{2}}(v + \phi - iz) \end{pmatrix}.$$

One can verify that Nambu-Goldstone bosons $\omega^{\pm}, z$ have zero masses and may be cancelled away by suitable choice of the $SU(2)\times U(1)$ rotation. The only physical component of the Higgs doublet is $\phi$, which acquires mass

$$m_H = \sqrt{2}\mu.$$

The Lagrangian $\mathcal{L}_{\mathcal{M}}$ generates fermion mass terms. Supposing the neutrinos to be massless, we write the Yukawa interaction of the fermions with Higgs doublet in the form

$$\mathcal{L}_{\mathcal{M}} = -f_e \bar{L}\Phi R - \sum_i f_i \bar{L}_i \Phi R_i - \sum_I f_I \bar{L}_I \left(i\sigma^2 \Phi^*\right) R_I + h.c. \tag{9.10}$$

Here we introduced doublet $L_i$ related with $L_I$ by

$$L_i = V_{i\,I}\, L_I,$$

and $f_{I,\,i}$ are the Yukawa coupling constants. Then the masses of fermions in the tree approximation are given by

$$m_{I,\,i} = \frac{f_{I,\,i}\, v}{\sqrt{2}}. \tag{9.11}$$

It is well known that quantization of dynamical systems is governed by Lagrangians having local gauge symmetry requires an additional care. Freedom of redefining gauge and matter fields without changing the Lagrangians leads to the vanishing of some components of the momenta, canonically conjugate to the gauge fields, say

$$\frac{\delta L}{\delta \partial_0 A_\mu} = -F^{0\,\mu} = 0 \quad (\text{for } \mu = 0).$$

To perform the quantization procedure, one should add to the Lagrangian a gauge fixing terms, breaking explicitly the local symmetry. In the functional integral formulation it leads, in the case of non-Abelian gauge symmetry, to modification of the path integral measure [19]. As a result, the measure of the path integral will be multiplied by functional determinant $\Delta(W^a_\mu)$. In order to apply the well known methods of perturbation theory, one may exponentiate $\Delta(W^a_\mu)$ and redefine the initial Lagrangian. It can be made by introducing auxiliary fields $c^a$ and $\bar{c}^a$ which are scalar fields anticommuting with themselves and belonging to the adjoint representation of the Lie algebra. The fields $c^a$ and $\bar{c}^a$ are called Faddeev-Popov ghosts (FP ghosts).

The gauge fixing terms are usually chosen in the form

$$L_{GF} = B^a F^a(W) + \frac{\xi}{2}\left(B^a\right)^2,$$

where $B^a$ are auxiliary fields introduced to linearize this expression, $\xi$ is the gauge parameter, $F^a = \partial^\mu W^a_\mu$. Then FP ghosts enter in the Lagrangian in the following way

$$L_{FP} = -\bar{c}^a \frac{\partial F^a}{\partial W^c_\mu} D^{cb}_\mu(W) c^b. \tag{9.12}$$

As it was pointed above, these additional terms violate local gauge invariance, but the final Lagrangian becomes invariant under the global transformations mixing the gauge fields and FP ghosts. This symmetry, found by by Becchi, Rouet, and Stora, was called BRS symmetry. The BRS infinitesimal transformations are defined by the following relations

$$\begin{aligned}
\delta^{BRS}\psi(x) &= i\beta g c^a(x) t^a \psi(x), \;\; \delta^{BRS} W^a_\mu(x) = \beta D^{ab}_\mu c^b(x), \\
\delta^{BRS}\bar{c}^a(x) &= \beta B^a(x), \;\; \delta^{BRS} c^a(x) = -\frac{\beta}{2} g f^{abc} c^b(x) c^c(x), \\
\delta^{BRS} B^a(x) &= 0.
\end{aligned}$$

Here $\psi$ denotes any matter field, the parameter $\beta$ does not depend on $x$ and anticommutes with $c^a$ and $\bar{c}^a$, as well as with all fermion fields. Using these relations, the formula (9.12) can be written in the brief form:

$$L_{GF} = \frac{\delta}{\delta\beta}\left(\bar{c}^a \delta^{BRS}\left(\partial^\mu W^a_\mu\right)\right), \tag{9.13}$$

where $\delta/\delta\beta$ means left differentiation.

In our case we choose the gauge fixing part of the Lagrangian in the form

$$\begin{aligned}
\mathcal{L}_{\mathcal{GF}} &= B^+(\partial^\mu W^-_\mu + \xi_W M_W \omega^-) + B^-(\partial^\mu W^+_\mu + \xi_W M_W \omega^+) \qquad (9.14) \\
&+ B^Z(\partial^\mu Z_\mu + \xi_Z M_Z z) + B^A(\partial^\mu A_\mu) + B^a(\partial^\mu G^a_\mu) \\
&+ \xi_W B^+ B^- + \frac{\xi_Z}{2} B^Z B^Z + \frac{\xi_A}{2} B^A B^A + \frac{\xi_G}{2} B^a_G B^a_G,
\end{aligned}$$

then FP–ghost Lagrangian looks as follows:

$$\begin{aligned}
&\mathcal{L}_{\mathcal{FP}} = \qquad (9.15) \\
&\frac{\delta}{\delta\beta}\Big\{\bar{c}^{\,+}\delta^{BRS}\left(\partial^\mu W^-_\mu + \xi_W M_W \omega^-\right) + \bar{c}^{\,-}\delta^{BRS}\left(\partial^\mu W^+_\mu + \xi_W M_W \omega^+\right) \\
&+\bar{c}^{\,Z}\delta^{BRS}\left(\partial^\mu Z_\mu + \xi_Z M_Z\, z\right) + \bar{c}^{\,A}\delta^{BRS}\left(\partial^\mu A_\mu\right) + \bar{c}^{\,a}\delta^{BRS}\left(\partial^\mu G^a_\mu\right)\Big\},
\end{aligned}$$

where the fields $c^A, c^Z$ are constructed from original ghosts $c^0, c^3$ just like the bosons $Z_\mu$, $A_\mu$ from initial fields $W^0_\mu, W^3_\mu$.

Now, we are ready to present the total Lagrangian of the **Standard Model** rewritten in the terms of physical fields [17].

$$\begin{aligned}
\mathcal{L}_{\mathcal{G}} &= -\frac{1}{2}F^+_{\mu\nu}F^{-\,\mu\nu} - \frac{1}{4}(F^Z_{\mu\nu})^2 - \frac{1}{4}(F^A_{\mu\nu})^2 - \frac{1}{4}(G^a_{\mu\nu})^2 \\
&+ ie\cot\vartheta_W\left(g^{\alpha\gamma}g^{\beta\delta} - g^{\alpha\delta}g^{\beta\gamma}\right)\left(W^-_\gamma Z_\delta\partial_\alpha W^+_\beta + Z_\gamma W^+_\delta\partial_\alpha W^-_\beta\right. \\
&+ \left.W^+_\gamma W^-_\delta\partial_\alpha Z_\beta\right) + ie\left(g^{\alpha\gamma}g^{\beta\delta} - g^{\alpha\delta}g^{\beta\gamma}\right)\left(W^-_\gamma A_\delta\partial_\alpha W^+_\beta\right. \\
&+ \left.A_\gamma W^+_\delta\partial_\alpha W^-_\beta + W^+_\gamma W^-_\delta\partial_\alpha A_\beta\right) - g_s f^{abc}G^a_\mu G^b_\nu\partial^\mu G^{c\,\nu} \\
&+ e^2\left(g^{\alpha\gamma}g^{\beta\delta} - g^{\alpha\beta}g^{\gamma\delta}\right)W^+_\alpha W^-_\beta A_\gamma A_\delta \\
&+ e^2\cot^2\vartheta_W\left(g^{\alpha\gamma}g^{\beta\delta} - g^{\alpha\beta}g^{\gamma\delta}\right)W^+_\alpha W^-_\beta Z_\gamma Z_\delta) \\
&+ e^2\cot\vartheta_W\left(g^{\alpha\delta}g^{\beta\gamma} + g^{\alpha\gamma}g^{\beta\delta} - 2g^{\alpha\beta}g^{\gamma\delta}\right)W^+_\alpha W^-_\beta A_\gamma Z_\delta \\
&+ \frac{e^2}{2\sin^2\vartheta_W}\left(g^{\alpha\beta}g^{\gamma\delta} - g^{\alpha\gamma}g^{\beta\delta}\right)W^+_\alpha W^+_\beta W^-_\gamma W^-_\delta \\
&- \frac{1}{4}g_s^2 f^{rab}f^{rcd}G^a_\mu G^b_\nu G^{c\,\mu}G^{d\,\nu},
\end{aligned} \tag{9.16}$$

where the field sthrenghtes $G^a_{\mu\nu}, F^+_{\mu\nu}, \ldots$ are given by

$$\begin{aligned}
F^+_{\mu\nu} &= \partial_\mu W^+_\nu - \partial_\nu W^+_\mu, \\
G^a_{\mu\nu} &= \partial_\mu G^a_\nu - \partial_\nu G^a_\mu, \\
&\ldots
\end{aligned}$$

$$\begin{aligned}
\mathcal{L}_{\mathcal{F}} &= i\bar{e}\hat{\partial}e + i\bar{\nu}_L\hat{\partial}\nu_L + i\sum_n \bar{q}_n\hat{\partial}q_n \\
&+ \frac{e}{\sqrt{2}\sin\vartheta_W}\left(\bar{\nu}_L\hat{W}^+e_L + \bar{e}_L\hat{W}^-\nu_L\right) + \frac{e}{\sin 2\vartheta_W}\bar{\nu}_L\hat{Z}\nu_L \\
&+ \frac{e}{\sin 2\vartheta_W}\left(\bar{e}\hat{Z}(2\sin^2\vartheta_W - \frac{1-\gamma_5}{2})e\right) - e\bar{e}\hat{A}e \\
&+ \frac{e}{\sqrt{2}\sin\vartheta_W}\sum_{I,i}\left(\bar{q}_I\hat{W}^+q_{i\,L}(V^+)_{Ii} + \bar{q}_i\hat{W}^-q_{I\,L}V_{iI}\right) \\
&+ \frac{e}{\sin 2\vartheta_W}\sum_I\left(\bar{q}_I\hat{Z}(\frac{1-\gamma_5}{2} - 2Q_I\sin^2\vartheta_W)q_I\right)
\end{aligned} \tag{9.17}$$

$$
\begin{aligned}
&+ \quad \frac{e}{\sin 2\vartheta_W} \sum_i \left( \bar{q}_i \hat{Z} ( \frac{-1+\gamma_5}{2} - 2Q_i \sin^2\vartheta_W ) q_i \right) \\
&+ \quad e \sum_n Q_n \bar{q}_n \hat{A} q_n + g_s \sum_n \bar{q}_n G^a_\mu \gamma^\mu t^a q
\end{aligned}
$$

$$
\begin{aligned}
\mathcal{L}_{\mathcal{H}} \;=\; & \frac{1}{2}(\partial_\mu \phi)^2 - \frac{m_H^2}{2}\phi^2 + \frac{1}{2}(\partial_\mu z)^2 + \partial_\mu \omega^+ \partial^\mu \omega^- \qquad (9.18) \\
+ \; & M_W^2 W^+_\mu W^{-\,\mu} + \frac{1}{2} M_Z^2 Z_\mu^2 - M_W \left( W^-_\mu \partial^\mu \omega^+ + W^+_\mu \partial^\mu \omega^- \right) \\
- \; & M_Z Z_\mu \partial^\mu z + \frac{e M_W}{\sin\vartheta_W} \phi W^+_\mu W^{-\,\mu} + \frac{e M_Z}{\sin 2\vartheta_W} \phi Z_\mu^2 \\
+ \; & \frac{e}{2\sin\vartheta_W} W^{+\,\mu} \left( \omega^- \overleftrightarrow{\partial}_\mu (\phi - iz) \right) + \frac{e}{2\sin\vartheta_W} W^{-\,\mu} \left( \omega^+ \overleftrightarrow{\partial}_\mu (\phi + iz) \right) \\
+ \; & ie(A^\mu + \cot 2\vartheta_W Z^\mu) \left( \omega^- \overleftrightarrow{\partial}_\mu \omega^+ \right) + \frac{e}{\sin 2\vartheta_W} Z^\mu \left( z \overleftrightarrow{\partial}_\mu \phi \right) \\
+ \; & ie M_Z \sin\vartheta_W Z^\mu (W^+_\mu \omega^- - W^-_\mu \omega^+) + ie M_W A^\mu (W^-_\mu \omega^+ - W^+_\mu \omega^-) \\
+ \; & \frac{e^2}{4\sin^2\vartheta_W} \phi^2 (W^+_\mu W^{-\,\mu} + 2\, Z^2_\mu) + \frac{ie^2}{2\cos\vartheta_W} \phi Z^\mu (W^+_\mu \omega^- - W^-_\mu \omega^+) \\
+ \; & \frac{ie^2}{2\sin\vartheta_W} \phi A^\mu (W^-_\mu \omega^+ - W^+_\mu \omega^-) + \frac{e^2}{4\sin^2\vartheta_W} z^2 (W^+_\mu W^{-\,\mu} + 2\, Z^2_\mu) \\
+ \; & \frac{e^2}{2\cos\vartheta_W} z Z^\mu (W^+_\mu \omega^- + W^-_\mu \omega^+) - \frac{e^2}{2\sin\vartheta_W} z A^\mu (W^+_\mu \omega^- + W^-_\mu \omega^+) \\
+ \; & \frac{e^2}{2\sin^2\vartheta_W} \omega^+ \omega^- W^+_\mu W^{-\,\mu} + e^2 \cot^2 2\vartheta_W \omega^+ \omega^- Z^2_\mu + e^2 \omega^+ \omega^- A^2_\mu \\
+ \; & 2e^2 \cot(2\vartheta) \omega^+ \omega^- A^\mu Z_\mu - \frac{e m_H^2}{4 M_W \sin\vartheta_W} \phi^3 - \frac{e m_H^2}{2 M_W \sin\vartheta_W} \omega^+ \omega_- \phi \\
- \; & \frac{e m_H^2}{4 M_W \sin\vartheta_W} z^2 \phi - \frac{e^2 m_H^2}{32 M_W^2 \sin^2\vartheta_W} \phi^4 - \frac{e^2 m_H^2}{32 M_W^2 \sin^2\vartheta_W} z^4 \\
- \; & \frac{e^2 m_H^2}{8 M_W^2 \sin^2\vartheta_W} \omega^+ \omega^- (\phi^2 + z^2) - \frac{e^2 m_H^2}{16 M_W^2 \sin^2\vartheta_W} z^2 \phi^2 \\
- \; & \frac{e^2 m_H^2}{8 M_W^2 \sin^2\vartheta_W} (\omega^+ \omega^-)^2
\end{aligned}
$$

Here symbol $f \overleftrightarrow{\partial}_\mu g$ is used as usual: $f \overleftrightarrow{\partial}_\mu g \equiv f \partial_\mu g - (\partial_\mu f) g$.

$$
\begin{aligned}
\mathcal{L}_{\mathcal{M}} &= -\frac{em_e}{M_Z \sin 2\vartheta_W}\phi\bar{e}e - \frac{e}{M_Z \sin 2\vartheta_W}\sum_n m_n\phi\bar{q}_n q_n \\
&+ \frac{ie\sqrt{2}m_e}{M_Z \sin 2\vartheta_W}\left(\omega^-\bar{e}\nu_L - \omega^+\bar{\nu}_L e\right) + \frac{iem_e}{M_Z \sin 2\vartheta_W} z\bar{e}\gamma_5 e \\
&+ \frac{ie}{\sqrt{2}M_Z \sin 2\vartheta_W}\omega^+ \sum_{I,i}(V^+)_{Ii}\bar{q}_I \left(m_I - m_i - (m_I + m_i)\gamma_5\right) q_i \\
&+ \frac{ie}{\sqrt{2}M_Z \sin 2\vartheta_W}\omega^- \sum_{I,i}(V)_{iI}\bar{q}_i \left(m_i - m_I - (m_I + m_i)\gamma_5\right) q_I \\
&- \frac{ie}{M_Z \sin 2\vartheta_W}\sum_I m_I\bar{q}_I\gamma_5 q_I + \frac{ie}{M_Z \sin 2\vartheta_W}\sum_i m_i\bar{q}_i\gamma_5 q_i
\end{aligned} \tag{9.19}
$$

$$
\begin{aligned}
\mathcal{L}_{\mathcal{FP}} &= -\bar{c}^{\,+}(\partial^2 + \xi_W M_W^2)c^- - \bar{c}^{\,-}(\partial^2 + \xi_W M_W^2)c^+ - \bar{c}^{\,A}\partial^2 c^A \\
&- \bar{c}^{\,Z}(\partial^2 + \xi_Z\ M_Z^2)c^Z - \bar{c}^{\,a}\partial^2 c^a + ie\cot\vartheta_W W^{+\,\mu}\left(\partial_\mu\bar{c}^{\,-}c^Z - \partial_\mu\bar{c}^{\,Z}c^-\right) \\
&+ ieW^{+\,\mu}\left(\partial_\mu\bar{c}^{\,-}c^A - \partial_\mu\bar{c}^{\,A}c^-\right) - ie\cot\vartheta_W W^{-\,\mu}\left(\partial_\mu\bar{c}^{\,+}c^Z - \partial_\mu\bar{c}^{\,Z}c^+\right) \\
&- ieW^{-\,\mu}\left(\partial_\mu\bar{c}^{\,+}c^A - \partial_\mu\bar{c}^{\,A}c^+\right) + ie\cot\vartheta_W Z^\mu\left(\partial_\mu\bar{c}^{\,+}c^- - \partial_\mu\bar{c}^{\,-}c^+\right) \\
&+ ieA^\mu\left(\partial_\mu\bar{c}^{\,+}c^- - \partial_\mu\bar{c}^{\,-}c^+\right) \\
&+ i\omega^+\left(-\xi_W eM_W\cot 2\vartheta_W\bar{c}^{\,-}c^Z - \xi_W eM_W\bar{c}^{\,-}c^A + \frac{\xi_Z e}{2\sin\vartheta_W}M_Z\bar{c}^{\,Z}c^-\right) \\
&+ i\omega^-\left(\xi_W eM_W\cot 2\vartheta_W\bar{c}^{\,+}c^Z + \xi_W eM_W\bar{c}^{\,+}c^A - \frac{\xi_Z e}{2\sin\vartheta_W}M_Z\bar{c}^{\,Z}c^+\right) \\
&+ \frac{i\xi_W e}{2}M_Z\cot\vartheta_W z\left(\bar{c}^-c^+ - \bar{c}^{\,+}c^-\right) - \frac{\xi_W e}{2\sin\vartheta_W}M_W\phi\left(\bar{c}^{\,-}c^+ + \bar{c}^{\,+}c^-\right) \\
&- \frac{\xi_Z e}{\sin 2\vartheta_W}M_Z\phi\bar{c}^{\,Z}c^Z
\end{aligned} \tag{9.20}
$$

# 10 FEYNMAN RULES

## 10.1 *General Remarks*

In this Section we present the complete list of Feynman rules corresponding to the Lagrangian of SM (see (9.16 – 9.20)).

First of all we define the propagators by the relation

$$\Delta_{ij}(k) = i \int d^4x \; e^{-ikx} < 0|T(\phi_i(x)\phi_j(0)|0 >, \qquad (10.1)$$

where $\phi_i$ presents any field. Curly, wavy and zigzag lines denote gluons, photons and weak bosons respectively, while full, dashed and dot lines stand for fermions (leptons and quarks), Higgs particles and ghosts fields, respectively. Arrows on the propagator lines show : for the $W^+$ and $\omega^+$ fields the flow of the positive charge, for the fermion that of the fermion number, and for the ghost that of the ghost number.

The vertices are derived using $L_I$, instead of usual usage of $i\ L_I$. All the momenta of the particles are supposed to flow in. The only exception was made for the ghost fields, where direction of momentum coincides with the direction of ghost number flow. This convention permits to minimize the number of times when the imaginary unit $i$ appears.

It should be noted ones more, that all fields can be "divided" into two parts:

- *physical fields*:
  $A$ (photon), $W^{\pm}$, $Z$, $G$ (gluon), $\psi$, $H$ (Hiigs).
- *non–physical fields*:
  $\omega^{\pm}$, $z$ (pseudogoldstones), $c^{\pm}$, $c^{z}$, $c^{A}$, $c_a$ ( ghosts).

Charged fermions have the electric charge (in the positron charge $e$ units) as follows:

$$Q(e^-) = Q(\mu^-) = Q(\tau^-) = -e,$$
$$Q(e^+) = Q(\mu^+) = Q(\tau^+) = +e,$$
$$Q(u) = Q(c) = Q(t) = +\frac{2}{3}e,$$
$$Q(d) = Q(s) = Q(b) = -\frac{1}{3}e.$$

The electric charge $e$ (or strong coupling constant $g_s$ in QCD) is related to the fine structure constant $\alpha$ (or $\alpha_s$ in QCD) as follows:

$$\alpha_{QED} \equiv \alpha = \frac{e^2}{4\pi}, \quad \alpha_{QCD} \equiv \alpha_s = \frac{g_s^2}{4\pi}.$$

The electric charge, the $\sin\vartheta_W$, and Fermi constant $G_F$ are related as follows:

$$\frac{e}{2\sqrt{2}\sin\vartheta_W} = M_W \sqrt{\frac{G_F}{\sqrt{2}}}. \tag{10.2}$$

Finally, every loop integration is performed by the rule

$$\int \frac{d^d\, k}{i\,(2\pi)^d}, \tag{10.3}$$

and with every fermion or ghost loop we associate extra factor $(-1)$.

## 10.2 *Propagators*

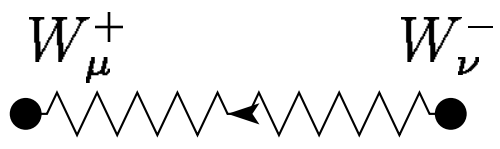

$$\frac{1}{k^2 - M_W^2 + i\varepsilon} \left( g_{\mu\nu} - (1 - \xi_W)\, \frac{k_\mu k_\nu}{k^2 - \xi_W\ M_W^2 + i\varepsilon} \right)$$

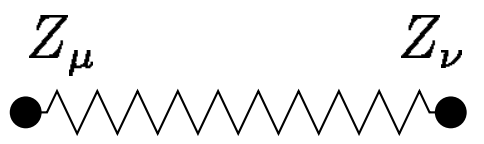

$$\frac{1}{k^2 - M_Z^2 + i\varepsilon} \left( g_{\mu\nu} - (1 - \xi_Z)\, \frac{k_\mu k_\nu}{k^2 - \xi_Z M_Z^2 + i\varepsilon} \right)$$

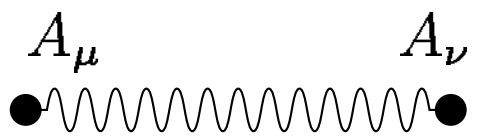

$$\frac{1}{k^2 + i\varepsilon} \left( g_{\mu\nu} - (1 - \xi_A)\, \frac{k_\mu k_\nu}{k^2 + i\varepsilon} \right)$$

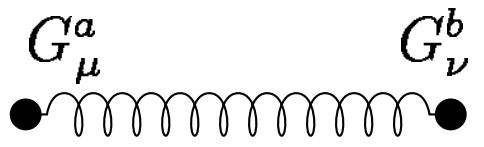

$$\delta^{ab} \frac{1}{k^2 + i\varepsilon} \left( g_{\mu\nu} - (1 - \xi_G)\, \frac{k_\mu k_\nu}{k^2 + i\varepsilon} \right)$$

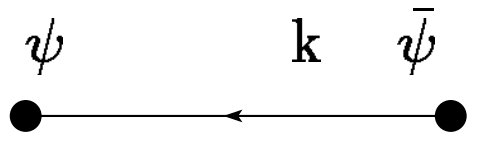

$$- \frac{\hat{k} + m}{k^2 - m^2 + i\varepsilon} = \frac{\hat{k} + m}{m^2 - k^2 + i\varepsilon}$$

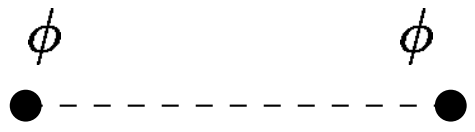

$$-\frac{1}{k^2 - m_H^2 + i\varepsilon}$$

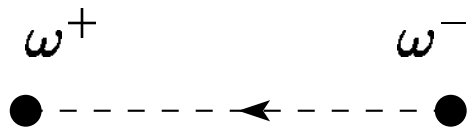

$$-\frac{1}{k^2 - \xi_W\, m_W^2 + i\varepsilon}$$

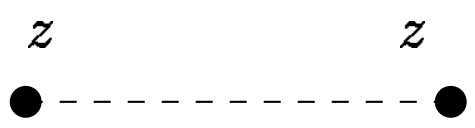

$$-\frac{1}{k^2 - \xi_Z m_Z^2 + i\varepsilon}$$

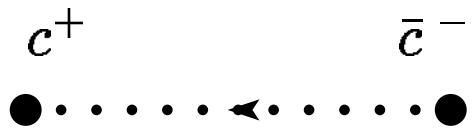

$$-\frac{1}{k^2 - \xi_W\, m_W^2 + i\varepsilon}$$

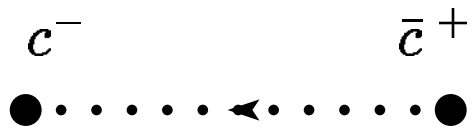

$$-\frac{1}{k^2 - \xi_W\, m_W^2 + i\varepsilon}$$

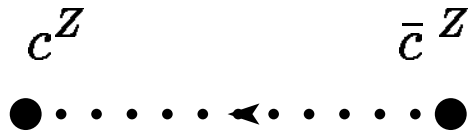

$$-\frac{1}{k^2 - \xi_Z m_Z^2 + i\varepsilon}$$

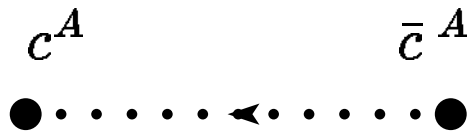

$$-\frac{1}{k^2 + i\varepsilon}$$

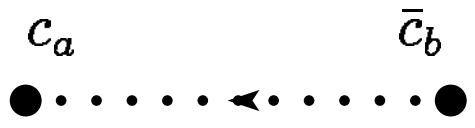

$$-\delta_{ab}\frac{1}{k^2 + i\varepsilon}$$

## 10.3 *Some Popular Gauges*

Here we discuss the explicit forms of the propagators for some popular gauges. Let us consider a theory with free boson Lagrangian:

$$\mathcal{L} = -\frac{1}{4}F_{\mu\nu}^2, \quad F_{\mu\nu} = \partial_\mu A_\nu - \partial_\nu A_\mu.$$

One can fix a gauge in one of three ways [3, 20]:

- **i** to impose a gauge condition,
- **ii** to add a *Gauge Fixing Term* (GFT) to the Lagrangian
- **iii** to fix a form of the Hamiltonian.

In a rigorous theory one should impose two gauge conditions. However, as it is usually accepted, we write only one condition. It should be considered rather as a symbol which denotes acceptable for a given gauge procedure of quantization, described somewhere in literature.

In practical calculations one needs an explicit form of a propagator with satisfactory prescription for poles (which plays a key role in the loop calculations). For this purposes it is sufficient to fix a gauge as mentioned in **ii** and **iii**. Polarization vectors of physical bosons and ghosts should be chosen in accordance with a detailed quantization procedure applicable for a given gauge.

**Covariant gauges.**

1. *Generalized Lorentz gauge.*

- Notation $\partial^\mu A_\mu(x) = B(x)$
- GFT $L_{GF} = -\frac{1}{2\xi}(\partial^\mu A_\mu)^2$
- Propagator
$$D^{\mu\nu} = \frac{1}{k^2 + i\varepsilon}[g^{\mu\nu} - (1-\xi)\frac{k^\mu k^\nu}{k^2 + i\varepsilon}].$$
- Comments

  $\xi = 1$ is Feynman gauge, while $\xi = 0$ is Landau gauge. For the photon (gluon) propagator one should write $\xi_G(\xi_A)$ (see Subsection 10.2).

2. *'t Hooft gauges ($R_\xi$-gauges).*

- Notation $\partial^\mu A^a_\mu(x) - i\xi(v, \tau^a\phi) = B^a(x)$

- GFT $L_{GF} = -\frac{1}{2\xi}(\partial^\mu A^a_\mu)^2$

- Propagator

$$D^{ab}_{\mu\nu} = \frac{\delta^{ab}}{k^2 - M^2 + i\varepsilon}[g_{\mu\nu} - (1-\xi)\frac{k_\mu k_\nu}{k^2 - \xi M^2 + i\varepsilon}].$$

- Comments

  The gauge parameter $\xi = \xi_W(\xi_Z)$ for the case of $W(Z)$ boson (see Subsection 10.2), $v/\sqrt{2}$ is the vacuum expectation value of the gauge field, $\tau^a$ are generators, $M$ is the vector boson mass. $\xi = 1$ is 't Hooft–Feynman gauge, $\xi = 0$ is Landau gauge, $\xi \to \infty$ corresponds to *unitary* gauge. Non–physical gauge bosons should also be taken into account in loop calculations. They also have gauge–dependent propagator, see Subsection 10.2.

**Non-covariant gauges**

3. *Coulomb gauge.*

- Notation $\vec{\partial}\vec{A} = 0, \;\; k = 1, 2, 3.$

- GFT $L_{GF} = -\frac{1}{2\xi}(\partial_k A_k)^2.$

- Propagator

$$D_{\mu\nu} = \frac{1}{k^2 + i\varepsilon}[g_{\mu\nu} - \frac{k_\mu k_\nu - k_0 k_\mu g_{\nu 0} - k_0 k_\nu g_{\mu 0}}{|\vec{k}|^2} - \frac{\xi k^2 k_\mu q_\nu}{|\vec{k}|^4}].$$

- Comments

  The proper *Coulomb gauge* corresponds to the case $\xi = 0$.

4. *The general axial gauge.*

- Notation $n^\mu A_\mu(x) = B(x)$.
- GFT $L_{GF} = -\frac{1}{4\xi}[n^* \cdot \partial \; n \cdot A]^2$.
- Propagator

$$D_{\mu\nu} = \frac{1}{k^2 + i\varepsilon}[g_{\mu\nu} - \frac{(n_\mu k_\nu + k_\mu n_\nu)\; n^* \cdot k}{n \cdot k \; n^* \cdot k + i\varepsilon} - \frac{(\xi k^2 - n^2)\;(n^* \cdot k)^2}{(n \cdot k \; n^* \cdot k + i\varepsilon)^2} k_\mu k_\nu].$$

- Comments

  Feynman rules in this gauge usually do not contain ghosts. As it has been shown in [20] one has to consider an additional gauge vector $n^{*\mu}$ in order to have a correct prescription for poles. The quantization in this gauge was considered, for example, in [21, 22].

  The gauge vector $n^\mu$ has the form:

$$n^\mu \;=\; (n_0; \vec{n}) \;=\; (n_0; \vec{n}_\perp, n_3) \;=\; (n_0; n_1, n_2, n_3).$$

  The explicit form of the component structure of $n^\mu$ and $n^{*\mu}$ should be considered separately in the cases $n^2 > 0$, $n^2 = 0$ and $n^2 < 0$. The following widely used gauges are obtained in the limit $\xi = 0$:

4a. *Temporal gauge:* $n^2 > 0$.

$$D_{\mu\nu} = \frac{1}{k^2 + i\varepsilon}[g_{\mu\nu} - \frac{(n_\mu k_\nu + k_\mu n_\nu)\; n^* \cdot k}{n \cdot k \; n^* \cdot k + i\varepsilon} + \frac{n^2\;(n^* \cdot k)^2}{(n \cdot k \; n^* \cdot k + i\varepsilon)^2} k_\mu k_\nu],$$

$$n^\mu \;=\; (n_0; \vec{n}_\perp, -i|\vec{n}_\perp|); \quad n^{*\mu} \;=\; (n_0; \vec{n}_\perp, i|\vec{n}_\perp|).$$

4b. *Light–cone gauge:* $n^2 = 0$.

$$D_{\mu\nu} = \frac{1}{k^2 + i\varepsilon}[g_{\mu\nu} - \frac{(n_\mu k_\nu + k_\mu n_\nu)\; n^* \cdot k}{n \cdot k \; n^* \cdot k + i\varepsilon}],$$

$$n^\mu \;=\; (|\vec{n}|; \vec{n}); \quad n^{*\mu} \;=\; (|\vec{n}|; -\vec{n}).$$

4c. *Proper axial gauge:* $n^2 < 0$.

$$D_{\mu\nu} = \frac{1}{k^2 + i\varepsilon}[g_{\mu\nu} - \frac{(n_\mu k_\nu + k_\mu n_\nu)\; n^* \cdot k}{n \cdot k \; n^* \cdot k + i\varepsilon} + \frac{n^2\;(n^* \cdot k)^2}{(n \cdot k \; n^* \cdot k + i\varepsilon)^2} k_\mu k_\nu],$$

$$n^\mu \;=\; (|\vec{n}_\perp|; \vec{n}); \quad n^{*\mu} \;=\; (|\vec{n}_\perp|; -\vec{n}).$$

5. *Planar gauge.*

- Notation $n^\mu A_\mu(x) = B(x),\ n^2 \neq 0$.
- GFT $L_{GF} = \frac{1}{2n^2}[\partial_\mu(n\cdot A)]^2$.
- Propagator

$$D_{\mu\nu} = \frac{1}{k^2+i\varepsilon}\left[g_{\mu\nu} - \frac{(n_\mu k_\nu + k_\mu n_\nu)\ n^*\cdot k}{n\cdot k\ n^*\cdot k + i\varepsilon}\right],$$

$$\begin{aligned} n^\mu &= (n_0;\vec{n}_\perp, -i|\vec{n}_\perp|), \quad n^{*\mu} = (n_0;\vec{n}_\perp, i|\vec{n}_\perp|), \quad \text{if} \quad n^2 > 0; \\ n_\mu &= (|\vec{n}_\perp|;\vec{n}), \quad n^*_\mu = (|\vec{n}_\perp|;-\vec{n}) \quad \text{if} \quad n^2 < 0. \end{aligned}$$

- Comments

  Yang-Mills theory is not multiplicatively renormalizable in this gauge. Quantization in this gauge is also poorly understood. This gauge has the same denotation as the axial gauge, that is not suitable. However, that should not lead to confusion (see the beginning of this Subsection).

## 10.4 *Vertices*

### 10.4.1 *Gauge Boson Three–vertices*

$ZW^+W^-$

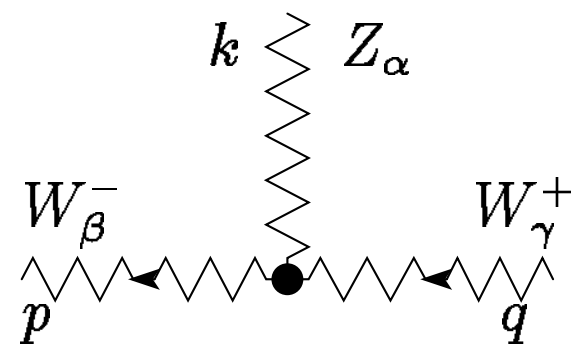

$$e\cot\vartheta_W\left((k-p)^\gamma g^{\alpha\beta}+(p-q)^\alpha g^{\beta\gamma}+(q-k)^\beta g^{\gamma\alpha}\right)$$

$AW^+W^-$

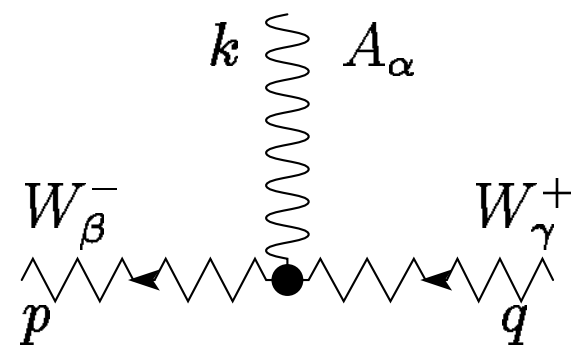

$$e\left((k-p)^\gamma g^{\alpha\beta}+(p-q)^\alpha g^{\beta\gamma}+(q-k)^\beta g^{\gamma\alpha}\right)$$

$GGG$

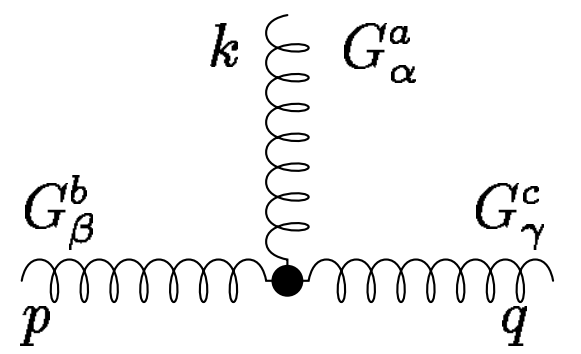

$$ig_s f^{abc}\left((p-k)^\gamma g^{\alpha\beta}+(q-p)^\alpha g^{\beta\gamma}+(k-q)^\beta g^{\gamma\alpha}\right)$$

### 10.4.2 *Gauge Boson Four–vertices*

$AAW^+W^-$

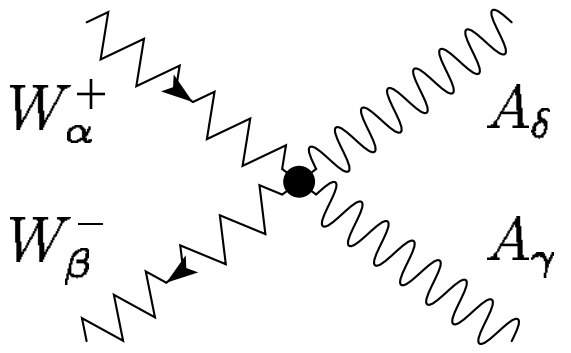

$AZW^+W^-$

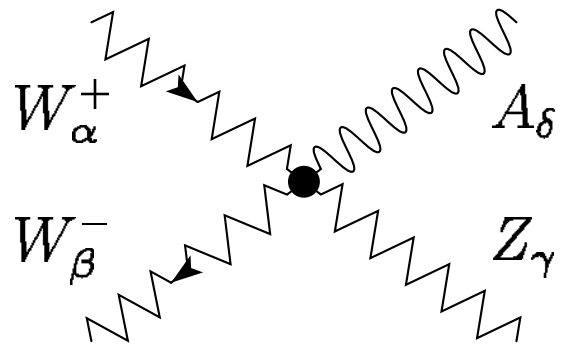

$$e^2(g^{\alpha\gamma}g^{\beta\delta}+g^{\alpha\delta}g^{\beta\gamma}-2g^{\alpha\beta}g^{\gamma\delta})$$

$$e^2\cot\vartheta_W(g^{\alpha\gamma}g^{\beta\delta}+g^{\alpha\delta}g^{\beta\gamma}-2g^{\alpha\beta}g^{\gamma\delta})$$

$\boxed{ZZW^+W^-}$

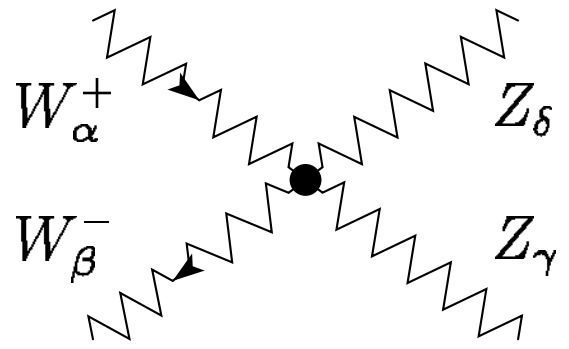

$$e^2\cot^2\vartheta_W(g^{\alpha\gamma}g^{\beta\delta}+g^{\alpha\delta}g^{\beta\gamma}-2g^{\alpha\beta}g^{\gamma\delta})$$

$\boxed{W^+W^-W^+W^-}$

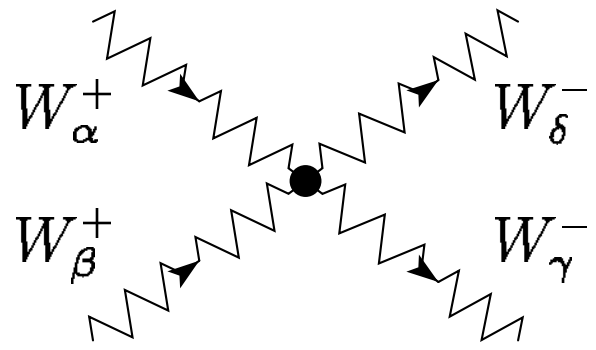

$$-\frac{e^2}{\sin^2\vartheta_W}(g^{\alpha\gamma}g^{\beta\delta}+g^{\alpha\delta}g^{\beta\gamma}-2g^{\alpha\beta}g^{\gamma\delta})$$

$\boxed{GGGG}$

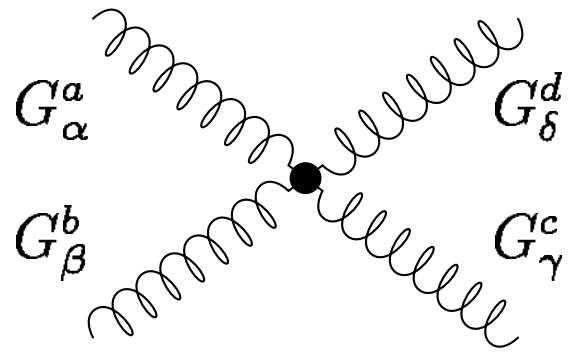

$$\begin{aligned}-g_s^2\ \big(&f^{rab}f^{rcd}(g^{\alpha\gamma}g^{\delta\beta}-g^{\alpha\delta}g^{\beta\gamma})\\ &+f^{rac}f^{rdb}(g^{\alpha\delta}g^{\beta\gamma}-g^{\alpha\beta}g^{\gamma\delta})\\ &+f^{rad}f^{rbc}(g^{\alpha\beta}g^{\gamma\delta}-g^{\alpha\gamma}g^{\delta\beta})\ \big)\end{aligned}$$

### 10.4.3 *Gauge–boson–fermion Vertices*

$\boxed{\bar\psi_I\psi_iW^+}$

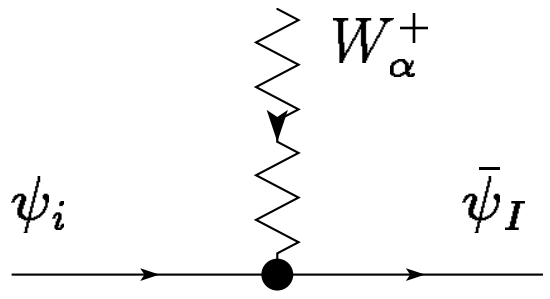

$$\frac{e}{2\sqrt{2}\sin\vartheta_W}V^+_{iI}\gamma^\alpha(1-\gamma^5)$$

$\boxed{\bar\psi_i\psi_IW^-}$

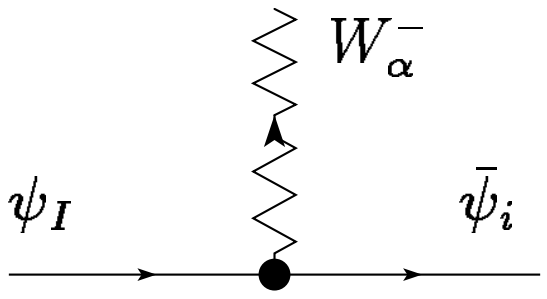

$$\frac{e}{2\sqrt{2}\sin\vartheta_W}V_{iI}\gamma^\alpha(1-\gamma^5)$$

$\bar{\psi}_I \psi_I Z$

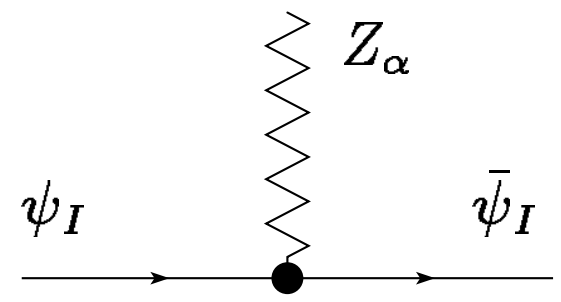

$$\frac{e}{\sin 2\vartheta_W}\gamma^\alpha \left(\frac{1}{2}(1-\gamma^5) - 2Q_I \sin^2\vartheta_W\right)$$

$\bar{\psi}_i \psi_i Z$

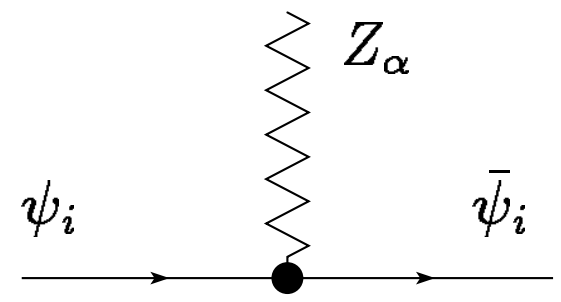

$$\frac{e}{\sin 2\vartheta_W}\gamma^\alpha \left(-\frac{1}{2}(1-\gamma^5) - 2Q_i \sin^2\vartheta_W\right)$$

$\bar{\psi}_n \psi_n A$

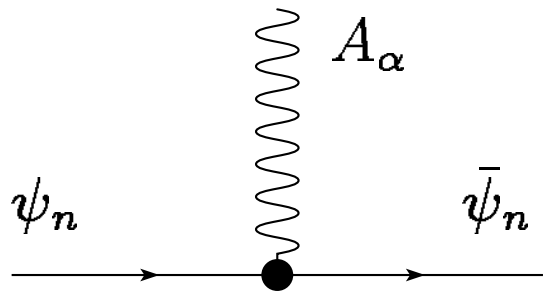

$$eQ_n\gamma^\alpha \quad (n = i, I)$$

$\bar{q} q G$

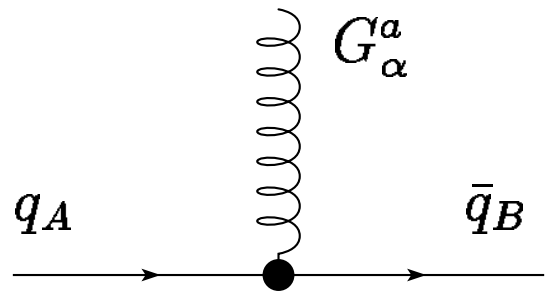

$$g_s\,(t^a)_{BA}\,\gamma^\alpha$$

### 10.4.4 *Gauge–boson–Higgs Three–vertices*

$\phi\,\omega^- W^+$

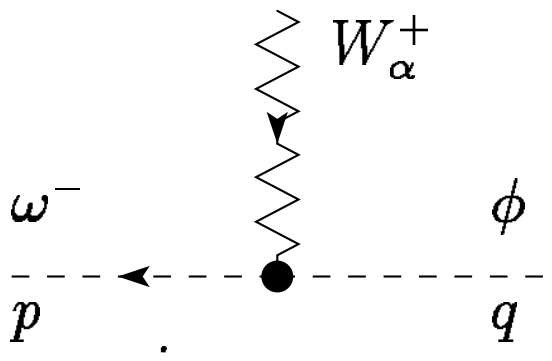

$$\frac{i\,e}{2\sin\vartheta_W}(p-q)^\alpha$$

$\phi\,\omega^+ W^-$

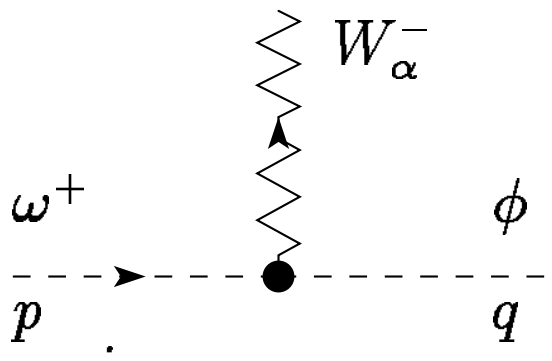

$$\frac{i\,e}{2\sin\vartheta_W}(p-q)^\alpha$$

$z\,\omega^- W^+$

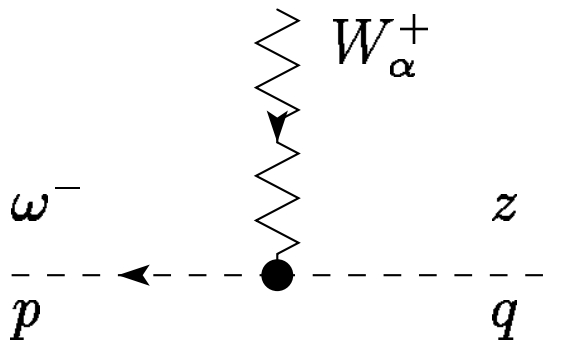

$$\frac{e}{2\sin\vartheta_W}(p-q)^\alpha$$

$z\,\omega^+ W^-$

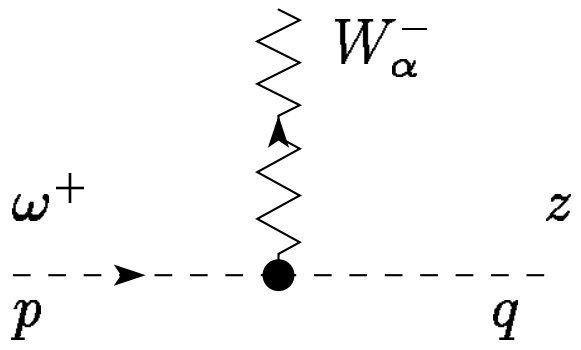

$$-\frac{e}{2\sin\vartheta_W}(p-q)^\alpha$$

$\omega^+\,\omega^- Z$

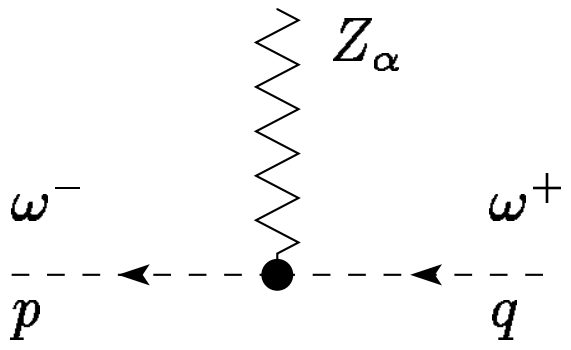

$$-e\cot 2\vartheta_W(p-q)^\alpha$$

$\omega^+\,\omega^- A$

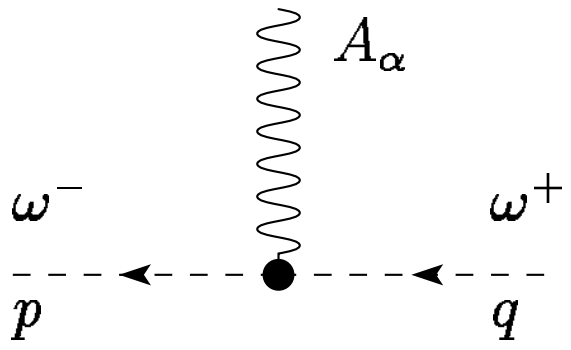

$$-e(p-q)^\alpha$$

$\phi\, zZ$

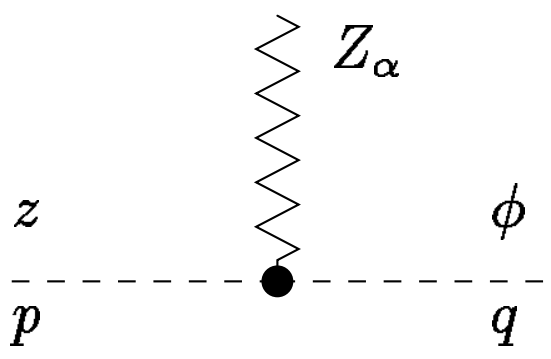

$$\frac{i\ e}{sin2\vartheta_W}(p-q)^\alpha$$

$\phi\, W^+\, W^-$

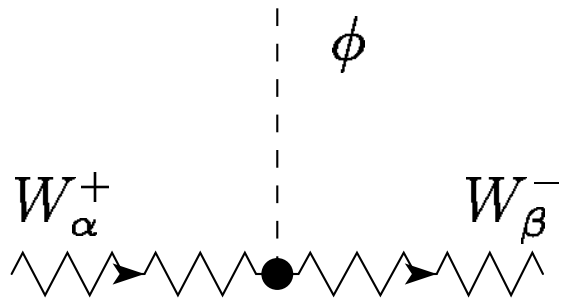

$$e\cot\vartheta_W\ M_Z\ g^{\alpha\beta}$$

$\phi\ Z\ Z$

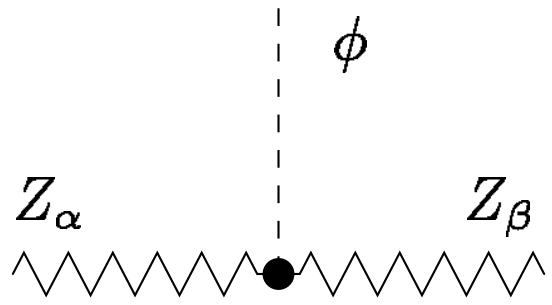

$$\frac{2\ e}{\sin 2\vartheta_W}M_Z\ g^{\alpha\beta}$$

$\omega^- \; W^+ \; Z$

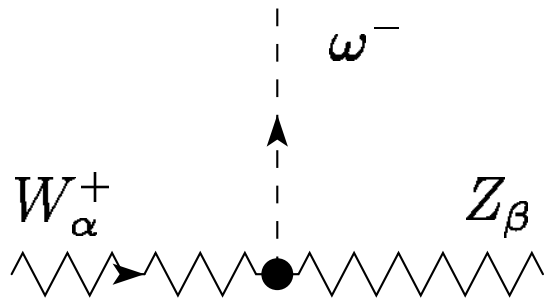

$$i\, e \sin\vartheta_W \; M_Z \; g^{\alpha\beta}$$

$\omega^- \; W^+ \; A$

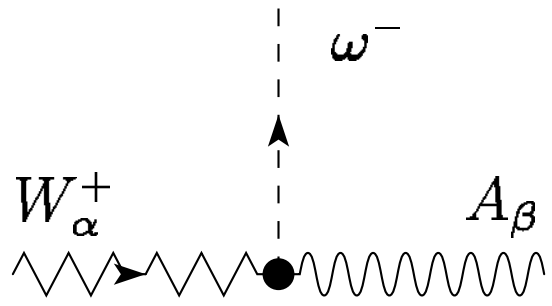

$$-i\; e\; M_W \; g^{\alpha\beta}$$

$\omega^+ \; W^- \; Z$

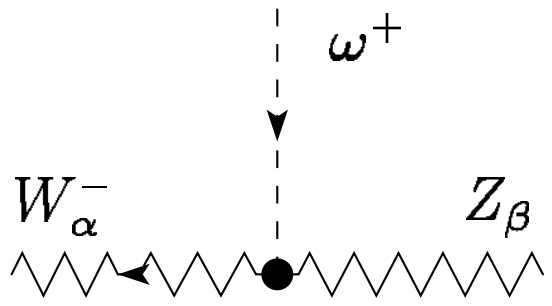

$$-i\, e \sin\vartheta_W \; M_Z \; g^{\alpha\beta}$$

$\omega^+ \; W^- \; A$

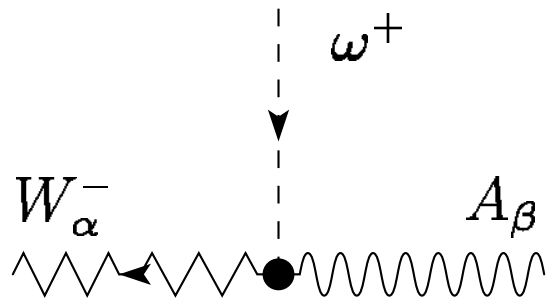

$$i\; e\; M_W \; g^{\alpha\beta}$$

### 10.4.5 *Gauge–boson–Higgs Four–vertices*

$\phi\,\phi\; W^+ \; W^-$

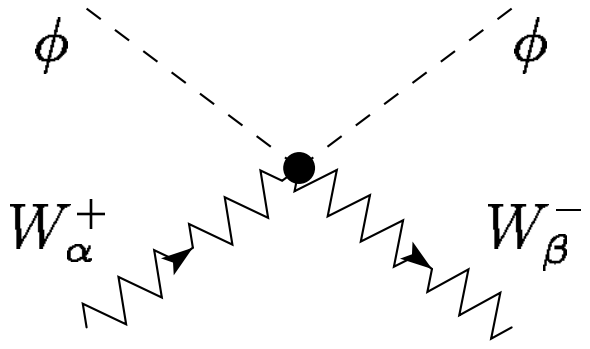

$$\frac{e^2}{2\sin^2\vartheta_W} g^{\alpha\beta}$$

$\phi\,\phi\; Z\; Z$

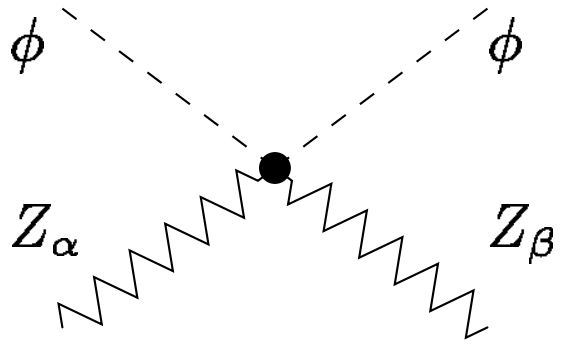

$$\frac{2\; e^2}{\sin^2 2\vartheta_W} g^{\alpha\beta}$$

$\phi\,\omega^- \; W^+ \; Z$

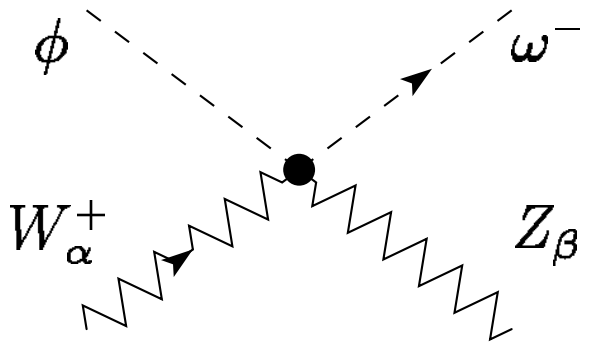

$$\frac{i\; e^2}{2\cos\vartheta_W} g^{\alpha\beta}$$

$\phi\,\omega^+ \; W^- \; Z$

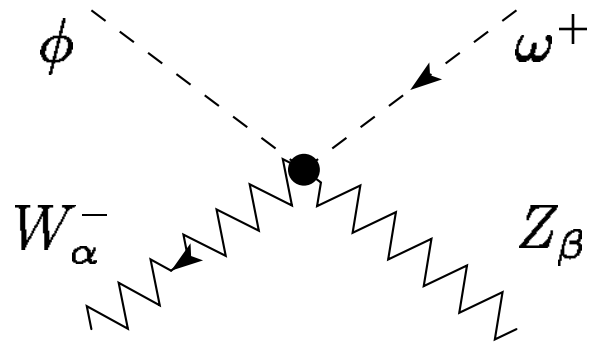

$$-\frac{i\; e^2}{2\cos\vartheta_W} g^{\alpha\beta}$$

$\boxed{\phi\,\omega^-\;W^+\;A}$

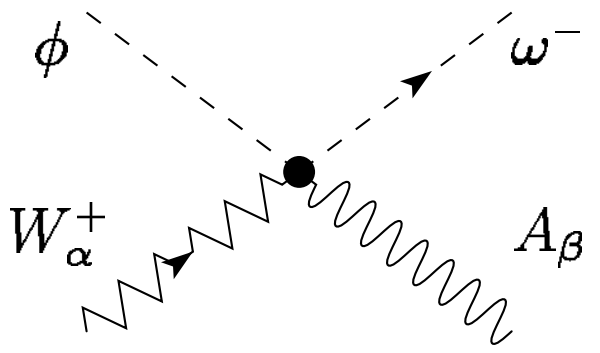

$$-\frac{i\;e^2}{2\sin\vartheta_W}g^{\alpha\beta}$$

$\boxed{\phi\,\omega^+\;W^-\;A}$

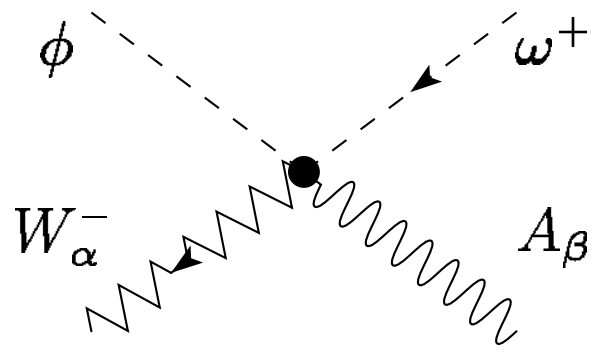

$$\frac{i\;e^2}{2\sin\vartheta_W}g^{\alpha\beta}$$

$\boxed{z\;z\;W^+\;W^-}$

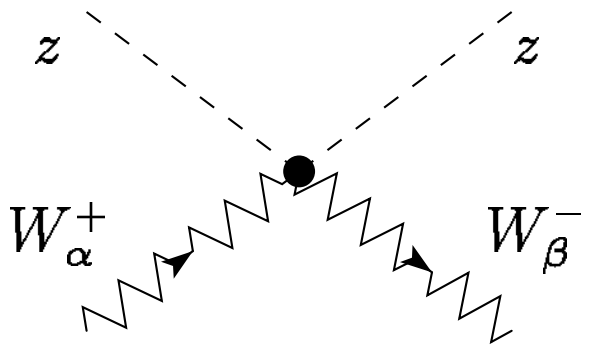

$$\frac{e^2}{2\sin^2\vartheta_W}g^{\alpha\beta}$$

$\boxed{z\;z\;Z\;Z}$

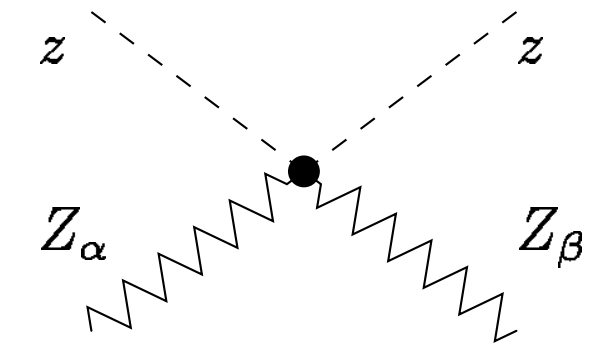

$$\frac{2\;e^2}{\sin^2 2\vartheta_W}g^{\alpha\beta}$$

$\boxed{z\;\omega^-\;W^+\;Z}$

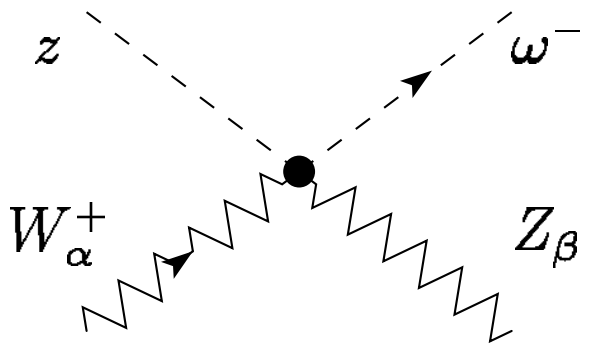

$$\frac{e^2}{2\cos\vartheta_W}g^{\alpha\beta}$$

$\boxed{z\;\omega^+\;W^-\;Z}$

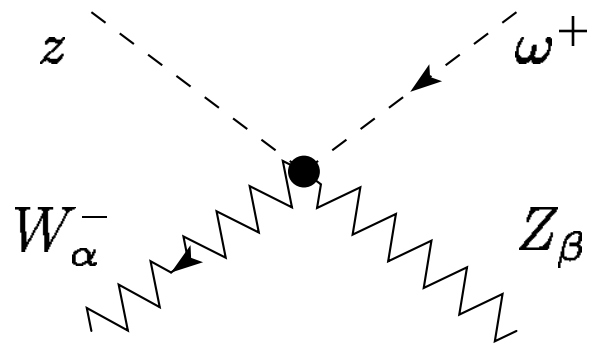

$$\frac{e^2}{2\cos\vartheta_W}g^{\alpha\beta}$$

$\boxed{z\;\omega^-\;W^+\;A}$

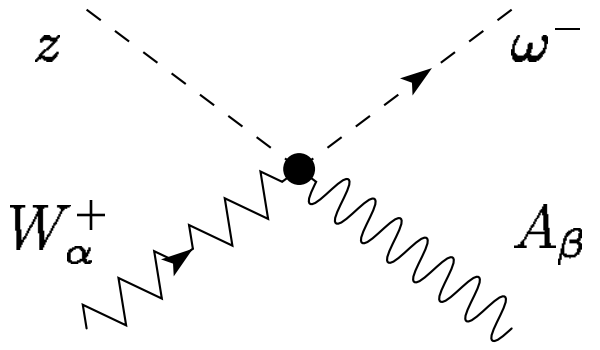

$$-\frac{e^2}{2\sin\vartheta_W}g^{\alpha\beta}$$

$\boxed{z\;\omega^+\;W^-\;A}$

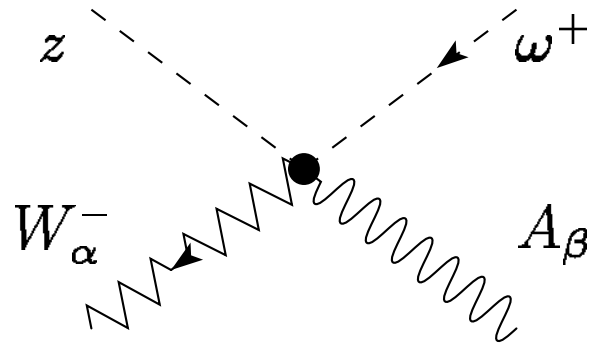

$$-\frac{e^2}{2\sin\vartheta_W}g^{\alpha\beta}$$

$\omega^+ \ \omega^- \ W^+ \ W^-$

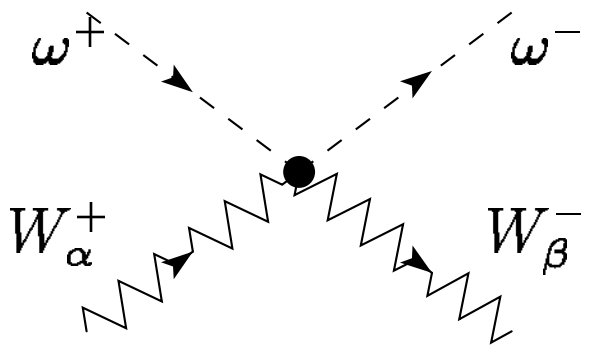

$$\frac{e^2}{2\sin^2\vartheta_W} g^{\alpha\beta}$$

$\omega^+ \ \omega^- \ Z \ Z$

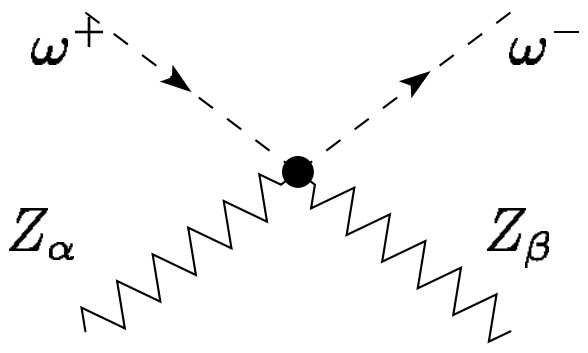

$$2\ e^2 \ \cot^2 2\vartheta_W \ g^{\alpha\beta}$$

$\omega^+ \ \omega^- \ A \ A$

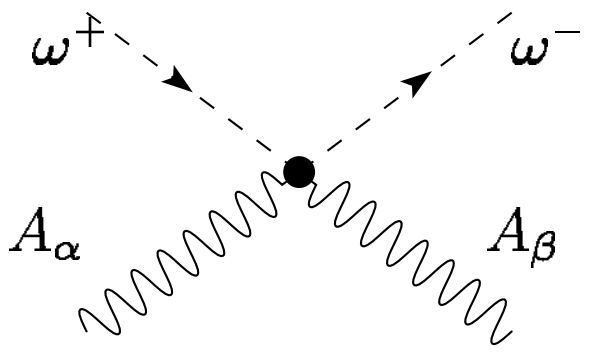

$$2\ e^2 \ g^{\alpha\beta}$$

$\omega^+ \ \omega^- \ A \ Z$

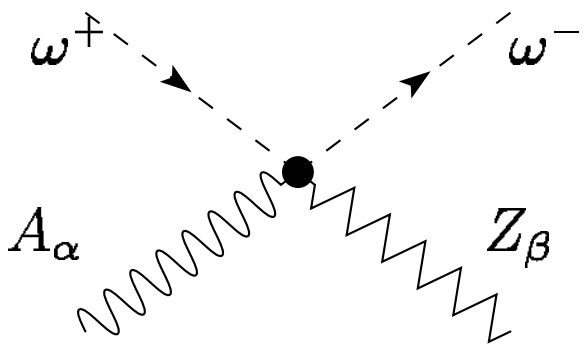

$$2\ e^2 \ \cot 2\vartheta_W \ g^{\alpha\beta}$$

### 10.4.6 *Higgs Three–vertices*

$\phi \ \phi \ \phi$

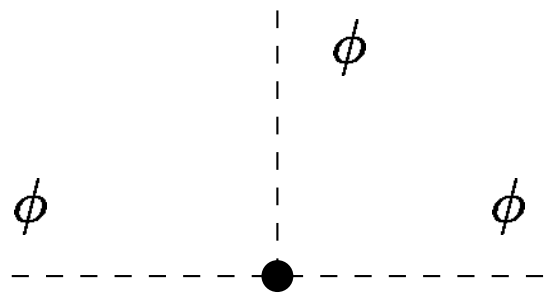

$$-\frac{3\ e}{\sin 2\vartheta_W}\frac{m_H^2}{M_Z}$$

$z \ z \ \phi$

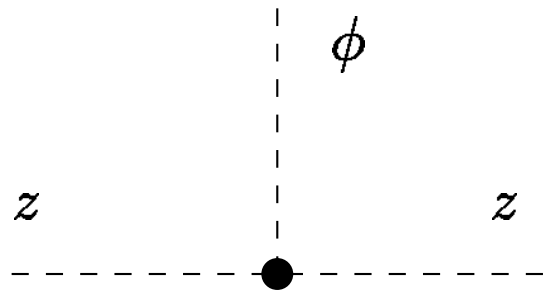

$$-\frac{e}{\sin 2\vartheta_W}\frac{m_H^2}{M_Z}$$

$\omega^+ \ \omega^- \ \phi$

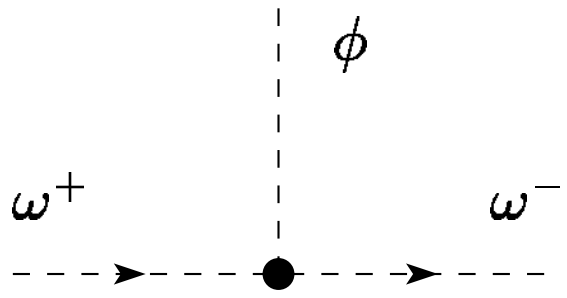

$$-\frac{e}{\sin 2\vartheta_W}\frac{m_H^2}{M_Z}$$

### 10.4.7 *Higgs Four–vertices*

$\boxed{\phi\ \phi\ \phi\ \phi}$

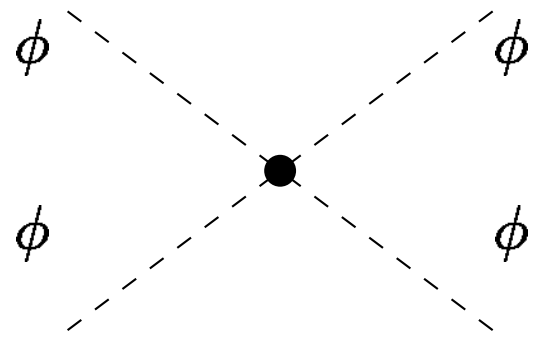

$$-\frac{3\ e^2}{\sin^2 2\vartheta_W}\frac{m_H^2}{M_Z^2}$$

$\boxed{z\ z\ z\ z}$

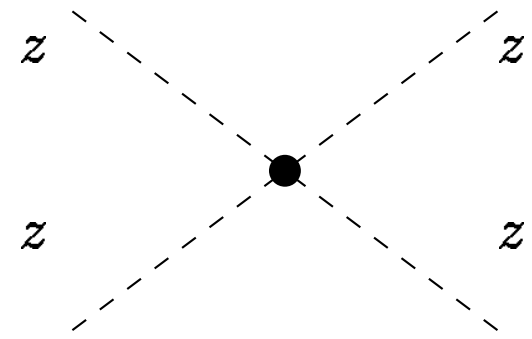

$$-\frac{3\ e^2}{\sin^2 2\vartheta_W}\frac{m_H^2}{M_Z^2}$$

$\boxed{\omega^+\ \omega^-\ \phi\ \phi}$

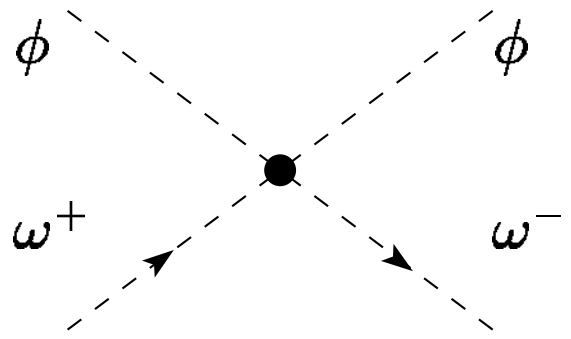

$$-\frac{e^2}{\sin^2 2\vartheta_W}\frac{m_H^2}{M_Z^2}$$

$\boxed{\omega^+\ \omega^-\ z\ z}$

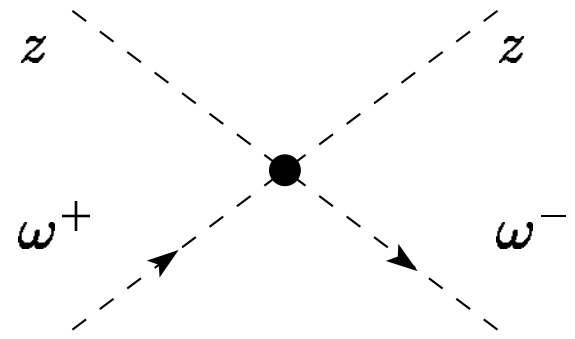

$$-\frac{e^2}{\sin^2 2\vartheta_W}\frac{m_H^2}{M_Z^2}$$

$\boxed{\phi\ \phi\ z\ z}$

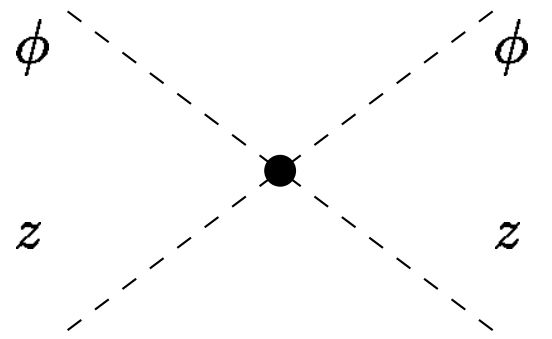

$$-\frac{e^2}{\sin^2 2\vartheta_W}\frac{m_H^2}{M_Z^2}$$

$\boxed{\omega^+\ \omega^-\ \omega^+\ \omega^-}$

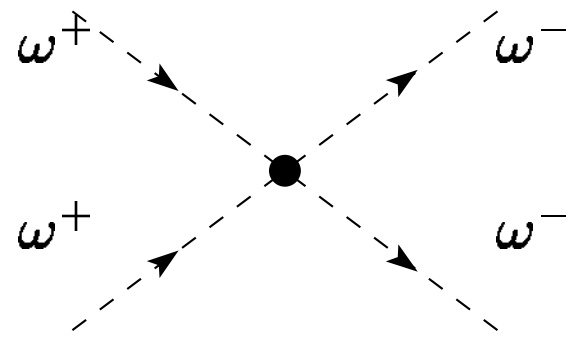

$$-\frac{2\ e^2}{\sin^2 2\vartheta_W}\frac{m_H^2}{M_Z^2}$$

### 10.4.8 *Higgs–boson–fermion Vertices*

$\bar{\psi}_n \psi_n \phi$

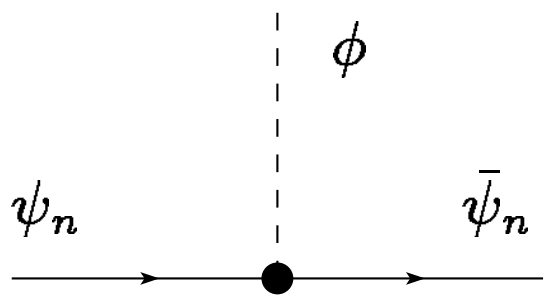

$$-\frac{e}{\sin 2\vartheta_W}\frac{m_n}{M_Z}$$

$\bar{\psi}_I \psi_i \omega^+$

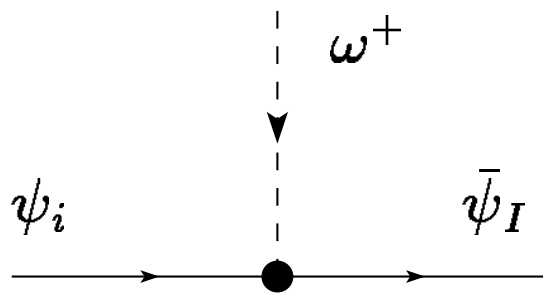

$$-\frac{i\,e}{\sqrt{2}\sin 2\vartheta_W \;\; M_Z} V_{Ii}^{+} \Big( (m_i - m_I) + (m_i + m_I)\,\gamma^5 \Big)$$

$\bar{\psi}_i \psi_I \omega^-$

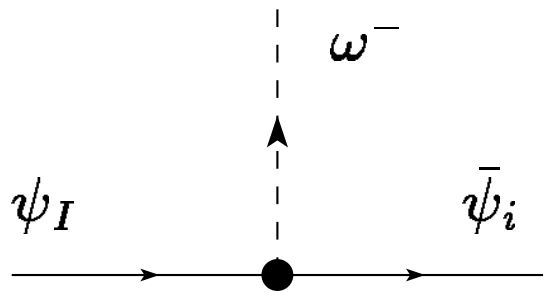

$$-\frac{i\,e}{\sqrt{2}\sin 2\vartheta_W \;\; M_Z} V_{iI} \Big( (m_I - m_i) + (m_i + m_I)\,\gamma^5 \Big)$$

$\bar{\psi}_I \psi_I z$

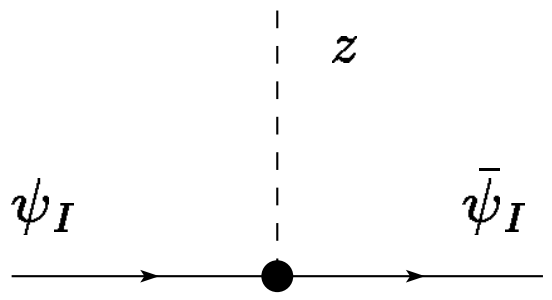

$$-\frac{i\,e}{\sin 2\vartheta_W}\frac{m_I}{M_Z}\gamma^5$$

$\bar{\psi}_i \psi_i z$

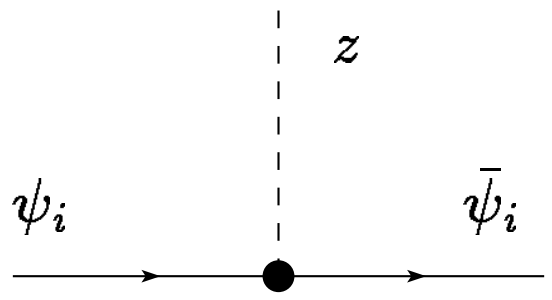

$$\frac{i\,e}{\sin 2\vartheta_W}\frac{m_i}{M_Z}\gamma^5$$

### 10.4.9 *Gauge–boson–ghost Vertices*

$\boxed{\bar{c}^{\,-}\ c^{Z}\ W^{+}}$

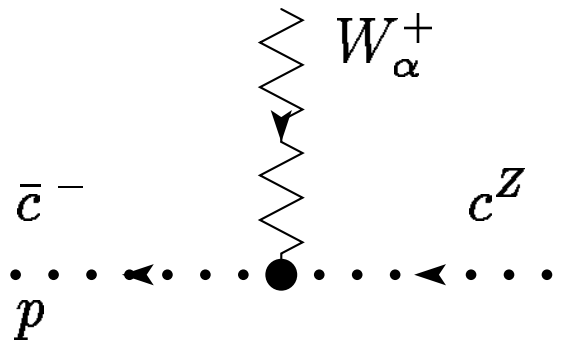

$-e\ \cot\vartheta_W\ p^{\alpha}$

$\boxed{\bar{c}^{\,Z}\ c^{-}\ W^{+}}$

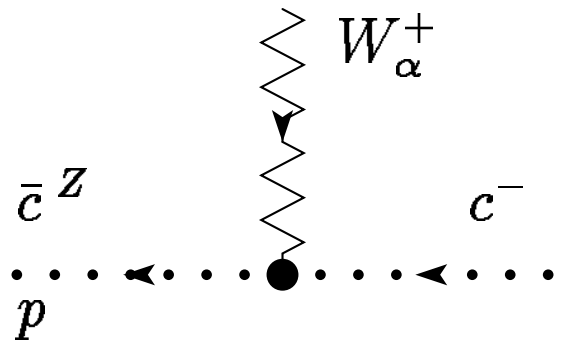

$e\ \cot\vartheta_W\ p^{\alpha}$

$\boxed{\bar{c}^{\,-}\ c^{A}\ W^{+}}$

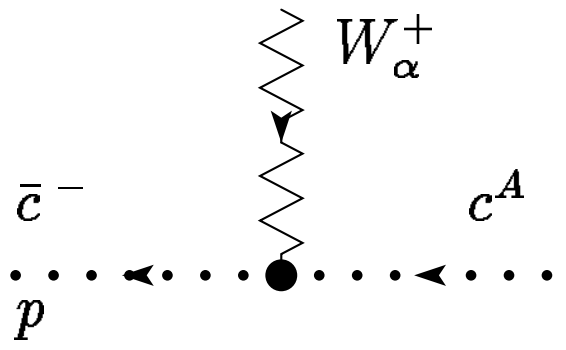

$-e\ p^{\alpha}$

$\boxed{\bar{c}^{\,A}\ c^{-}\ W^{+}}$

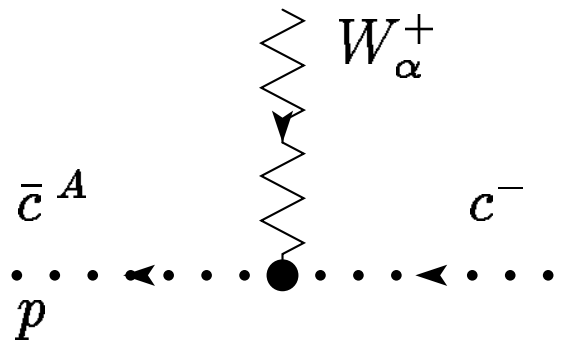

$e\ p^{\alpha}$

$\boxed{\bar{c}^{\,+}\ c^{Z}\ W^{-}}$

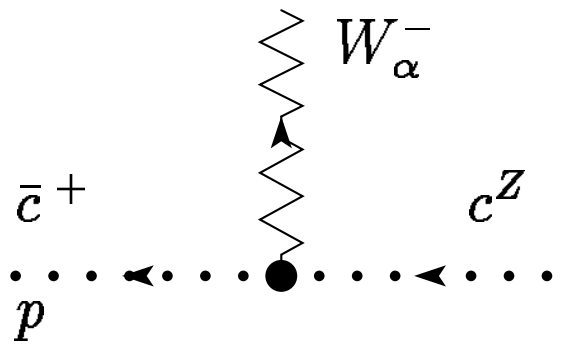

$e\ \cot\vartheta_W\ p^{\alpha}$

$\boxed{\bar{c}^{\,Z}\ c^{+}\ W^{-}}$

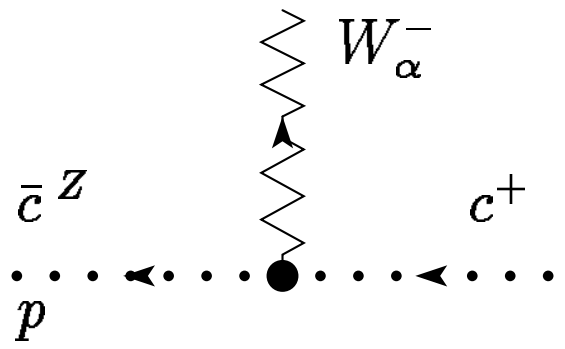

$-e\ \cot\vartheta_W\ p^{\alpha}$

$\bar{c}^{\,+}\ c^{A}\ W^{-}$

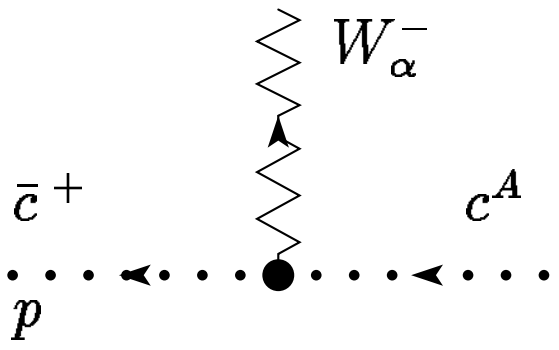

$e\ p^\alpha$

$\bar{c}^{\,A}\ c^{+}\ W^{-}$

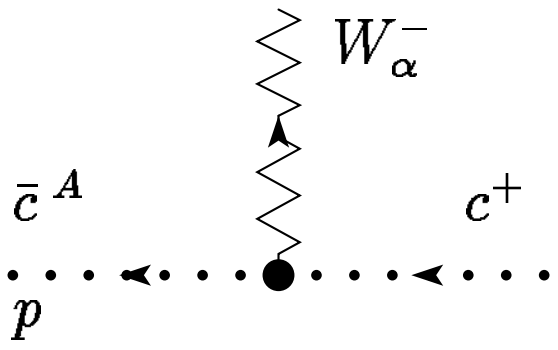

$-e\ p^\alpha$

$\bar{c}^{\,+}\ c^{-}\ Z$

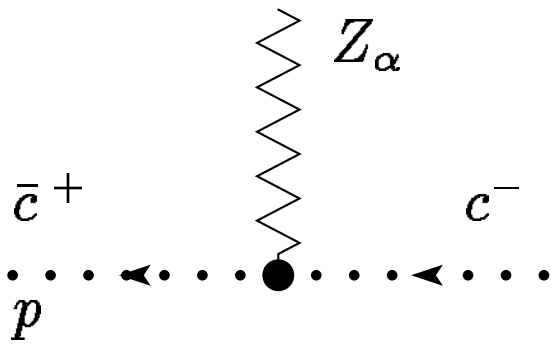

$-e\ \cot\vartheta_W\ p^\alpha$

$\bar{c}^{\,-}\ c^{+}\ Z$

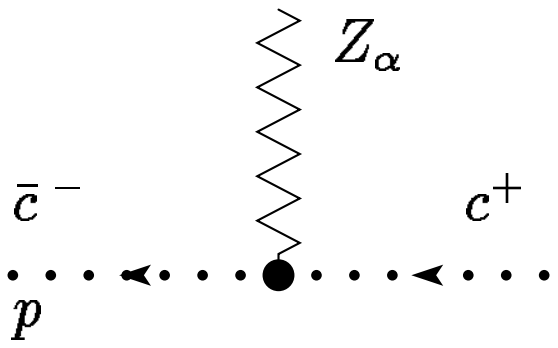

$e\ \cot\vartheta_W\ p^\alpha$

$\bar{c}^{\,+}\ c^{-}\ A$

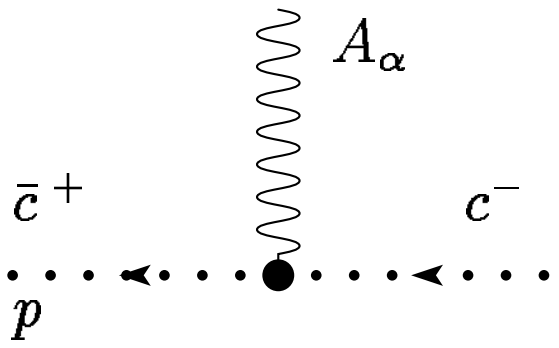

$-e\ p^\alpha$

$\bar{c}^{\,-}\ c^{+}\ A$

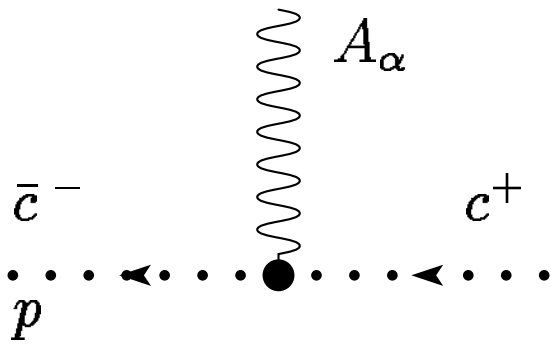

$e\ p^\alpha$

$\bar{c}\ c\ G$

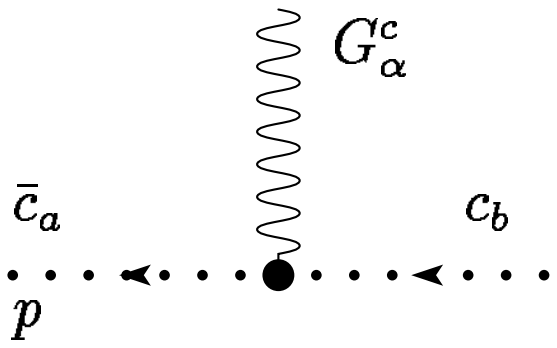

$-ig_s f_{abc} p^\alpha$

### 10.4.10 *Higgs–ghost Vertices*

$\bar{c}^{\,-}\ c^{Z}\ \omega^{+}$

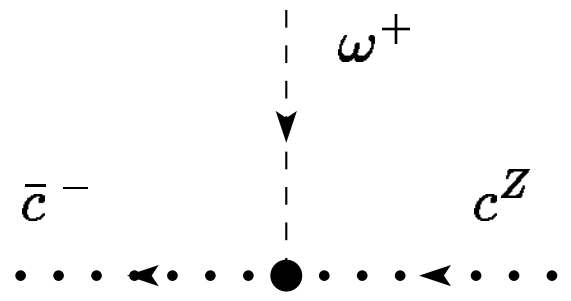

$$-\frac{i\xi_W e\cos 2\vartheta_W}{2\sin\vartheta_W} M_Z$$

$\bar{c}^{\,-}\ c^{A}\ \omega^{+}$

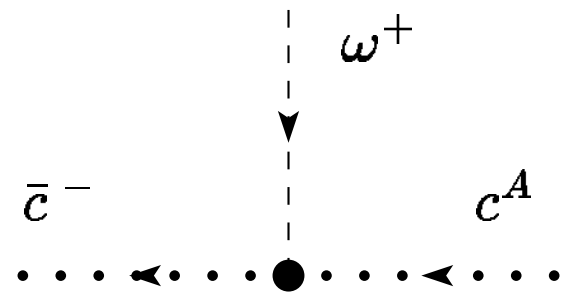

$$-i\xi_W e\ M_W$$

$\bar{c}^{\,Z}\ c^{-}\ \omega^{+}$

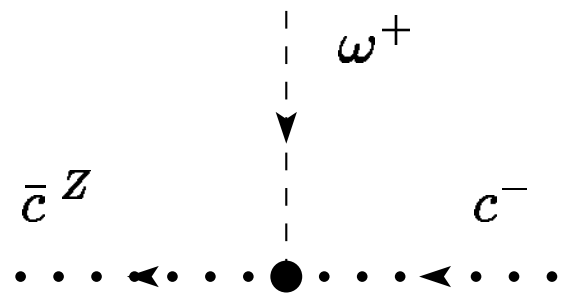

$$\frac{i\xi_Z e}{2\sin\vartheta_W} M_Z$$

$\bar{c}^{\,+}\ c^{Z}\ \omega^{-}$

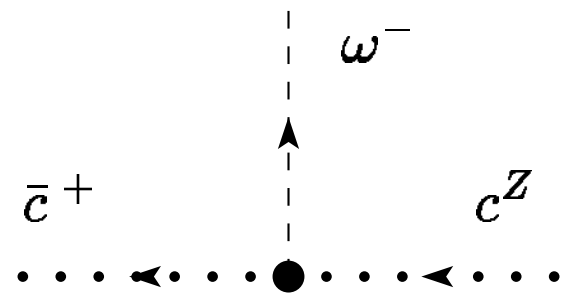

$$\frac{i\xi_W e\ \cos 2\vartheta_W}{2\sin\vartheta_W} M_Z$$

$\bar{c}^{\,+}\ c^{A}\ \omega^{-}$

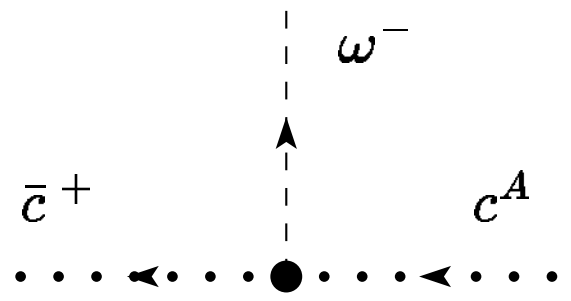

$$i\ \xi_W e\ M_W$$

$\bar{c}^{\,Z}\ c^{+}\ \omega^{-}$

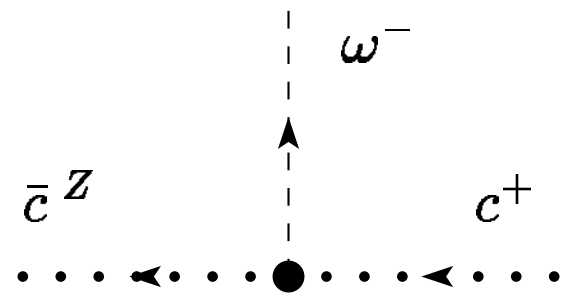

$$-\frac{i\xi_Z e}{2\sin\vartheta_W} M_Z$$

$\boxed{\bar{c}^{\,+}\ c^{-}\ z}$

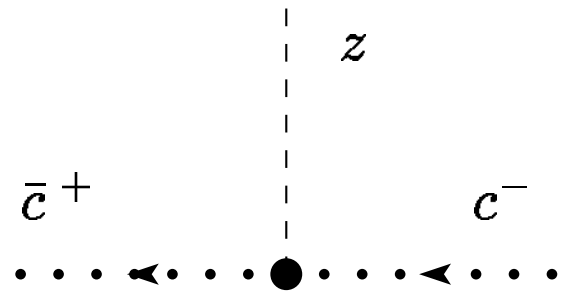

$$-\frac{i}{2}\xi_W\ e\ \cot\vartheta_W M_Z$$

$\boxed{\bar{c}^{\,-}\ c^{+}\ z}$

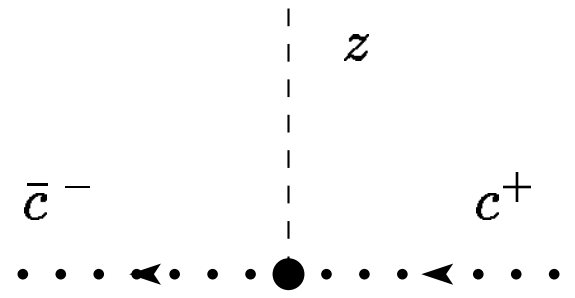

$$\frac{i}{2}\xi_W\ e\ \cot\vartheta_W M_Z$$

$\boxed{\bar{c}^{\,+}\ c^{-}\ \phi}$

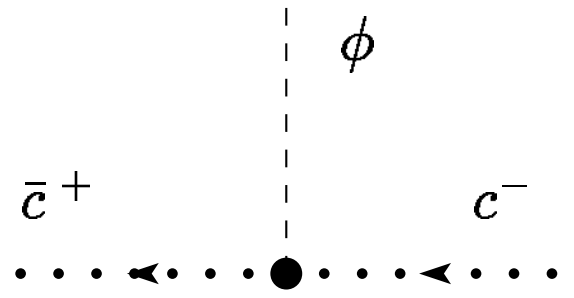

$$-\frac{1}{2}\xi_W\ e\ \cot\vartheta_W M_Z$$

$\boxed{\bar{c}^{\,-}\ c^{+}\ \phi}$

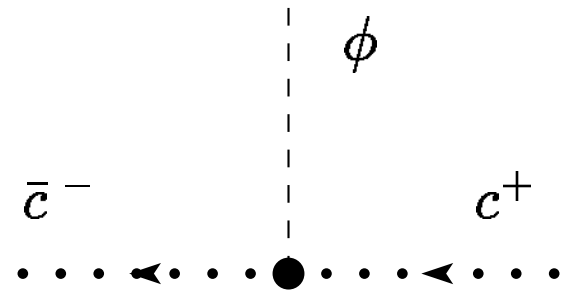

$$-\frac{1}{2}\xi_W\ e\ \cot\vartheta_W M_Z$$

$\boxed{\bar{c}^{\,Z}\ c^{Z}\ \phi}$

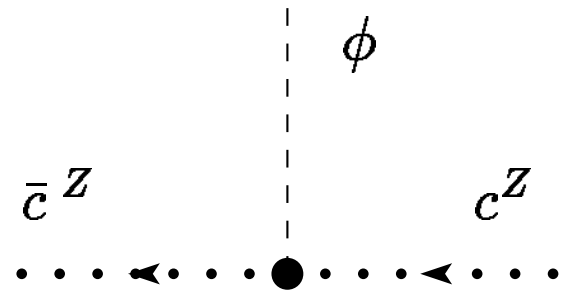

$$-\frac{\xi_Z\ e}{\sin 2\vartheta_W}M_Z$$

# 11 INTEGRATION IN $N$-DIMENSIONS

## 11.1 *Dimensional Regularization*

A powerful method of the evaluation of the loop integrals (which very often are divergent) is *Dimensional Regularization* (DR) [23]. The idea of DR is to consider the loop integral as an *analytic function* of $n$ – number of dimensions. Then one can calculate this integral in that region of the complex $n$ plane, where this function is convergent.

A typical loop integral looks as follows:

$$\int \frac{d^4p}{(2\pi)^4} \frac{P(q_i^\nu, m_i, p^\nu)}{\prod_{i=1}^{l}(m_i^2 - (p-k_i)^2)},$$

where $q_i$ ($m_i$) are 4–momenta (masses) of external particles; $P(q_i^\nu, m_i, p^\nu)$ is a function of masses $m_i$ and momenta $q_i$ and $p$.

To use the DR method one needs to transform the product of denominators into expression such as : $p^2 + (pk) + M^2$, where $k^\nu$ is the linear combination of $q_i$ momenta and $M$ is a combination of $q_i^2$, $(q_iq_j)$, and $m_i^2$. That can be done by using of *Feynman parameterization*:

$$\begin{aligned}
\frac{1}{a^\alpha b^\beta} &= \frac{\Gamma(\alpha+\beta)}{\Gamma(\alpha)\Gamma(\beta)} \int_0^1 dx \frac{x^{\alpha-1}(1-x)^{\beta-1}}{[ax+b(1-x)]^{\alpha+\beta}},\\
\frac{1}{a^n} - \frac{1}{b^n} &= \int_0^1 \frac{n(a-b)dx}{[(a-b)x+b]^{n+1}},\\
\frac{1}{b_1^{\alpha_1} b_2^{\alpha_2} \ldots b_m^{\alpha_m}} &= \frac{\Gamma(\alpha_1+\ldots+\alpha_m)}{\Gamma(\alpha_1)\Gamma(\alpha_2)\ldots\Gamma(\alpha_m)} \int_0^1 dx_1 \int_0^{x_1} dx_2 \ldots \int_0^{x_{m-2}} dx_{m-1}\\
&\quad \frac{x_{m-1}^{\alpha_1-1}(x_{m-2}-x_{m-1})^{\alpha_2-1}\ldots(1-x_1)^{\alpha_m-1}}{[b_1x_{m-1}+b_2(x_{m-2}-x_{m-1})+\ldots+b_m(1-x_1)]^{\alpha_1+\ldots+\alpha_m}},
\end{aligned}$$

where $\Gamma(z)$ is the Euler *Gamma function.*

Using the Wick rotation $p_0 \to ip_0$ and replacement $4 \to n$, one can obtain a typical integral in $n$-dimensional Euclidean space:

$$J = \int d^np \frac{P(q_i^\nu, m_i, p^\nu)}{(p^2+2(pk)+M^2)^\alpha}, \qquad \mathrm{Re}\,\alpha > 0.$$

The differential $d^np$ has the form:

$$d^np \;=\; p^{n-1}dp\, d\Omega_n, \quad \int d\Omega_n = \Omega_n = 2\pi^{\frac{n}{2}}/\Gamma(\frac{n}{2}), \tag{11.1}$$

$$d\Omega_n = (\sin^{n-2}\vartheta_{n-1} d\vartheta_{n-1}) d\Omega_{n-1} \qquad (11.2)$$
$$= (\sin^{n-2}\vartheta_{n-1} d\vartheta_{n-1})(\sin^{n-3}\vartheta_{n-2} d\vartheta_{n-2}) \ldots d\vartheta_1,$$

where $0 \le \vartheta_i \le \pi, \quad 0 \le \vartheta_1 \le 2\pi$. (The last equality in (11.2) obeys for the integer $n$.)

## 11.2 *Integrals*

Let us introduce the following notation:

$$\hat{J}\, f(p) \equiv \int d^n p \, \frac{1}{(p^2 + 2(pk) + M^2)^\alpha} f(p), \qquad (11.3)$$

$$J_0 \equiv \frac{i\pi^{n/2}\; i^n}{(M^2 - k^2)^{\alpha - \frac{n}{2}}\Gamma(\alpha)}. \qquad (11.4)$$

Then :

$$I_0 = \hat{J}\, 1 = \frac{i\pi^{n/2}}{(M^2 - k^2)^{\alpha - n/2}} \frac{\Gamma(\alpha - \frac{n}{2})}{\Gamma(\alpha)} = \Gamma(\alpha - \frac{n}{2}) J_0.$$

$$I^\mu = \hat{J}\, p^\mu = (-k^\mu) I_0,$$

$$I_2 = \hat{J}\, p^2 = J_0\{k^2\Gamma(\alpha - \frac{n}{2}) + \frac{n}{2}\Gamma(\alpha - 1 - \frac{n}{2})(M^2 - k^2)\},$$

$$I^{\mu\nu} = \hat{J}\, p^\mu p^\nu = J_0\{k^\mu k^\nu \Gamma(\alpha - \frac{n}{2}) + \frac{1}{2}\Gamma(\alpha - 1 - \frac{n}{2}) g^{\mu\nu}(M^2 - k^2)\},$$

$$I^{\mu\nu\lambda} = \hat{J}\, p^\mu p^\nu p^\lambda = J_0\{-k^\mu k^\nu k^\lambda \Gamma(\alpha - \frac{n}{2})$$
$$-\frac{1}{2}\Gamma(\alpha - 1 - \frac{n}{2})(M^2 - k^2)(g^{\mu\nu}k^\lambda + g^{\mu\lambda}k^\nu + g^{\nu\lambda}k^\mu)\},$$

$$I_2^\mu = \hat{J}\, p^2 p^\mu = -J_0\, k^\mu \,\{k^2\Gamma(\alpha - \frac{n}{2}) + \frac{n+2}{2}\Gamma(\alpha - 1 - \frac{n}{2})(M^2 - k^2)\}.$$

For calculation of the basic integral $I_0$ one can use the well–known relation [24]:

$$\int_0^\infty \frac{x^\beta}{(x^2 + M^2)^\alpha} dx = \frac{\Gamma(\frac{\beta+1}{2})\Gamma(\frac{2\alpha - \beta - 1}{2})}{2\Gamma(\alpha)(M^2)^{\alpha - \frac{\beta+1}{2}}}.$$

## 11.3 *Spence Integral (Dilogarithm)*

As a rule the final expressions for the loop integrals include so–called *Spence integral* or *Euler dilogarithm* [24, 25, 26]:

$$\mathrm{Li}_2(z) = \mathrm{Li}(z) \equiv -\int_0^z \frac{\ln(1-t)}{t}dt = \int_0^1 \frac{\ln t}{t - z^{-1}}dt \quad [arg(1-z) < \pi]. \quad (11.5)$$

Dilogarithm is a special case of the polylogarithm [24, 25, 26]:

$$\mathrm{Li}_\nu(z) \equiv \sum_{k=1}^{\infty} \frac{z^k}{k^\nu} \quad [|z| < 1, \quad \text{or } |z| = 1, \ \mathrm{Re}\nu > 1]. \quad (11.6)$$

The main properties of $\mathrm{Li}(z)$ are as follows:

$$\mathrm{Li}_n(z) + \mathrm{Li}_n(-z) = 2^{1-n}\mathrm{Li}_n(z^2),$$
$$\mathrm{Li}_n(iz) + \mathrm{Li}_n(-iz) = 4^{1-n}\mathrm{Li}_n(z^4) - 2^{1-n}\mathrm{Li}_n(z^2),$$
$$\mathrm{Li}_n(iz) - \mathrm{Li}_n(-iz) = 2i\sum_{k=0}^{\infty}\frac{(-1)^k z^{2k+1}}{(2k+1)^n} \quad [|z| < 1],$$
$$\mathrm{Li}_n(z) = \int_0^z \frac{\mathrm{Li}_{n-1}(t)}{t}dt \quad (n = 1, 2, \ldots),$$
$$\mathrm{Li}_0(z) = \frac{z}{1-z}, \quad \mathrm{Li}_1(z) = -\ln(1-z).$$

The Riemann sheet of the $\mathrm{Li}_2(z)$ has a cut along the real axes when $z > 1$, and

$$\mathrm{Im}\ \mathrm{Li}_2(z \pm i\varepsilon) = \pm\pi\Theta(z-1)\ln(z),$$

where the $\Theta(x)$ is the step function (see Subsection 16.1).
The equation $\mathrm{ReLi}_2(z) = 0$ has two solutions on the real axes

$$z_1 = 0, \quad \text{and} \quad z_2 \approx 12.6.$$

$\mathrm{ReLi}_2(z)$ achieves its maximum at $z = 2$:

$$\mathrm{Li}_2(2) = \frac{\pi^2}{4},$$

and at this point the $\mathrm{Li}_2(z)$ has the expansion as follows [9]:

$$\mathrm{Li}_2(2-\delta) = \frac{\pi^2}{4} - \frac{\delta^2}{4} - \frac{\delta^3}{6} - \frac{5\delta^4}{48} - \frac{\delta^5}{15} - \ldots$$

One easily gets:

$$\mathrm{Li}_2(0) = 0, \quad \mathrm{Li}_2(1) = \frac{\pi^2}{6}, \quad \mathrm{Li}_2(-1) = -\frac{\pi^2}{12},$$
$$\mathrm{Li}_2(\frac{1}{2}) = \frac{\pi^2}{12} - \frac{1}{2}\ln^2 2,$$
$$\mathrm{Li}_2(\pm i) = -\frac{\pi^2}{48} \pm i\mathbf{G}, \quad \mathbf{G} = \sum_{k=0}^{\infty} \frac{(-1)^k}{(2k+1)^2} = 0.915965594\ldots$$

The various relations with $\mathrm{Li}_2$ are as follows [24, 25, 26]:

$$\mathrm{Li}_2(z) = -\mathrm{Li}_2(1-z) + \frac{\pi^2}{6} - \ln z\, \ln(1-z) \ \ [|argz|,\, |arg(1-z)| < \pi],$$
$$\mathrm{Li}_2(z) = -\mathrm{Li}_2(\frac{1}{z}) - \frac{1}{2}\ln^2 z + i\pi \ln z + \frac{\pi^2}{3} \quad [|arg(-z)| < \pi],$$
$$\mathrm{Li}_2(z) = \mathrm{Li}_2(\frac{1}{1-z}) + \frac{1}{2}\ln^2(1-z) - \ln(-z)\ln(1-z) - \frac{\pi^2}{6}$$
$$[|arg(-z)| < \pi].$$

The Hill identity has the form [9, 25]:

$$\begin{aligned}
\mathrm{Li}_2(\omega z) &= \mathrm{Li}_2(\omega) + \mathrm{Li}_2(z) - \mathrm{Li}_2\left(\frac{\omega - \omega z}{1-\omega z}\right) - \mathrm{Li}_2\left(\frac{z - \omega z}{1-\omega z}\right) \\
&- \ln\left(\frac{1-\omega}{1-\omega z}\right)\ln\left(\frac{1-z}{1-\omega z}\right) \\
&- \eta\left[1-\omega,\, \frac{1}{1-\omega z}\right]\ln\omega \ - \eta\left[1-z,\, \frac{1}{1-\omega z}\right]\ln z,
\end{aligned}$$

where the function $\eta$ compensates for the cut in the Riemann sheet of the logarithm [9]:

$$\ln xy = \ln x + \ln y + \eta(x,y).$$

A typical integral, which can be expressed via the dilogarithm, is, for example:

$$\int_a^b \frac{\ln(p+qt)}{t}dt = \ln p \ln\frac{b}{a} - \mathrm{Li}_2(-b\frac{q}{p}) + \mathrm{Li}_2(-a\frac{q}{p}).$$

The Euler *Gamma function* $\Gamma(z)$ is given by the integral representation [24]:

$$\Gamma(z) \equiv \int_0^\infty dt\, t^{z-1}\, e^{-t}, \qquad \mathrm{Re}\, z > 0.$$

The main properties of the $\Gamma(z)$ are as follows [24]:

$$\begin{aligned}
&\Gamma(1+z) = z\Gamma(z), \quad \Gamma(n+1) = n!, \\
&\Gamma(z)\Gamma(-z) = -\frac{\pi}{z\ \sin(\pi z)}, \quad \Gamma(z)\Gamma(1-z) = \frac{\pi}{\sin(\pi z)}, \\
&\Gamma(\frac{1}{2}+z)\Gamma(\frac{1}{2}-z) = \frac{\pi}{\cos(\pi z)}, \quad \Gamma(2z) = \frac{2^{(2z-1)}}{\sqrt{\pi}}\Gamma(z)\Gamma(\frac{1}{2}+z), \\
&\Gamma(1) = \Gamma(2) = 1, \quad \Gamma\left(\frac{1}{2}\right) = \sqrt{\pi}, \\
&\Gamma(z)|_{z\to 0} \simeq \frac{1}{z} + \Gamma'(1); \quad \Gamma'(1) = \Gamma(1)\Psi(1) = \Psi(1) = -\gamma = -0.577\ldots,
\end{aligned}$$

where $\gamma$ is Euler constant.

# 12 KINEMATICS

The nice book by E. Byckling and K. Kajantie [27] contains a lot of information about relativistic kinematics. Here we present a brief description of relativistic kinematics following the Review of Particle Properties [18].

## 12.1 *Variables*

Initial (final) particles total momentum (energy) squared will be denoted by:

$$s \equiv \Big( \sum_{initial} p_i \Big)^2 = \Big( \sum_{final} p_j \Big)^2. \tag{12.1}$$

Let $E$ and $\vec{p}$ be energy and momentum of a particle. The energy and momentum of this particle $(E', \vec{p}')$ in the frame moving with the velocity $\vec{\beta}$ are given by the Lorentz transformation:

$$E' = \gamma(E + \beta p_{||}), \quad p'_{||} = \gamma(p_{||} + \beta E), \quad \vec{p}\,'_{\rm T} = \vec{p}_{\rm T}, \tag{12.2}$$

where $\gamma = 1/\sqrt{1-\beta^2}$ and $\vec{p}_{\rm T}(p_{||})$ are the components of $\vec{p}$ perpendicular (parallel) to $\vec{\beta}$.
The beam direction choose along the $z$–axes. 4–momentum of a particle $p^\mu = (E, \vec{p})$ can be written as:

$$\begin{aligned} &E = p_0, \ \vec{p}_{\rm T} = (p_x, p_y), \ p_z, \\ &p_x = |\vec{p}| \cos\phi \sin\vartheta, \ p_y = |\vec{p}| \sin\phi \sin\vartheta, \ p_z = |\vec{p}| \cos\vartheta, \end{aligned} \tag{12.3}$$

where $\phi$ is the azimuthal angle ($0 \le \phi \le 2\pi$); $\vartheta$ is the polar angle ($0 \le \vartheta \le \pi$).
Another parameterization of $p^\mu$ looks as follows:

$$E = m_{\rm T} \cosh y, \ p_x, \ p_y, \ p_z = m_{\rm T} \sinh y, \tag{12.4}$$

where $m_{\rm T}^2 = m^2 + p_{\rm T}^2$ is the transverse mass ("old" definition), $y$ is the rapidity.
*Rapidity* $y$ is defined by

$$y \equiv \frac{1}{2} \ln \Big( \frac{E + p_z}{E - p_z} \Big) = \ln \Big( \frac{E + p_z}{m_{\rm T}} \Big) = \tanh^{-1} \Big( \frac{p_z}{E} \Big). \tag{12.5}$$

Under a boost along $z$–direction to a frame with velocity $\beta$,

$$y \rightarrow y + \tanh^{-1}\beta.$$

*Pseudorapidity* $\eta$ is defined by:

$$\eta \equiv -\ln(\tan(\vartheta/2)), \tag{12.6}$$
$$\sinh\eta = \cot\vartheta, \ \ \cosh\eta = \frac{1}{\sin\vartheta}, \ \ \tanh\eta = \cos\vartheta.$$

For $p \gg m$ and $\vartheta \gg 1/\gamma$ one has : $\eta \approx y$.
Feynman's $x_F = x$ variable is given by

$$x = \frac{p_z}{p_z\max} \approx \frac{(E+p_z)}{(E+p_z)\max}, \ \text{in cms} \ x = \frac{2p_z}{\sqrt{s}}. \tag{12.7}$$

The last equation is valid for two particles collisions, and here $s$ is total energy squared (see (12.1)).
In the collider's experiments the following additional variables are used:

$$\begin{array}{lcll} E_\perp & = & E\sin\vartheta & \text{– transverse energy,} \\ \vec{p}_{\perp mis} & = & -(\Sigma\vec{p}_\perp) & \text{– "missing" transverse momentum,} \\ \vec{E}_{\perp mis} & = & -(\Sigma\vec{E}_\perp) & \text{– "missing" transverse energy} \end{array}$$

where sum is performed over all detected particles.
The "distance" in $(\eta,\phi)$–plane between two particles (clusters) 1 and 2 is given by

$$\Delta R \equiv \sqrt{(\Delta\phi)^2 + (\Delta\eta)^2}, \ \Delta\phi = \phi_1 - \phi_2, \ \Delta\eta = \eta_1 - \eta_2.$$

The "transverse" mass of the particle (cluster) $c$ with momentum $\vec{p}_c$ and the "missing" transverse momentum (energy) $\vec{p}_{\perp\, mis}$ ($\vec{E}_{\perp\, mis}$) is given by:

$$M_\perp^2(c, \vec{p}_{\perp\, mis}) \equiv (\sqrt{m_c^2 + p_{\perp c}^2} + p_{\perp\, mis})^2 - (\vec{p}_{\perp c} + \vec{p}_{\perp\, mis})^2.$$

## 12.2 *Event Shape Variables*

In this Subsection we describe in brief event shape variables for $n$–particle final state (for details, see, for example [28]). None of the variables presented in this Subsection are Lorentz invariant.

• **Sphericity**
The *sphericity* tensor is defined as [28, 29]:

$$S^{ab} \equiv \frac{\sum_i p_i^a p_i^b}{\sum_i |\vec{p}_i|^2}, \tag{12.8}$$

where $a, b = 1, 2, 3$ corresponds to the $x, y$ and $z$ components. By standard diagonalization of $S^{ab}$ one can find three eigenvalues

$$\lambda_1 \geq \lambda_2 \geq \lambda_3, \quad \text{with} \quad \lambda_1 + \lambda_2 + \lambda_3 = 1.$$

Then, the *sphericity* is defined as:

$$S \equiv \frac{3}{2}(\lambda_2 + \lambda_3), \quad 0 \leq S \leq 1. \tag{12.9}$$

Eigenvectors $\vec{s}_i$ can be found that correspond to the three eigenvalues $\lambda_i$. The $\vec{s}_1$ eigenvector is called the *sphericity axes*, while the *sphericity event plane* is spanned by $\vec{s}_1$ and $\vec{s}_2$.
Sphericity is essentially a measure of the summed $\vec{p}_{\mathrm{T}}$ with respect to sphericity axes. So, one can use another definition of the sphericity:

$$S = \frac{3}{2} \min_{\vec{n}} \frac{\sum_i \vec{p}_{\mathrm{T}i}^{\,2}}{\sum_i |\vec{p}_i|^2}, \tag{12.10}$$

where $\vec{p}_{\mathrm{T}i}$ is a component of $\vec{p}_i$ perpendicular to $\vec{n}$. So, the sphericity axes $\vec{s}_i$ given (12.10) by the $\vec{n}$ vector for which minimum is attained. A 2–jet event corresponds to $S \approx 0$ and isotropic event to $S \approx 1$.

Sphericity is not an infrared safe quality in QCD perturbation theory. Sometimes one can use the generalization of the sphericity tensor, given by

$$S^{(r)ab} \equiv \frac{\sum_i |\vec{p}_i|^{r-2} p_i^a p_i^b}{\sum_i |\vec{p}_i|^r}, \tag{12.11}$$

• **Aplanarity**
The *aplanarity* $A$ is define as [28, 30]:

$$A \equiv \frac{3}{2}\lambda_2, \quad 0 \leq A \leq \frac{1}{2}. \tag{12.12}$$

The aplanarity measures the transverse momentum component out of the event plane. A planar event has $A \approx 0$ and isotropic one $A \approx \frac{1}{2}$.

• **Thrust**
The *thrust* $T$ is given by [28, 31]

$$T \equiv \max_{|\vec{n}|=1} \frac{\sum_i |(\vec{n}\vec{p}_i)|}{\sum_i |\vec{p}_i|}, \quad \frac{1}{2} \leq T \leq 1. \tag{12.13}$$

and the *thrust axes* $\vec{t}_i$ is given by the $\vec{n}$ vector for which maximum is attained. 2–jet event corresponds to $T \approx 1$ and isotropic event to $T \approx \frac{1}{2}$.
• **Major and minor values**
In the plane perpendicular to the thrust axes, a *major axes* $\vec{m}_a$ and *major value* $M_a$ may be defined in just the same fashion as thrust [28], i.e.

$$M_a \equiv \max_{|\vec{n}|=1,\,(\vec{n}\vec{t}_1)=0} \frac{\sum_i |(\vec{n}\vec{p}_i)|}{\sum_i |\vec{p}_i|}. \tag{12.14}$$

Finally, a third axes, the *minor axes*, is defined perpendicular to the thrust ($\vec{t}_1$) and major ($\vec{m}_a$) axes. The *minor value* $M_i$ is calculated just as thrust and major values.
• **Oblatness**
The *oblatness* $O$ is given by [28]

$$O \equiv M_a - M_i.$$

In general, $O \approx 0$, corresponds to an event symmetrical around the thrust axes $\vec{t}_1$ and high $O$ to aplanar event.
• **Fox–Wolfram moments**
The *Fox–Wolfram moments* $H_l$, $l = 0, 1, 2, \ldots$, are defined by [28, 32]:

$$H_l \equiv \sum_{i,j=1} \frac{|\vec{p}_i||\vec{p}_j|}{E_{vis}^2} P_l(\cos\vartheta_{ij}), \tag{12.15}$$

where $\vartheta_{ij}$ is the opening angle between hadron $i$ and $j$, and $E_{vis}$ is the total visible energy of the event. $P_l(z)$ are the Legendre polynomials [24]:

$$P_0(z) = 1,\ P_1(z) = z,\ P_2(z) = \frac{1}{2}(3z^2 - 1), \ldots$$
$$P_k(z) = \frac{1}{k}[(2k-1)zP_{k-1}(z) - (k-1)P_{k-2}(z)].$$

Neglecting the masses of all the particles, one gets $H_0 = 1$. If momentum is balanced, then $H_1 = 0$. 2–jet events tend to give $H_l \approx 1$ for $l$ even and $H_l \approx 0$ for $l$ odd.

The summary of the discussed quantities are presented in Table 12.1.

**Table 12.1.**
Summary of event shape variables.

| | $S$ | $A$ | $T$ | $O$ | $H_0$ all $m_i = 0$ | $H_l$ |
|---|---|---|---|---|---|---|
| isotropic | 1 | $\frac{1}{2}$ | $\frac{1}{2}$ | - | 1 | - |
| 2–jet | 0 | - | 1 | 0 | 1 | $H_1 = 0$ <br> $H_{2k} \approx 1,\ H_{2k+1} \approx 0$ |
| planar | - | 0 | - | $\gg 0$ | 1 | - |

## 12.3 *Two–body Final State*

In the collision of two particles of mass $m_1$ and $m_2$ and momenta $p_1$ and $p_2$

$$s = (p_1 + p_2)^2 = m_1^2 + m_2^2 + 2E_{1\,Lab} m_2,$$

where the last equation is valid in the frame, where one particle (second one) is at rest (Lab frame).

The energies and momenta of the particles 1 and 2 in their center–of–mass system (cms) are equal to:

$$E_1^* = \frac{s + m_1^2 - m_2^2}{2\sqrt{s}}, \quad E_2^* = \frac{s - m_1^2 + m_2^2}{2\sqrt{s}}, \tag{12.16}$$

$$\vec{p}_1^{\,*} = -\vec{p}_2^{\,*}, \quad |\vec{p}_1^{\,*}| = \frac{\sqrt{[s - (m_1 + m_2)^2][s - (m_1 - m_2)^2]}}{2\sqrt{s}}, \tag{12.17}$$

or

$$|\vec{p}_1^{\,*}| = \frac{1}{2\sqrt{s}} \lambda^{1/2}(s, m_1^2, m_2^2),$$

where $\lambda(x, y, z)$ is the so–called *kinematical function* [27]:

$$\begin{aligned} \lambda(x, y, z) &\equiv (x - y - z)^2 - 4yz & (12.18) \\ &= x^2 + y^2 + z^2 - 2xy - 2yz - 2zx & \\ &= \{x - (\sqrt{y} + \sqrt{z})^2\}\{x - (\sqrt{y} - \sqrt{z})^2\}. & (12.19) \end{aligned}$$

Let us now consider the two–body reaction (4–momenta of the particles are presented in the parentheses):

$$\begin{aligned} a(p_a) + b(p_b) &\rightarrow 1(p_1) + 2(p_2) \\ p_a + p_b &= p_1 + p_2 \end{aligned}$$

The Lorentz–invariant Mandelstam variables for reaction $2 \rightarrow 2$ are defined by:

$$\begin{aligned} s &= (p_a + p_b)^2 = (p_1 + p_2)^2, \quad t = (p_a - p_1)^2 = (p_b - p_2)^2, \\ u &= (p_a - p_2)^2 = (p_b - p_1)^2, \end{aligned} \tag{12.20}$$

and they satisfy

$$s + t + u = m_a^2 + m_b^2 + m_1^2 + m_2^2.$$

Two limits of t (corresponding to $\vartheta_{cm} = 0$ and $\pi$) equal:

$$\begin{aligned} t_\pm &= m_a^2 + m_1^2 - 2E_a^* E_1^* \pm 2|\vec{p}_a^{\,*}||\vec{p}_1^{\,*}| = \\ &= m_a^2 + m_1^2 - \frac{1}{2s}(s + m_a^2 - m_b^2)(s + m_1^2 - m_2^2) \\ &\pm \frac{1}{2s}\lambda^{1/2}(s, m_a^2, m_b^2)\lambda^{1/2}(s, m_1^2, m_2^2). \end{aligned} \tag{12.21}$$

## 12.4 *Three–body Final State*

Let us consider three–body decay of particle $a$ with mass $M$

$$a(P) \rightarrow 1(p_1) + 2(p_2) + 3(p_3).$$

Defining

$$p_{ij} \equiv p_i + p_j, \quad m_{ij}^2 \equiv p_{ij}^2, \tag{12.22}$$

then

$$\begin{aligned} & m_{12}^2 + m_{23}^2 + m_{13}^2 = M^2 + m_1^2 + m_2^2 + m_3^2, \\ \text{and} \quad & m_{ij}^2 = (P - p_k)^2 = M^2 + m_k^2 - 2ME_k. \end{aligned}$$

The $1 \rightarrow 3$ decay is described by two variables (for example, $m_{12}$ and $m_{13}$). If $m_{12}$ is fixed, then limits of $m_{13}^2$ variation are equal to:

$$
\begin{aligned}
\left(m_{13}^2\right)_\pm &= m_1^2 + m_3^2 - \frac{1}{2m_{12}^2}(m_{12}^2 - M^2 + m_3^2)(m_{12}^2 + m_1^2 - m_2^2) \\
&\pm \frac{1}{2m_{12}^2}\lambda^{1/2}(m_{12}^2, M^2, m_3^2)\lambda^{1/2}(m_{12}^2, m_1^2, m_2^2) = \\
&= (E_1^* + E_3^*)^2 - (\sqrt{E_1^{*\,2} - m_1^2} \mp (\sqrt{E_3^{*\,2} - m_3^2})^2,
\end{aligned}
$$

where $E_1^* = \frac{m_{12}^2 + m_1^2 - m_2^2}{2m_{12}}$ and $E_3^* = \frac{M^2 - m_{12}^2 - m_3^2}{2m_{12}}$.
$2 \to 3$ scattering is described by five independent variables. For example,

$$
s = (p_a + p_b)^2,\ m_{12}^2,\ m_{23}^2,\ t_1 = (p_q - p_1)^2,\ t_2 = (p_b - p_3)^2.
$$

## 12.5 *Lorentz Invariant Phase Space*

Lorentz invariant phase space (LIPS) of $n$ particles with 4–momenta $p_j$ ($j = 1, 2, \ldots n$) and the total momentum $P = \sum_{j=1}^n p_j$ is given by:

$$
dR_n(P;\ p_1, p_2, \ldots p_n) \equiv \delta^{(4)}(P - \sum_{j=1}^n p_j) \prod_{j=1}^n \frac{d^3 p_j}{(2\pi)^3 2E_j}. \tag{12.23}
$$

Through of this Subsection we use the following notation:

$$
s \equiv P^2.
$$

This LIPS can be generated recursively as follows [18, 27]:

$$
dR_n = dR_2(P;\ p_n, q)(2\pi)^3 dq^2\, dR_{n-1}(q;\ p_1, \ldots p_{n-1}), \tag{12.24}
$$

where $q = \sum_{i=1}^{n-1} p_i$ and $(m_1 + m_2 + \ldots + m_{n-1})^2 \le q^2 \le (\sqrt{P^2} - m_n)^2$, or:

$$
dR_n = dR_{n-j+1}(P;\ q, p_{j+1}, \ldots p_n)(2\pi)^3 dq^2\, dR_j(q;\ p_1, \ldots p_j), \tag{12.25}
$$

here $q = \sum_{l=1}^j p_l$ and

$$
(m_1 + \ldots + m_j)^2 \le q^2 \le (\sqrt{P^2} - \sum_{l=j+1}^n m_l)^2.
$$

The integrated LIPS for $m_1 = m_2 = \ldots = m_n = 0$ equals:

$$R_n(0) = \frac{1}{(2\pi)^{3n}} \frac{(\pi/2)^{n-1}}{(n-1)!(n-2)!}(P^2)^{n-2}.$$

Two–particle LIPS equals:

$$R_2 = \frac{1}{(2\pi)^6} \frac{p_1^*}{4\sqrt{P^2}} \int d\Omega_1^* = \frac{1}{(2\pi)^6} \frac{\pi p_1^*}{\sqrt{P^2}} = \frac{1}{(2\pi)^6} \frac{\pi p_1^*}{\sqrt{s}},$$

where $p_1^*$ is momentum of first (second) particle in cms (see (12.17)).

Different choice of $m_1$ and $m_2$ leads to:

$$\begin{aligned}
R_2 &= \frac{1}{(2\pi)^6} \frac{\pi\sqrt{[s-(m_1+m_2)^2][s-(m_1-m_2)^2]}}{2s}, \quad (m_1+m_2) \le \sqrt{s}, \\
R_2 &= \frac{1}{(2\pi)^6} \frac{\pi}{2}\sqrt{1-\frac{4m^2}{s}}, \quad m_1 = m_2 = m, \\
R_2 &= \frac{1}{(2\pi)^6} \frac{\pi}{2}(1-\frac{m_1^2}{s}), \quad m_2 = 0, \\
R_2 &= \frac{1}{(2\pi)^6} \frac{\pi}{2}, \quad m_1 = m_2 = 0.
\end{aligned}$$

Three body decay final state LIPS equals:

$$dR_3 = \frac{1}{(2\pi)^9} \frac{\pi^2}{4s} dm_{12}^2 dm_{13}^2 = \frac{1}{(2\pi)^9} \pi^2 dE_1\, dE_2,$$

where $m_{12}$ and $m_{13}$ are defined in (12.22), $E_1(E_2)$ is the energy of the first (second) particle in $P$ rest frame. This is the standard form of the Dalitz plot.

## 12.6 *Width and Cross Section*

The partial decay rate (*partial width*) of a particle of mass $M$ into $n$ bodies in its rest frame is given in terms of the Lorentz–invariant matrix element $M_{fi}$ by:

$$d\Gamma = \frac{(2\pi)^4}{2M} |M_{fi}|^2 dR_n(P;\ p_1, p_2, \ldots, p_n). \tag{12.26}$$

The differential cross section of the reaction $a + b \to 1 + 2 + \ldots + n$ $(p_a + p_b \equiv P)$ is given by:

$$d\sigma = \frac{(2\pi)^4}{2I} |M_{fi}|^2 dR_n(P;\ p_1, p_2, \ldots, p_n), \tag{12.27}$$
$$I^2 = [s - (m_a + m_b)^2][s - (m_a - m_b)^2] = 4[(p_a p_b)^2 - m_a^2 m_b^2].$$

# 13 DECAYS

## 13.1 *Standard Model Higgs Decays Rates*

Standard Model Higgs is expected to have a mass between 45 Gev and 1 TeV, and, since it couples directly to the masses of other particles, to decay into heaviest available particles. The SM Higgs decay rates, calculated without radiative corrections are as follows (see [33] and references therein):

$$H \to f\bar{f}, \quad \Gamma = \frac{N_c G_F m_f^2}{4\sqrt{2}\pi} m_H \beta^3,$$

where $\beta = \sqrt{1 - 4m_f^2/m_H^2}$ and $N_c = 1(3)$ for $f =$ lepton (quark).

$$H \to W^+W^-(ZZ), \quad \Gamma = \frac{G_F^2 M^2 m_H}{8\sqrt{2}\pi} \frac{\sqrt{1-x}}{x}(3x^2 - 4x + 4),$$

where $x = 4M^2/m_H^2$, $M$ is $W^\pm(Z)$–boson mass.

Higgs decay into two photons or two gluons proceeds via loops. Its decay rates are equal [34]:

$$H \to \gamma\gamma, \quad \Gamma = \frac{\alpha^2 G_F}{8\sqrt{2}\pi^3} m_H^3 |I|^2,$$

where $I = I_{lepton} + I_{hadron} + I_W + \ldots$, and $|I| \approx O(1)$.

$$H \to gg, \quad \Gamma = \frac{\alpha_s^2 G_F}{4\sqrt{2}\pi^3} \frac{m_H^3}{9} |N|^2,$$

where $N \equiv 3\sum_j N_j$ is the sum of contributions $N_j$ from quarks $j = 1, 2, \ldots,$ given by [35]:

$$N_j = \int_0^1 dx \int_0^{1-x} dy \frac{1 - 4xy}{1 - xy\frac{m_H^2}{m_j^2} - i\varepsilon} = 2\lambda_j + \lambda_j(4\lambda_j - 1)G(\lambda_j),$$

where $\lambda_j \equiv m_j^2/m_H^2$, and

$$G(z) = -2[\arcsin(\frac{1}{2\sqrt{z}})]^2, \quad z \geq \frac{1}{4},$$

$$G(z) = \frac{1}{2}\ln^2\left[\frac{1+\sqrt{1-4z}}{1-\sqrt{1-4z}}\right] - \frac{\pi^2}{2} + i\pi\ln\left[\frac{1+\sqrt{1-4z}}{1-\sqrt{1-4z}}\right], \quad z \leq \frac{1}{4}.$$

$N_q$ vanishes for $m_q \ll m_H$ and $N_q \to 1/3$ for $m_q \gg m_H$.

$$\begin{aligned}
H \to W^\pm f\bar{f}', \quad \Gamma &= \frac{g^4 m_H}{307\pi^3} F(\epsilon), \quad \epsilon = \frac{m_W}{m_H}, \\
H \to W^\pm \sum f\bar{f}', \quad & \text{(except } W^+ \to t\bar{b}\text{)} \\
\Gamma &= \frac{3g^4 m_H}{512\pi^3} F(\epsilon), \quad \epsilon = \frac{m_W}{m_H}, \\
H \to Z \sum f\bar{f}, \quad \Gamma &= \frac{g^4 m_H}{2048\pi^3 \cos^4 \vartheta_W} \\
&\times \left(7 - \frac{40}{3}\sin^2\vartheta_W + \frac{160}{9}\sin^4\vartheta_W\right) F(\epsilon'), \quad \epsilon' = \frac{m_Z}{m_H}, \\
F(z) &= \frac{3(1 - 8z^2 + 20z^4)}{\sqrt{4z^2 - 1}} \arccos\left(\frac{3z^2 - 1}{2z^3}\right) \\
&- (1 - z^2)\left(\frac{27}{2}z^2 - \frac{13}{2} + \frac{1}{z^2}\right) - 3(1 - 6z^2 + 4z^4)\ln z.
\end{aligned}$$

## 13.2 *W and Z Decays*

The partial decay widthes for gauge bosons to decay into massless fermions $f_1\bar{f}_2$ are equal to [4, 18]:

$$\begin{aligned}
W^+ \to e^+\nu_e, \quad \Gamma &= \frac{G_F M_W^3}{6\sqrt{2}\pi} \approx 227 \pm 1 \text{ MeV}, \\
W^+ \to u_i\bar{d}_i, \quad \Gamma &= C\frac{G_F M_W^3}{6\sqrt{2}\pi}|V_{ij}|^2 \approx (707 \pm 3)|V_{ij}|^2 \text{ MeV}, \\
Z \to \psi_i\bar{\psi}_i, \quad \Gamma &= C\frac{G_F M_Z^3}{6\sqrt{2}\pi}[g_{iV}^2 + g_{iA}^2] \approx \\
&= \begin{cases} 167.1 \pm 0.3 \text{ MeV } (\nu\bar{\nu}), & 83.9 \pm 0.2 \text{ MeV } (e^+e^-), \\ 298.0 \pm 0.6 \text{ MeV } (u\bar{u}), & 384.5 \pm 0.8 \text{ MeV } (d\bar{d}), \\ 375.2 \pm 0.4 \text{ MeV } (b\bar{b}), & \end{cases}
\end{aligned}$$

For lepton $C = 1$, while for quarks $C = 3(1 + \frac{\alpha_s(M_V)}{\pi} + 1.409\frac{\alpha_s^2}{\pi^2} - 12.77\frac{\alpha_s^3}{\pi^3})$, where 3 is due to color and the factor in parentheses is a QCD correction.

## 13.3 *Muon Decay*

In the SM the total muon decay width is equal (up to 100% accuracy) to the width of the decay

$$\mu^- \to e^- \bar{\nu}_e \nu_\mu.$$

The matrix element squared for this decay equals [4]:

$$|M|^2 = 128 G_f^2 (p_\mu p_{\nu_e})(p_e p_{\nu_\mu}).$$

Then the total muon width is given by [36]:

$$\Gamma_\mu^{tot} = \frac{G_F^2 m_\mu^5}{192\pi^3} F(\frac{m_e^2}{m_\mu^2})(1 + \frac{3}{5}\frac{m_\mu^2}{M_W^2})[1 + \frac{\alpha(m_\mu)}{2\pi}(\frac{25}{4} - \pi^2)], \tag{13.1}$$

where $F(x) = 1 - 8x + 8x^3 - x^4 - 12x^2 \ln x$, and

$$\alpha(m_\mu)^{-1} = \alpha^{-1} - \frac{2}{3\pi}\ln(\frac{m_\mu}{m_e}) + \frac{1}{6\pi} \approx 136.$$

For pure $V - A$ coupling (and neglecting of the electron mass) in the rest frame of the polarized muon ($\mu^\mp$) the differential decay rate is:

$$d\Gamma(\mu^\mp) = \frac{G_F^2 m_\mu^5}{192\pi^3}[3 - 2x \pm (1 - 2x)\cos\vartheta]x^2 dx d(\cos\vartheta),$$

where $\vartheta$ is the angle between the electron momentum and the muon spin, and $x = 2E_\mu/m_\mu$.

## 13.4 *Charged Meson Decay*

The decay constant $f_P$ for pseudoscalar meson $P$ is defined by [18]

$$< 0|A_\mu(0)|P(k) >= i f_P k_\mu.$$

The state vector is normalized by $< P(k)|P(k') >= (2\pi)^3 2E_q \delta^3(\vec{k} - \vec{k}')$. The annihilation rate of the $P(q_1\bar{q}_2') \to f\bar{f}'$ decay is given by

$$\Gamma(P \to f\bar{f}') = C\frac{G_F^2 |V_{q_1 q_2'}|^2}{8\pi} f_P^2 m_f^2 M_P (1 - \frac{m_f^2}{M_P^2}), \tag{13.2}$$

where $C = 1$ for $P \to l\nu_l$ decay and $C = (3|V_{q_3 q_4'}|^2)$ for $P \to q_3\bar{q}_4'$ one, and $m_f$ is the heaviest final fermion mass.

## 13.5 *Quark Decay*

In the region $m_q \ll M_W$ the total quark width is given by [4]:

$$\Gamma(Q \to q_2 q_3 \bar{q}_4') = \frac{G_F^2 m_Q^5}{64\pi^3} |V_{Qq_2}|^2 |V_{q_3 q_4'}|^2. \tag{13.3}$$

For the case of $m_Q \gg m_W + m_q$ the width of the heavy quark decay $Q \to W + q$ equals [37]:

$$\begin{aligned} \Gamma(Q \to W + q) &= \frac{G_F m_Q^3}{8\pi\sqrt{2}} |V_{Qq}|^2 \frac{2k}{m_Q} f_Q(\frac{m_q^2}{m_Q^2}, \frac{M_W^2}{m_Q^2}) \\ &\approx 180 \quad \text{MeV} \quad |V_{Qq}|^2 \left(\frac{m_Q}{m_W}\right)^3, \end{aligned} \tag{13.4}$$

where

$$\begin{aligned} f_Q(x,y) &= (1-x)^2 + (1+x)y - 2y^2, \\ k &= \frac{1}{2m_Q}\sqrt{[m_Q^2 - (m_W + m_q)^2][m_Q^2 - (m_W - m_q)^2]}, \end{aligned}$$

here $k$ is $W$ (or $q$) momentum in the $Q$–quark rest frame.
The width of the heavy $Q$ decay

$$Q \to q + W(\to l\nu)$$

is given by [37]:

$$\Gamma(Q \to q + W(\to l\nu) = \frac{G_F^2 m_Q^5}{192\pi^3} |V_{Qq}| F(\frac{m_Q^2}{m_W^2}; \frac{m_q^2}{m_Q^2}; \frac{\Gamma_W^2}{m_W^2}),$$

where

$$\begin{aligned} &F(a,b,c) = 2\int_0^{(1-\sqrt{b})^2} dt \frac{f_Q(b,t)\sqrt{1+b^2+t^2-2(b+bt+t)}}{[(1-at)^2+c]}, \\ &F(a,0,c) = \\ &\frac{2}{a^4}[c - 3(1-a)]A + 2a(1-a) - a[3(2-a)c - (2+a)(1-a)^2]B, \\ &A = \ln\frac{c+1}{c+(1-a)^2}, \quad B = \frac{1}{a\sqrt{c}}[\arctan(\frac{1}{\sqrt{c}}) - \arctan(\frac{1-a}{\sqrt{c}})]. \end{aligned}$$

## 13.6 *Heavy Quarkonia* $(Q\bar{Q})$ *Decays*

Suppose that the matrix element of the vector state $V$ decay $V \to l^+l^-$ is given by

$$M = g_V e_V^\nu \bar{u}(l^+)\gamma^\nu u(-l^-).$$

Then

$$\Gamma(V \to l^+l^-) = \frac{g_V^2}{12\pi}M_V, \quad g_V = \sqrt{\frac{12\pi\Gamma(V \to l^+l^-)}{M_V}}.$$

Denote $R_0^2 \equiv 4\pi|\psi(0)|^2$, where $\psi(0)$ is bound state wave function in the origin.

The width of the decay of the quark antiquark vector state $1^{--}$ equals:

$$\Gamma(1^{--} \to l^+l^-) = N_c\frac{4}{3}\frac{\alpha^2Q_q^2}{M^2}R_0^2.$$

where $N_c = 1(3)$ for colorless (color) quarks, $Q_q$ is the effective charge:

$$\begin{array}{rcccccccc}
\rho & = & \frac{1}{\sqrt{2}}(u\bar{u} - d\bar{d}) & \Rightarrow & Q_q^2 & = & |\frac{1}{\sqrt{2}}(\frac{2}{3} + \frac{1}{3})|^2 & = & \frac{1}{2}, \\
\omega & = & \frac{1}{\sqrt{2}}(u\bar{u} + d\bar{d}) & \Rightarrow & Q_q^2 & = & |\frac{1}{\sqrt{2}}(\frac{2}{3} - \frac{1}{3})|^2 & = & \frac{1}{18}, \\
\phi & = & s\bar{s} & \Rightarrow & Q_q^2 & = & \frac{1}{9}, & & \\
J/\psi & = & c\bar{c} & \Rightarrow & Q_q^2 & = & \frac{4}{9}, & & \\
\Upsilon & = & b\bar{b} & \Rightarrow & Q_q^2 & = & \frac{1}{9}. & &
\end{array}$$

For positron annihilation (with $Q_e = 1$) one has:

$$\Gamma(0^- \to \gamma\gamma) = \frac{4\alpha^2}{M^2}R_0^2,$$

$$\Gamma(1^{--} \to \gamma\gamma\gamma) = \frac{16}{9\pi}(\pi^2 - 9)\frac{\alpha^3}{M^2}R_0^2.$$

For quarkonia annihilation one gets:

$$\Gamma(0^- \to \gamma\gamma) = \frac{12\alpha^2Q_q^4}{M^2}R_0^2.$$

For the two (three) gluon annihilation one need to change : $\alpha^2Q_q^4 \to 2\alpha_s^2/9$ $(\alpha^3 \to 5\alpha_s^3/18)$:

$$\Gamma(0^- \to gg) = \frac{8\alpha_s^2}{3M^2}R_0^2,$$

$$\Gamma(1^{--} \to ggg) = \frac{40}{81\pi}(\pi^2 - 9)\frac{\alpha_s^3}{M^2}R_0^2.$$

# 14 CROSS SECTIONS

## 14.1 $e^+e^-$ *Annihilation*

For pointlike spin$-\frac{1}{2}$ fermions the differential cross section in the cms for $e^+e^- \to f\bar{f}$ via single photon and $Z$–boson (with mass $M_Z$ and total width $\Gamma_Z$) is given by [18]:

$$\begin{aligned}
\frac{d\sigma}{d\Omega} &= \frac{\alpha^2}{4s}\beta Q_f^2\left\{1+\cos^2\vartheta+(1-\beta^2)\sin^2\vartheta\right\} && (14.1)\\
&+ \frac{\alpha^2}{4s}\beta\chi_2\Big\{V_f^2(1+V^2)[1+\cos^2\vartheta+(1-\beta^2)\sin^2\vartheta] && (14.2)\\
&\quad +\beta^2 a_f^2(1+V^2)[1+\cos^2\vartheta]-8\beta V V_f a_f\cos\vartheta\Big\}\\
&- \frac{\alpha^2}{4s}\beta 2Q_f\chi_1\Big\{VV_f[1+\cos^2\vartheta+(1-\beta^2)\sin^2\vartheta] && (14.3)\\
&\quad -2a_f\beta\cos\vartheta\Big\},
\end{aligned}$$

where $\beta=\sqrt{1-4m_f^2/s}$ is the velocity of the final state fermion in the center of mass, $Q_f$ is the charge of the fermion in units of the proton charge,

$$\begin{aligned}
\chi_1 &= \frac{1}{16\sin^2\vartheta_W\cos^2\vartheta_W}\frac{s(s-M_Z^2)}{(s-M_Z^2)^2+\Gamma_Z^2M_Z^2},\\
\chi_2 &= \frac{1}{256\sin^4\vartheta_W\cos^4\vartheta_W}\frac{s^2}{(s-M_Z^2)^2+\Gamma_Z^2M_Z^2},\\
V &= -1+4\sin^2\vartheta_W, \quad V_f=2T_{3f}-4Q_f\sin^2\vartheta_W, \quad a_f=2T_{3f},
\end{aligned}$$

here the subscript $f$ refers to the particular fermion and

$$\begin{aligned}
T_3 &= +\frac{1}{2} \quad \text{for} \quad \nu,u,c,t,\\
T_3 &= -\frac{1}{2} \quad \text{for} \quad l^-,d,s,b.
\end{aligned}$$

The first (14.1), second (14.2), and third (14.3) terms correspond to the $e^+e^- \to f\bar{f}$ process via single photon annihilation, via $Z$–boson exchange, and photon – $Z$–boson interference, respectively.
For $s \gg m_f^2$ (i.e. $\beta \to 1$) the annihilation via single photon exchange (14.1)

tends to:

$$\sigma = \frac{4\pi\alpha^2}{3s} Q_f^2 \approx \frac{86.3 Q_f^2}{s\,(\mathrm{GeV}^2)}\ \mathrm{nb}. \tag{14.4}$$

## 14.2 *Two–photon Process at $e^+e^-$ Collisions*

When an $e^+$ and $e^-$ collide with energies $E_1$ and $E_2$, they emit $dn_1$ and $dn_2$ virtual photons with energies $\omega_1$ and $\omega_2$ and 4–momenta $q_1$ and $q_2$. In the equivalent photon approximation (EPA) [38], the cross section for the reaction

$$e^+e^- \to e^+e^- X \tag{14.5}$$

is related to the cross section for $\gamma\gamma \to X$ by:

$$d\sigma_{EPA}(s) \equiv d\sigma_{e^+e^-\to e^+e^-X}(s) = dn_1\, dn_2 d\sigma_{\gamma\gamma\to X}(W^2), \tag{14.6}$$

where $s = 4E_1E_2$, $\quad W^2 = 4\omega_1\omega_2$ and

$$dn_i = \frac{\alpha}{\pi}\Big[1 - \frac{\omega_i}{E_i} + \frac{\omega_i^2}{2E_i^2} - \frac{m_e^2\omega_i^2}{(-q_i^2)E_i^2}\Big]\frac{d\omega_i}{\omega_i}\frac{dq_i^2}{q_i^2}.$$

After integration (including that over $q_i^2$ in the region $m_e^2\omega_i^2/E_i(E_i\omega_i) \le -q_i^2 \le (-q^2)_{max}$), the cross section is

$$\sigma_{EPA}(s) = \frac{\alpha^2}{\pi^2}\int_{z_{th}}^{1}\frac{dz}{z}\left[f(z)\left(\ln\frac{(-q^2)_{max}}{m_e^2 z} - 1\right)^2 - \frac{\ln^3 z}{3}\right]\sigma_{\gamma\gamma\to X}(zs), \tag{14.7}$$

where $z = W^2/s$, and

$$f(z) = (1 + \frac{z}{2})^2 \ln\frac{1}{z} - \frac{1}{2}(1-z)(3+z).$$

The value $(-q^2)_{max}$ depends on properties of the produced system $X$. For example, $(-q^2)_{max} \sim m_\rho^2$ for hadron production ($X = h$), and $(-q^2)_{max} \sim M_{ll}^2$ for the lepton pair production ($X = l^+l^-$).
For the production of a resonance of mass $M_R$ and spin $J \neq 1$ one has:

$$\begin{aligned}\sigma_{EPA}(s) &= (2J+1)\frac{8\alpha^2\Gamma(R\to\gamma\gamma)}{M_R^3} \\ &\times \left[f(\frac{M_R^2}{s})(\ln\frac{sM_0^2}{m_e^2M_R^2} - 1)^2 - \frac{1}{3}(\ln\frac{s}{M_R^2})^3\right],\end{aligned} \tag{14.8}$$

where $M_0$ is the mass that enters into the from factor of the $\gamma\gamma \to R$ transition: $M_0 \sim m_\rho$ for $R = \pi^0, \rho^0, \omega, \phi, \ldots$ and $M_0 \sim M_R$ for $R = c\bar{c}$ or $b\bar{b}$ resonances.

## 14.3 *l h Reactions*

The reaction of the lepton hadron deep inelastic scattering (DIS)

$$l(k, m_l) \quad h(P, M) \;\to\; l'(k', m_{l'}) \quad X, \tag{14.9}$$

is described by the following invariant kinematic variables (the 4–momenta and masses of the particles are denoted in the parentheses) [18]:

$q = k - k'$ is four–momentum transferred by exchanged particle ($\gamma$, $Z$, or $W^\pm$) to the target,

$\nu = \frac{q \cdot P}{M} = E - E'$ is the lepton's energy loss in the lab frame, $E$ and $E'$ are the initial and final lepton energies in the lab,

$Q^2 = -q^2 = 2(EE' - \vec{k} \cdot \vec{k}') - m_l^2 - m_{l'}^2$, if $EE' \sin^2(\vartheta/2) \gg m_l^2,\ m_{l'}^2$, then $Q^2 \approx 4EE' \sin^2(\vartheta/2)$, where $\vartheta$ is the lepton's scattering angle in the lab,

$x = \frac{Q^2}{2M\nu} = \frac{Q^2}{2q \cdot P}$, in the parton model, $x$ is the fraction of the target hadron's momentum carried by the struck quark,

$y = \frac{q \cdot P}{k \cdot P} = \frac{\nu}{E}$, is the fraction of the lepton's energy lost in the lab,

$W^2 = (P + q)^2 = M^2 + 2M\nu - Q^2$, is the mass squared of the system recoiling against the lepton,

$s = (P + k)^2 = M^2 + \frac{Q^2}{xy}$.

The differential cross section of the reaction (14.9) as a function of the different variables is given by

$$\frac{d^2\sigma}{dx dy} = \nu(s - M^2)\frac{d^2\sigma}{d\nu dQ^2} = \frac{2\pi M\nu}{E'}\frac{d^2\sigma}{d\Omega_{lab} dE'} = x(s - M^2)\frac{d^2\sigma}{dx dQ^2}.$$

Parity conserving neutral current process, $l^{\pm}h \rightarrow l^{\pm}X$, can be written in terms of two structure functions $F_1^{NC}(x,Q^2)$ and $F_2^{NC}(x,Q^2)$:

$$\begin{aligned} \frac{d^2\sigma}{dxdy} &= \frac{4\pi\alpha^2(s-M^2)}{Q^4} \\ &\times \Big[(1-y)F_2^{NC} + y^2xF_1^{NC} - \frac{M^2}{(s-M^2)}xyF_2^{NC}\Big]. \end{aligned} \tag{14.10}$$

Parity violating charged current processes, $lh \rightarrow \nu X$ and $\nu h \rightarrow lX$, can be written in terms of three structure functions $F_1^{CC}(x,Q^2)$, $F_2^{CC}(x,Q^2)$, and $F_3^{CC}(x,Q^2)$:

$$\begin{aligned} \frac{d^2\sigma}{dxdy} &= \frac{G_F^2(s-M^2)}{2\pi}\frac{M_W^4}{(Q^2+M_W^2)^2} \\ &\times \Big\{[(1-y-\frac{M^2xy}{(s-M^2)}]F_2^{CC} + y^2xF_1^{CC} \pm (y-\frac{y^2}{2})xF_3^{CC}\Big\}, \end{aligned} \tag{14.11}$$

where the last term is positive for $l^-$ and $\nu$ reactions and negative for $l^+$ and $\bar{\nu}$ reaction.

## 14.4 *Cross Sections in the Parton Model*

In the *parton model* framework the reaction

$$h_1 \quad h_2 \quad \rightarrow \quad C \quad X, \tag{14.12}$$

where $C$ is a particle (or group of the particles) with large mass (invariant mass) or with high $p_{\rm T}$ can be considered as a result of the hard interaction of the one $i$–parton from $h_1$ hadron with $j$–parton from $h_2$ hadron. Then the cross section of the reaction (14.12) can be written as follows:

$$\sigma(h_1h_2 \rightarrow CX) = \sum_{ij}\int f_i^{h_1}(x_1,Q^2)f_j^{h_2}(x_2,Q^2)\hat{\sigma}(ij \rightarrow C)dx_1dx_2, \tag{14.13}$$

where sum is performed over all partons, participating in the subprocess $ij \rightarrow C$; $f_i^h(x,Q^2)$ is *parton distribution* in $h$–hadron; $Q$ is a typical momentum transfer in partonic process $ij \rightarrow C$ and $\hat{\sigma}$ is partonic cross section.

## 14.5 *Vector Boson Polarization Vectors*

Let us consider a vector boson with mass $m$ and 4–momentum $p^\mu$ ($p^2 = m^2$). Three polarization vectors of this boson can expressed in terms of $p^\mu$,

$$p^\mu = (E, p_x, p_y, p_z), \ p_\mathrm{T} = \sqrt{p_x^2 + p_y^2}$$

as folows [39]:

$$\left.\begin{array}{l} \varepsilon^\mu(p, \lambda = 1) = \frac{1}{|\vec{p}|p_\mathrm{T}}(0,\ p_x p_z,\ p_y p_z,\ -p_\mathrm{T}^2), \\ \varepsilon^\mu(p, \lambda = 2) = \frac{1}{p_\mathrm{T}}(0,\ -p_y,\ p_x, 0), \\ \varepsilon^\mu(p, \lambda = 3) = \frac{E}{m|\vec{p}|}(\frac{\vec{p}^{\,2}}{E},\ p_x,\ p_y,\ p_z). \end{array}\right\} \tag{14.14}$$

It is easy to verify that

$$p^\mu \varepsilon_\mu(p, \lambda) = 0, \quad \varepsilon^\mu(p, \lambda)\varepsilon_\mu(p, \lambda') = -\delta^{\lambda\lambda'}. \tag{14.15}$$

For $p_\mathrm{T} = 0$ (i.e. $p^\mu = (E, 0, 0, p)$) these polarization vectors can be chosen as follows:

$$\left.\begin{array}{l} \varepsilon^\mu(p, \lambda = 1) = (0,\ 0,\ 1,\ 0), \\ \varepsilon^\mu(p, \lambda = 2) = (0,\ 1,\ 0,\ 0), \\ \varepsilon^\mu(p, \lambda = 3) = \frac{1}{m}(p,\ 0,\ 0,\ E). \end{array}\right\} \tag{14.16}$$

Massless vector boson has only two polarization states, $\lambda = 1$ and 2, on its mass-shell.

In the axial gauge for the polarization vectors of the gluon $\epsilon_g^\mu$ there appears an additional condition (see Subsection 10.3):

$$\epsilon_g^\mu(p, \lambda) n^\mu = 0,$$

where $n$ is axial gauge fixing vector.

For this case polarization vectors $\epsilon_g^\mu(p, \lambda = 1, 2)$ can be chosen as follows:

$$\epsilon_g^\mu(p, \lambda) = \varepsilon^\mu(p, \lambda) - \frac{\varepsilon(p, \lambda) \cdot n}{p \cdot n} p^\mu, \tag{14.17}$$

where $\varepsilon^\mu(p, \lambda)$ are given in (14.14) or (14.16).

• **Two Photons (Gluons) System**

For the system of two photons (gluons) with momenta $p_1$ and $p_2$ the polarization vectors $\varepsilon^{\mu}_{1(2)}$ can be written in the explicitly covariant form:

$$\varepsilon^{\mu}_{i}(\pm) = \frac{1}{\sqrt{2\Delta_3}}[(p_1p_2)q^{\mu} - (qp_2)p_1^{\mu} - (qp_1)p_2^{\mu} \pm i\varepsilon^{\mu\nu\alpha\beta}q^{\nu}p_{1\alpha}p_{2\beta}]. \quad (14.18)$$

where sign $+(-)$ corresponds to positive (negative) helicity, $q$ is any arbitrary vector, which is independent on $p_1$ and $p_2$ (it may be a momentum of some particle), and

$$\Delta_3 = \delta^{qp_1p_2}_{qp_1p_2} = (p_1p_2)(2\,(qp_1)(qp_2) - q^2(p_1p_2)).$$

These vectors were considered also in Subsection 6.6.
Projectors on various combinations of the helicity states look as follows:

$$\frac{1}{2}\,(\varepsilon_1^{\mu}(+)\varepsilon_2^{\nu}(-) + \varepsilon_1^{\mu}(-)\varepsilon_2^{\nu}(+)) = \frac{1}{2(p_1p_2)}(p_1^{\nu}\,p_2^{\mu} - (p_1p_2)\,g^{\mu\nu}),$$
$$\frac{1}{2}\,(\varepsilon_1^{\mu}(+)\varepsilon_2^{\nu}(-) - \varepsilon_1^{\mu}(-)\varepsilon_2^{\nu}(+)) = -\frac{i}{2\,(p_1p_2)}\varepsilon^{p_1p_2\mu\nu},$$
$$\frac{1}{2}\,(\varepsilon_1^{\mu}(+)\varepsilon_2^{\nu}(+) + \varepsilon_1^{\mu}(-)\varepsilon_2^{\nu}(-)) = \frac{1}{2\Delta_3}\{2[(p_1p_2)(qp_1)(qp_2)g^{\mu\nu}$$
$$+q^{\mu}q^{\nu}(p_1p_2)^2 - (p_1p_2)((qp_1)p_2^{\mu}q^{\nu} + (qp_2)p_1^{\nu}q^{\mu})]$$
$$+q^2(p_1p_2)(p_1^{\nu}p_2^{\mu} - (p_1p_2)g^{\mu\nu})\},$$
$$\frac{1}{2}\,(\varepsilon_1^{\mu}(+)\varepsilon_2^{\nu}(+) - \varepsilon_1^{\mu}(-)\varepsilon^{\nu}u_2(-)) =$$
$$\frac{i}{2\Delta_3}\{((p_1p_2)q^{\mu} - (qp_1)p_2^{\mu})\varepsilon^{\nu qp_1p_2} + ((p_1p_2)q^{\nu} - (qp_2)p_1^{\nu})\varepsilon^{\mu qp_1p_2}\},$$
$$= \frac{i\,(p_1p_2)}{2\Delta_3}\,(q^{\mu}\varepsilon^{\nu qp_1p_2} + q^{\nu}\varepsilon^{\mu qp_1p_2}\,(qp_1)\varepsilon^{p_2q\mu\nu} + (qp_2)\,\varepsilon^{p_1q\mu\nu})\,.$$

# 15 MATRIX ELEMENTS

## 15.1 *General Remarks*

In this Section we present the matrix elements squared $|M|^2$ for various processes in the Standard Model. Almost all of these $|M|^2$ were presented in the book by R. Gastmans and Tai Tsun Wu [14]. The symbol $|M|^2$ is used to denote the square of the absolute value of the matrix element $M$ summed over the **initial** and **final** degrees of freedom (polarization and color), but **without** averaging over the **initial** state degrees of freedom.

So, one can use the well–known crossing relations to obtain $\overline{|M|^2}$ for processes differing from each other by repositioning the final and/or initial particles. The averaged over the initial state degrees of freedom matrix element squared $\overline{|M|^2}$ can be obtained from $|M|^2$ by trivial procedure:

$$
\begin{aligned}
e^+e^-,\ e^\pm\gamma,\ \gamma\gamma &: \ \frac{1}{2}\cdot\frac{1}{2}\,(\text{spin}) && \Rightarrow \overline{|M|^2} = \frac{1}{4}|M|^2,\\
q\bar{q},\ qq,\ \bar{q}\bar{q} &: \ \frac{1}{2}\cdot\frac{1}{2}\,(\text{spin})\frac{1}{3}\cdot\frac{1}{3}\,(\text{color}) && \Rightarrow \overline{|M|^2} = \frac{1}{36}|M|^2,\\
gq,\ g\bar{q} &: \ \frac{1}{2}\cdot\frac{1}{2}\,(\text{spin})\frac{1}{8}\cdot\frac{1}{3}\,(\text{color}) && \Rightarrow \overline{|M|^2} = \frac{1}{96}|M|^2,\\
gg &: \ \frac{1}{2}\cdot\frac{1}{2}\,(\text{spin})\frac{1}{8}\cdot\frac{1}{8}\,(\text{color}) && \Rightarrow \overline{|M|^2} = \frac{1}{256}|M|^2.
\end{aligned}
$$

For the $2 \to 2$ processes the differential cross section is related to the $\overline{|M|^2}$ as follows:

$$\frac{d\sigma(2 \to 2)}{dt} = \frac{\overline{|M|^2}}{16\pi I^2}, \quad I^2 \approx s^2, \tag{15.1}$$

where $t$ and $I$ are defined in (12.20) and (12.27).

The notations, used through of this Section, are the same as in Section 10:

$e$ is the electric charge of the positron, $\alpha_{QED} \equiv \alpha = \frac{e^2}{4\pi} \approx \frac{1}{137}$,

$Q_f$ is the charge of the quark in units of the positron charge,

$g_s$ is the QCD coupling constant, $\alpha_{QCD} \equiv \alpha_s = \frac{g_s^2}{4\pi}$,

$G_F$ is the Fermi constant.

As in Section 12 for the reaction $2 \to 2$

$$\begin{aligned} a(p_1) + b(p_2) &\to 1(q_1) + 2(q_2) \\ p_1 + p_2 &= q_1 + q_2 \end{aligned}$$

the Lorentz–invariant Mandelstam variables for reaction are given by

$$\begin{aligned} &s = (p_1 + p_2)^2 = (q_1 + q_2)^2, \quad t = (p_1 - q_1)^2 = (p_2 - q_2)^2, \\ &u = (p_1 - q_2)^2 = (p_2 - q_1)^2, \\ &s + t + u = m_a^2 + m_b^2 + m_1^2 + m_2^2. \end{aligned}$$

## 15.2 *Matrix Elements*

### 15.2.1 $e^+e^- \to f\bar{f}$ *(no Z–boson exchange)*

- $e^+e^- \to l^+l^-$ ($l \neq e$, $l = \mu, \tau$).

$$\begin{aligned} |M_e|^2 &= 8e^4 \frac{1}{s^2}[t^2 + u^2 + (m_e^2 + m_f^2)(2s - m_e^2 - m_f^2)], \qquad (15.2) \\ &= 8e^4 \frac{t^2 + u^2}{s^2}, \quad \text{for } m_e = m_f = 0. \end{aligned}$$

- $e^+e^- \to q\bar{q}$

$$|M_q|^2 = 3Q_f^2 |M_e|^2.$$

The detailed description of the process $e^+e^- \to f\bar{f}$ with $Z$–boson exchange is presented in Subsection 14.1.

### 15.2.2 $e^+e^- \to e^+e^-$ *(no Z–boson exchange)*

$$\begin{aligned} |M|^2 &= 8e^4 \Big\{ \frac{1}{s^2}[t^2 + u^2 + 8m^2(s - m^2)] \qquad (15.3) \\ &+ \frac{2}{st}(u - 2m^2)(u - 6m^2) \Big\}, \\ &= 8e^4 \frac{s^4 + t^4 + u^4}{s^2 t^2}, \quad \text{for } m = 0. \end{aligned}$$

### 15.2.3 $e^+e^- \to \gamma\gamma\gamma$

• $e^+(p_1) + e^-(p_2) \to \gamma(k_1) + \gamma(k_2) + \gamma(k_3), \quad m_e = 0.$

$$|M|^2 = 8e^6 \; \frac{\sum\limits_{i=1}^{3} (p_1k_i)(p_2k_i)[(p_1k_i)^2 + (p_2k_i)^2]}{\prod\limits_{i=1}^{3} (p_1k_i)(p_2k_i)}.$$

• $e^+e^- \to \gamma\gamma\gamma, \quad m_e = m \neq 0.$
For the case of $s = (p_{e^+} + p_{e^-})^2 \to 4m^2$, i.e. in the limit

$$p_{e^+} = p_{e^-} = (m, 0),$$

the $|M|^2$ is given by [2]:

$$|M|^2 = 64e^6 \left[ \Big(\frac{m-\omega_1}{\omega_2\omega_3}\Big)^2 + \Big(\frac{m-\omega_2}{\omega_1\omega_3}\Big)^2 + \Big(\frac{m-\omega_3}{\omega_1\omega_2}\Big)^2 \right], \tag{15.4}$$

where $\omega_i$ is $i$–photon energy in cms.

### 15.2.4 $e^+e^- \to l^+l^-\gamma$

$$e^+(p_1) + e^-(p_2) \to l^+(q_1) + l^-(q_2) + \gamma(k), \quad m_e = m_l = 0.$$

Invariants:

$$\begin{aligned} s &= 2(p_1p_2), \quad t = -2(p_1q_1), \quad u = -2(p_1q_2), \\ s' &= 2(q_1q_2), \quad t' = -2(p_2q_2), \quad u' = -2(p_2q_1). \end{aligned} \tag{15.5}$$

• $l \neq e$, for example, $e^+e^- \to \mu^+\mu^-\gamma$

$$|M|^2 = -4e^6(v_p - v_q)^2 \frac{t^2 + t'^2 + u^2 + u'^2}{ss'}. \tag{15.6}$$

• $l = e$, i.e. $e^+e^- \to e^+e^-\gamma$

$$|M|^2 = -4e^6(v_p - v_q)^2 \frac{ss'(s^2 + s'^2) + tt'(t^2 + t'^2) + uu'(u^2 + u'^2)}{ss'tt'}. \tag{15.7}$$

where in (15.6) and (15.7) we use:

$$v_p^\mu \equiv \frac{p_1^\mu}{(p_1k)} - \frac{p_2^\mu}{(p_2k)}, \quad v_q^\mu \equiv \frac{q_1^\mu}{(q_1k)} - \frac{q_2^\mu}{(q_2k)}. \tag{15.8}$$

### 15.2.5 $e^+e^- \to q\bar{q}g$

For this reaction the invariants are the same as in (15.5).

$$|M|^2 = 16e^4Q_f^2g_s^2\frac{t^2+t'^2+u^2+u'^2}{s(q_1k)(q_2k)}. \qquad (15.9)$$

### 15.2.6 $e^+e^- \to q\bar{q}\gamma$

For this reaction the invariants are the same as in (15.5).

$$|M|^2 = -12e^6(v_p+Q_fv_q)^2\frac{t^2+t'^2+u^2+u'^2}{ss'}, \qquad (15.10)$$

where $v_p$ and $v_q$ are defined in (15.8).

### 15.2.7 $gg \to q\bar{q}$, $m_q = m \neq 0$

The final $q\bar{q}$–pair can be in color *singlet* or color *octet* final states.

$$\begin{aligned} |M_{singl}|^2 &= 16g_s^4\chi_0\left[\frac{1}{3}\right], \quad |M_{oct}|^2 = 16g_s^4\chi_0\left[\frac{7}{3}-6\chi_1\right], \\ |M_{tot}|^2 &= |M_{singl}|^2+|M_{oct}|^2 = 16g_s^4\chi_0\left[\frac{8}{3}-6\chi_1\right], \end{aligned} \qquad (15.11)$$

where

$$\begin{aligned} \chi_0 &= \frac{m^2-t}{m^2-u}+\frac{m^2-u}{m^2-t}+4\left(\frac{m}{m^2-t}+\frac{m^2}{m^2-u}\right) \qquad (15.12) \\ &\quad -4\left(\frac{m}{m^2-t}+\frac{m^2}{m^2-u}\right)^2, \\ \chi_1 &= \frac{(m^2-t)(m^2-u)}{s^2}. \qquad (15.13) \end{aligned}$$

For $m_q = 0$,

$$\chi_0 = \frac{t^2+u^2}{ut}, \quad \chi_1 = \frac{ut}{s^2}.$$

### 15.2.8 $\gamma g(\gamma\gamma) \rightarrow f\bar{f}$

- $g\gamma \rightarrow q\bar{q}$

$$|M|^2 = 32 g_s^2 e^2 Q_f^2 \chi_0.$$

- $\gamma\gamma \rightarrow q\bar{q}$

$$|M|^2 = 24 e^4 Q_f^4 \chi_0.$$

- $\gamma\gamma \rightarrow e^+e^-$

$$|M|^2 = 8 e^4 \chi_0.$$

### 15.2.9 $q\bar{q} \rightarrow Q\bar{Q}$, $m_q = 0$, $m_Q = m \neq 0$

$$|M|^2 = 16 g_s^4 \frac{t^2 + u^2 + 2m^2(2s - m^2)}{s^2}. \tag{15.14}$$

### 15.2.10 $qq \rightarrow qq$, $m_q = 0$

$$|M|^2 = 16 g_s^4 \left[ \frac{s^4 + t^4 + u^4}{t^2 u^2} - \frac{8}{3} \frac{s^2}{tu} \right]. \tag{15.15}$$

### 15.2.11 $q\bar{q} \rightarrow q\bar{q}$, $m_q = 0$

$$|M|^2 = 16 g_s^4 \frac{1}{s^2 t^2} [s^4 + t^4 + u^4 - \frac{8}{3} stu^2]. \tag{15.16}$$

### 15.2.12 $gg \rightarrow gg$

$$|M|^2 = 288 g_s^4 \frac{(s^4 + t^4 + u^4)(s^2 + t^2 + u^2)}{s^2 t^2 u^2}. \tag{15.17}$$

### 15.2.13 $f_1 \bar{f}_2 \rightarrow W^* \rightarrow f_3 \bar{f}_4$

$$f_1(p_1) + \bar{f}_2(p_2) \rightarrow f_3(p_3) + \bar{f}_4(p_4), \quad m_{1,2,3,4} \neq 0.$$

$$|M|^2 = C 128 G_F^2 M_W^4 \frac{(p_1 p_4)(p_2 p_3)}{(s - M_W^2)^2 + \Gamma_W^2 M_W^2}, \tag{15.18}$$

where $C = 1$ for $l^- \bar{\nu} \rightarrow l'^- \bar{\nu}'$, $C = 3$ for $l^- \bar{\nu} \rightarrow q\bar{q}'(q\bar{q}' \rightarrow l^- \bar{\nu})$, and $C = 9$ for $q_1 \bar{q}_2 \rightarrow q_3 \bar{q}_4$, $M_W$ and $\Gamma_W$ are the mass and total width of the $W$–boson.

### 15.2.14 $l^- \bar{\nu} \to d\bar{u}g$

$$l^-(p_1) + \bar{\nu}(p_2) \to d(p_3) + \bar{u}(p_4) + g(k), \quad m_{d,u} \neq 0.$$

$$\begin{aligned}
|M|^2 &= 256 G_F^2 M_W^4 g_s^2 \frac{A_1 - A_2 - A_3}{(s - M_W^2)^2 + \Gamma_W^2 M_W^2}, \qquad (15.19)\\
A_1 &= \frac{1}{(kp_3)(kp_4)} \Big\{ s\,[(p_1p_4)^2 + (p_2p_3)^2] \\
&- (m_u^2 + m_d^2)\,[\frac{s}{2}((p_1p_4) + (p_2p_3)) - (p_1p_3)(p_2p_3) - (p_1p_4)(p_2p_4)] \Big\}, \\
A_2 &= \frac{2m_u^2}{(kp_4)^2}(p_2p_3)\,[(p_1k) + (p_1p_4)], \\
A_3 &= \frac{2m_d^2}{(kp_3)^2}(p_1p_4)\,[(p_2k) + (p_2p_3)].
\end{aligned}$$

# 16 MISCELLANEA

## 16.1 *Miscellanea*

• Let us consider the recursion $A_n = aA_{n-1} + bA_{n-2}$ for given $A_0$ and $A_1$. Then

$$\begin{aligned}
&A_n = \alpha z_1^n + \beta z_2^n, \\
&z_{1,2} = \frac{a}{2}[1 \pm \sqrt{1 + 4b/a^2}], \quad \alpha = \frac{A_1 - z_2 A_0}{z_1 - z_2}, \quad \beta = \frac{z_1 A_0 - A_1}{z_1 - z_2}.
\end{aligned}$$

• Various representations of the *Dirac* $\delta$–function:

$$\begin{aligned}
&\delta(x) \equiv \frac{1}{(2\pi)} \int_{-\infty}^{\infty} e^{ixt} dt, \\
&\delta(x, \alpha) = \frac{\alpha}{\pi(\alpha^2 x^2 + 1)}, \ \alpha \to \infty; \quad \delta(x, \beta) = \frac{\beta}{\pi(x^2 + \beta^2)}, \ \beta \to 0, \\
&\delta(x, \alpha) = \frac{\alpha}{\sqrt{\pi}} e^{-\alpha^2 x^2}, \quad \alpha \to \infty, \quad \delta(x, \alpha) = \frac{\alpha}{\pi} \frac{\sin(\alpha x)}{(\alpha x)}, \ \alpha \to \infty, \\
&\frac{1}{x \pm i\varepsilon} = \mathcal{P}\frac{1}{x} \mp i\pi\delta(x).
\end{aligned}$$

• *Step*–functions $\Theta(x)$ and $\varepsilon(x)$

$$\begin{aligned}
\Theta(x) &\equiv \frac{1}{(2\pi i)} \int_{-\infty}^{\infty} \frac{e^{ixt}}{t - i\varepsilon} dt \ = \left\{ \begin{array}{ll} 1, & x > 0 \\ 0, & x < 0 \end{array} \right. \\
\varepsilon(x) &\equiv \frac{1}{(i\pi)} \mathcal{P} \int_{-\infty}^{\infty} \frac{e^{ixt}}{t} dt \ = \left\{ \begin{array}{ll} 1, & x > 0 \\ -1, & x < 0 \end{array} \right.
\end{aligned}$$

•

$$\begin{aligned}
\frac{1}{(a - i\varepsilon)^k} &= \frac{i^k}{\Gamma(k)} \int_0^{\infty} e^{i\alpha(-a + i\varepsilon)} \alpha^{k-1} d\alpha, \ k \geq 0, \\
\int_0^{\infty} (e^{ita} - e^{itb}) \frac{dt}{t} \ &= \ \ln\left(\frac{b + i\varepsilon}{a + i\varepsilon}\right).
\end{aligned}$$

## 16.2 *Properties of Operators*

The various properties of the operators can be found, for example, in [5, 40, 41]. Let $f(A)$ be any function from the operator (matrix) $A$, which can expanded into series with respect to operators (matrices) $A^n$:

$$f(A) = \sum_{n=0}^{\infty} c_n A^n.$$

• Let $\xi$ be a parameter, then:

$$\begin{aligned} e^{\xi A} e^{-\xi A} &= 1, \qquad e^{\xi A} A e^{-\xi A} = A, \qquad e^{\xi A} A^n e^{-\xi A} = A^n, \\ e^{\xi A} f(A) e^{-\xi A} &= f(A). \end{aligned}$$

• Let $A$ and $B$ be noncommuting operators, $\xi$ and $n$ be parameters ($n$ integer). Then:

$$\begin{aligned} & e^{\xi A} B^n e^{-\xi A} = (e^{\xi A} B e^{-\xi A})^n, \\ & e^{\xi A} F(B) e^{-\xi A} = F(e^{\xi A} B e^{-\xi A}), \\ & e^{\xi A} B e^{-\xi A} = B + \xi [A, B] + \frac{\xi^2}{2!}[A, [A, B]] + \frac{\xi^3}{3!}[A, [A, [A, B]]] + \cdots \end{aligned}$$

• Let $A$ be an operator and there exists the inverse operator $A^{-1}$. Then for any integer $n$ :

$$AB^n A^{-1} = (ABA^{-1})^n, \qquad Af(B)A^{-1} = f(ABA^{-1}).$$

• Let $A(x)$ be an operator, depending on the scalar variable $x$, then

$$\begin{aligned} \frac{dA^{-1}(x)}{dx} &= -A^{-1}(x) \frac{dA(x)}{dx} A^{-1}(x), \\ \frac{de^{A(x)}}{dx} &= \int_0^1 e^{(1-t)A(t)} \frac{dA(t)}{dt} e^{tA(t)} dt. \end{aligned}$$

## 16.3 *The Baker-Campbell-Hausdorff Formula*

Let $A$ and $B$ be non–commuting operators, then :

$$e^A \, e^B = e^{\sum_{i=1}^{\infty} Z_i}, \tag{16.1}$$

where

$$Z_1 = A + B; \tag{16.2}$$

$$Z_2 = \frac{1}{2}[A,B]; \tag{16.3}$$

$$Z_3 = \frac{1}{12}[A,[A,B]] + \frac{1}{12}[[A,B],B]; \tag{16.4}$$

$$Z_4 = \frac{1}{48}\Big[A,[[A,B],B]\Big] + \frac{1}{48}\Big[[A,[A,B]],B\Big]; \tag{16.5}$$

$$\begin{aligned} Z_5 &= \frac{1}{120}\Big[\Big[A,[[A,B],B]\Big],B\Big] + \frac{1}{120}\Big[A,\Big[[A,[A,B]],B\Big]\Big] \\ &- \frac{1}{360}\Big[A,\Big[[[A,B],B],B\Big]\Big] - \frac{1}{360}\Big[\Big[A,[A,[A,B]]\Big],B\Big] \\ &- \frac{1}{720}\Big[A,\Big[A,[A,[A,B]]\Big]\Big] - \frac{1}{720}\Big[\Big[[[A,B],B],B\Big],B\Big], \ldots \end{aligned} \tag{16.6}$$

The other terms can be evaluated from the relation (see [5, 41]):

$$\sum_{k=0}^{\infty} \frac{1}{(k+1)!} [\![ Z^k, Z' ]\!] = A + \sum_{j=0}^{\infty} \xi^j \frac{[\![ A^j, B ]\!]}{j!}, \tag{16.7}$$

where $e^Z = e^{\xi A} e^{\xi B}$; $Z = \sum_{n=1}^{\infty} \xi^n Z_n$; $Z' = \sum_{n=1}^{\infty} n \xi^{n-1} Z_n$. The repeated commutator bracket is defined as follows

$$[\![ A^0, B ]\!] = B, \quad [\![ A^{n+1}, B ]\!] = \Big[A, [\![ A^n, B ]\!]\Big].$$

Since relation (16.7) must be satisfied identically in $\xi$, one can equate the coefficients of $\xi^j$ on the two sides of this relation. In particular, $j = 0, 1, 2, 3, 4$ gives (16.2, 16.3, 16.4, 16.5, 16.6), respectively.

# REFERENCES

[1] Bogolyubov N.N. and Shirkov D.V., *Introduction to the Theory of Quantized Fields* (Chichester: Wiley–Interscience, 3rd edn, 1980).

[2] Berestetskii V.B., Lifshits E.M., and Pitaevskii L.P., *Quantum Electrodynamics* (Moscow: Nauka, 1980) (in Russian).

[3] Itzykson C. and Zuber J.-B., *Quantum Field Theory* (New York: McGraw-Hill, 1980).

[4] Okun' L.B., *Leptons and Quarks* (Moscow: Nauka, 1981) (in Russian).

[5] Veltman M., *Diagrammatica – The Path to Feynman Rules* (Cambridge: Cambridge University Press, 1994).

[6] Dubrovin B.A., Novikov S.P., and Fomenko A.T., in *Modern Geometry*, (Moscow: Nauka, 1979) (in Russian).

[7] Cartan E., *Theory of Spinors* (Paris: Hermann, 1966).

[8] Green M.B., Schwarz J., and Witten E., in *Superstring Theory* (Cambridge: Cambridge University Press, 1987).

[9] van Oldenborgh G. and Vermaseren J.A.M, *Z. Phys.* **C46**, 425 (1991).

[10] Penrose R. and Rindler B., *Spinors and Space-time* (Cambridge: Cambridge University Press, 1986).

[11] de Causmaecker P., Gastmans R., Troost W., and Wu T.T., *Phys. Lett.* **105B**, 215 (1981).

[12] Bogush A.A. et al., in *Proc. of the XI Workshop on Problems on HEP and FT* (Moscow: Nauka, 1989).

[13] Rogalyov R.N., *Teor. Mat. Fiz.* **101**, 384 (1994).

[14] Gastmans R. and Tai Tsun Wu, *The Ubiquitous Photon, Helicity Method for QED and QCD* (Oxford: Clarendon Press, 1990).

[15] Dittner P., *Comm. Math. Phys.* **22**, 238 (1971).

[16] Cvitanovic P., *Phys.Rev.* **D14**, 1536 (1976).

[17] Aoki K., Hioki Z., Kawabe R., Konumu M, and Muta T., *Supplement of the Progress of Theoretical Physics* **73**, 1 (1982).

[18] Montanet L. et al. (Review of Particle Properties), *Phys. Rev.* **D50**, Part I, (1994).

[19] Faddeev L.D., in *Methods in Field Theory* (Proc. Les Houches Session XXVIII, 1975), ed. R. Balian and J. Zinn–Justin.

[20] Liebbrandt G., *Noncovariant Gauges* (Singapore: World Scientific, 1994).

[21] Burnel A., *Phys. Rev.* **D40**, 1221 (1990).

[22] Bassetto A., *Phys. Rev.* **D31**, 2012 (1985).

[23] t'Hooft G. and Veltman M., *Nucl. Phys.* **B44**, 189 (1972).

[24] Bateman Manuscript Project, *Higher Transcendental Functions*, ed. A. Erdélyi (New York: McGraw-Hill, 1953) vol. 1.

[25] Prudnikov A.P., Brychkov Yu.A., and Marichev O.I., *Integrals and Series (Supplementary Chapters)* (Moscow: Nauka, 1986) (in Russian).

[26] Lewin L., *Dilogarithms and Associated Functions* (London: Macdonals, 1958).

[27] Byckling E. and Kajantie K., *Particle Kinematics* (New York: J.W. & Sons, 1973).

[28] Sjöstrand T., *PYTHIA 5.7 and JETSET 7.4, Physics and Manual*, CERN–TH.7112/93.

[29] Bjorken J.D. and Brodsky S.J., *Phys. Rev.* **D1**, 1416 (1970).

[30] MARK J Collaboration, Barber D.P. et al., *Phys. Rev. Lett.* **43**, 830 (1979).

[31] Brandt S., Peyrow Ch., Sosnowski R., and Wroblewski A., *Phys. Lett.* **12**, 57 (1964);
Fahri E., *Phys. Rev. Lett.* **39**, 1587 (1977).

[32] Fox G.C. and Wolfram S., *Nucl. Phys.* **B149**, 413 (1979).

[33] *Physics at LEP*, Ed. J. Ellis and R. Peccei, CERN-86-02, vol. 1, 1986.

[34] Ellis J., Gaillard M.K., and Nanopolous D.V., *Nucl. Phys.* **B106**, 292 (1976);
Gaillard M.K., *Commun. Nucl. Part. Phys.* **8**, 31 (1978);
Ansel'm A.A., Uraltsev N.G., and Khoze V.A., *Sov. Phys. Uspekhi* **28**, 113 (1985);
Resnick L., Sundaresan M.K., and Watson P.J.S., *Phys. Rev.* **D8**, 172 (1979);
Wilczek F., *Phys. Rev. Lett.* **39**, 1304 (1977).

[35] Georgi H., Glashow S.L., Macahek M.E., and Nanopolous D.V., *Phys. Rev. Lett.* **40**, 692 (1978).

[36] Marciano W.J. and Sirlin A., *Phys. Rev. Lett.* **61**, 1815 (1988).

[37] Bigi I.I. et al., *Phys. Lett.* **181B**, 157 (1986).

[38] Budnev V.M., Ginzburg I.F., Medelin G.V., and Serbo V.G., *Phys. Reports* **15C**, 181 (1975);
Brodsky S., Kinoshita T., and Terazawa H., *Phys. Rev.* **D4**, 1532 (1971).

[39] Hagiwara K. and Zeppenfeld D., *Nucl. Phys.* **B274**, 1 (1986).

[40] Louisell W. H., *Radiation and Noise in Quantum Electronics* (New York: McGraw-Hill, 1964, Chap. 3).

[41] Wilcox R. M., *J. Mat. Phys.* **8**, 962 (1967).